\documentclass[fleqn,usenatbib]{mnras}

\usepackage[T1]{fontenc}

\DeclareRobustCommand{\VAN}[3]{#2}
\let\VANthebibliography\thebibliography
\def\thebibliography{\DeclareRobustCommand{\VAN}[3]{##3}\VANthebibliography}

\usepackage{svg}
\usepackage{threeparttable}
\usepackage{tablefootnote}
\usepackage{multirow}
\usepackage{booktabs}

\newcommand{\angstrom}{\textup{\AA}}

\usepackage{graphicx}
\usepackage{amsmath}

\usepackage{epstopdf}
\AppendGraphicsExtensions{.pdf}

\title[Neon  abundances in Seyfert~2 nuclei]{Chemical abundances in  Seyfert galaxies -- VII.
Direct abundance determination of neon based on optical and infrared emission lines}

\author[Mark Armah et al.]{
Mark Armah,$^{1, 2}$\thanks{E-mail: armah@ufrgs.br}
O. L. Dors,$^1$\thanks{E-mail: olidors@univap.br}
C. P. Aydar,$^3$ 
M. V. Cardaci,$^{4,5}$
G. F. H\"agele,$^{4,5}$
\newauthor{Anna Feltre,$^{6,7}$
R. Riffel,$^{2}$
R. A. Riffel$^{8}$ and
A. C. Krabbe$^{1}$}
\\
$^1$ Universidade do Vale do Paraíba. Av. Shishima Hifumi, 2911, CEP: 12244-000, São José dos Campos, SP, Brazil\\
$^2$ Departamento de Astronomia, Universidade Federal do Rio Grande do Sul, Av. Bento Gonçalves 9500, Porto Alegre, RS, Brazil \\
$^3$ Universidade de São Paulo. R. do Matão, 1226, CEP: 05508-090, São Paulo, SP, Brazil \\
$^{4}$ Instituto de Astrof\'{i}sica de La Plata (CONICET-UNLP), Argentina \\
$^{5}$ Facultad de Ciencias Astron\'{o}micas y Geof\'{i}sicas, Universidad Nacional de La Plata, Paseo del Bosque s/n, 1900 La Plata, Argentina \\
$^{6}$ SISSA, Via Bonomea 265, 34136 Trieste, Italy \\
$^{7}$ INAF - Osservatorio di Astrofisica e Scienza dello Spazio di Bologna, Via P. Gobetti 93/3, 40129 Bologna, Italy\\
$^{8}$Departamento de F\'isica, Centro de Ci\^encias Naturais e Exatas, Universidade Federal de Santa Maria, 97105-900, Santa Maria, RS, Brazil\\}

\date{Accepted XXX. Received YYY; in original form ZZZ}

\pubyear{2021}

\begin{document}
\label{firstpage}
\pagerange{\pageref{firstpage}--\pageref{lastpage}}
\maketitle

\begin{abstract}
 For the first time, neon abundance has been derived in the narrow line
region from a sample of Seyfert~2 nuclei. In view of this,
we compiled from the literature  fluxes of optical and infrared (IR) narrow emission lines for 35 Seyfert 2 nuclei in the local universe ($z \:\la \:0.06$).  The relative intensities of emission lines were used to derive the ionic and total neon and oxygen abundances through electron temperature estimations ($T_{\rm e}$-method).  For the neon, abundance estimates were obtained by using  both $T_{\rm e}$-method and IR-method. 
Based on photoionization model results, we found a lower electron temperature [$t_{\rm e}({\rm \ion{Ne}{iii}})$] for the gas phase where the Ne$^{2+}$ is located in comparison  with $t_{3}$
for the O$^{2+}$ ion.
We find that the  differences (D) between Ne$^{2+}$/H$^{+}$ ionic abundances calculated
from IR-method and  $T_{\mathrm{e}}{-}$method (assuming $t_{3}$ in the Ne$^{2+}$/H$^{+}$ derivation) are similar to the derivations in star-forming regions (SFs) and they are reduced
by a mean factor of $\sim3$ when $t_{\rm e}({\rm \ion{Ne}{iii}})$ is considered. We propose a semi-empirical  Ionization Correction Factor (ICF)  for the neon, based on
[\ion{Ne}{ii}]12.81$\micron$, [\ion{Ne}{iii}]15.56$\micron$ and  oxygen ionic
abundance ratios. We find that the  average Ne/H abundance for the Seyfert 2s sample is nearly 2 times higher than similar estimate for SFs. Finally, for the very high metallicity regime (i.e. [$\rm 12+log(O/H)\: \ga \: 8.80$]) an increase in Ne/O with O/H is found, which likely indicates secondary stellar production for the neon.


\end{abstract}




\section{Introduction}

Active Galactic Nuclei (AGNs) present prominent emission lines in their spectra, whose relative intensities can be used to estimate the metallicity and elemental abundances of heavy elements (O, N, Ne, S, etc.) in the gas-phase of these objects. This feature, together with 
their high luminosity, has made these objects essential to chemical evolution studies of galaxies along the Hubble time.

The first chemical abundance study in AGNs,  based on direct determination of the electron temperature (hereafter $T_{\rm e}$-method), was carried out 
by \citet{1975ApJ...197..535O} for the radio galaxy 3C 405 (Cygnus~A). These authors derived the oxygen 
abundance relative to hydrogen (O/H) (among other elements) in the order of 12+log(O/H)=8.60. Most AGN studies have mainly been carried out following this aforementioned pioneering work. In fact, \citet{ferland1983shock} compared observational optical emission line 
ratios to photoionization model predictions built with 
the first version of the {\sc Cloudy} code \citep{ferland80} and found
that the metallicities of Seyfert~2s are in the range 
$0.1 \: \la \:  (Z/{\rm Z_{\odot}})\:  \la \: 1$, 
but the nitrogen abundance 
can have a relative enhancement in relation with oxygen, which is analogous to
\ion{H}{ii} regions. Thereafter, several studies have relied on the estimations of metallicities
for AGNs using photoionization models in the local universe  
(e.g. \citealt{grazina84, ferland86,  storchi1998chemical, groves2006emission, feltre2016nuclear, castro2017metallicity,  2019MNRAS.489.2652P, thomas2019metallicity, 2020MNRAS.492.5675C})
as well as at  high redshifts (e.g. \citealt{nagao2006, matsuoka09, 
matsuoka18,  nakajima18, dors2014metallicity, 
dors2018hemical, 2019MNRAS.486.5853D, mignoli19, guo20}).

Since oxygen  presents prominent emission lines (e.g. [\ion{O}{ii}]$\lambda3726$ {\AA}+$\lambda3729$ {\AA}, [\ion{O}{iii}]$\lambda4959,\lambda5007$ {\AA}) in the optical spectrum of gaseous nebulae, emitted by its most abundant ions ($\rm O^{+}$, $\rm O^{2+}$), 
 it has usually been used as metallicity tracer for the gas phase of 
line-emitting objects (e.g. \citealt{2021MNRAS.507..466D, Kewley2019}). Specifically,
\citet{2020MNRAS.496.2191F} and \citet{2020MNRAS.496.3209D} found that the $\rm O^{3+}$ abundance in AGNs is not larger than 20 per cent of the total O/H abundance. Therefore,
the oxygen abundance determination has usually been derived through only the lines emitted by 
$\rm O^{+}$ and $\rm O^{2+}$ ions (for a review, see \citealt{dors2020chemical}). On the other hand, the abundances of other heavy elements, e.g. N, Ne, S, etc., are poorly known
in AGNs. \cite{dors17}  presented the first quantitative nitrogen abundance determination
for a sample of 44 Seyfert 2 nuclei in the local universe ($z \: \la \: 0.1$; see also \citealt{contini2017abundance, 2019MNRAS.489.2652P}). 
Moreover, for the sulphur, only qualitative abundance determinations, based on the comparison between observational line ratios and photoionization model predictions were
performed by \citet{thaisa90}.

In galaxy evolution and stellar nucleosynthesis, the knowledge of neon abundance is relevant, especially among the heavy elements.  Neon is one of the noble gas elements which does not combine with
itself or with other chemical species 
in the formation of molecules and dust grains (e.g. \citealt{1987ASSL..134..533J, 1993MNRAS.261..306H, 1994ApJ...430..650S, 2004ASPC..309..393S, 2013MNRAS.432.2112B}). Therefore, the depletion of  abundance in the gas phase process is not expected in neon, conversely to such occurrence in the oxygen (e.g. \citealt{izotov2006chemical, 2007MNRAS.376..353P})
and refractory elements (e.g. Mg, Si, Fe; \citealt{1992ApJ...389..305O, 1992RMxAA..24..155P, 1993ApJ...418..760P, 1995ApJ...449L..77G, 2010ApJ...724..791P})
 trapped in dust. Regarding chemical galaxy evolution, the chemical abundances of neon and oxygen are expected to  closely trace each other \citep{2006ApJ...637..741C} due to the fact that both elements are produced
 in stars  more massive than 10 ${\rm M_{\odot}}$ (e.g. \citealt{1995ApJS..101..181W})
 and a constant Ne/O abundance ratio over a wide range of O/H abundance is supposed to be found. However, chemical abundance studies of star-forming regions have revealed a slight dependence of Ne/O on O/H (see \citealt{dors2013optical} and references therein), which brings forth a worthwhile means of cross-checking the stellar nucleosynthesis theory.

The study of neon and oxygen abundances in AGNs can also provide important
insights into the origin of heavy elements, mainly in the regime of high metallicities. Unfortunately, neon abundance in relation with hydrogen (Ne/H) in AGNs is rarely found in the literature, and only a few AGNs relative abundance of Ne with other heavy elements has been derived. For instance, \citet{1970ApJ...161..811N}, by using the $T_{\rm e}$-method, derived the Fe/Ne abundance ratio for NGC\,4151 to be 0.11. 
 Assuming a solar abundance ratio (Fe/Ne)$_{\odot}$ = 0.282 \citep{2001AIPC..598...23H} shows that AGNs have an overabundance
 of Fe, as found by \citet{1993ApJ...418...11H}. The above result indicates a very high and oversolar neon abundance. Furthermore, based on a comparison between observational soft X-ray spectrum of the Narrow Line Quasar PG1404+226 ($z$ = 0.098) and photoionization model predictions, 
 \citet{1999A&A...350..816U}
found that the abundances of 
oxygen and neon are about 0.2 and 4 times the solar value, respectively, which  again implies an overabundance value of neon. 
However, \citet{2010ApJ...721.1835S}, who compared AGNs spectra  from the Sloan Digital Sky Survey (SDSS, \citealt{york00}) in the redshift range of $0.2 \: < \: z \: < \:0.35$ with photoionization model predictions, found  no significant difference for the Fe/Ne abundance ratios in the sample of objects considered.

 With the foregoing in mind, the primary aim of this study is to derive neon abundance in relation with hydrogen (Ne/H) 
 in the NLRs of relatively large sample of Seyfert 2s at low redshift ($z \: \la \: 0.06)$ and compare the results with previous SFs findings. In view of this, we compiled from the literature narrow optical and infrared  (IR) emission  line  intensities for Seyfert~2 galaxies.
 These observational data will be
 used to derive the twice ionized $(\rm{ Ne^{2+}/H^{+}})$ and total (Ne/H and O/H) abundances through the $T_{\rm e}$-method and
 infrared emission lines method. Also, it is possible to derive the singly ionized neon abundance relative to hydrogen ($\rm{ Ne^{+}/H^{+}}$) through infrared emission lines. The use of $T_{\rm e}$-method, based on direct
 temperature determinations via optical lines (for a review see \citealt{2017PASP..129h2001P, 2017PASP..129d3001P}) can lead to non-negligible deviations in the estimations of abundances, in the sense that abundances can
 be underestimated in relation with other distinct methods. Therefore, we also consider
 Ne/H abundances derived from IR lines, which have weak dependence on the electron temperature
 \citep{simpson1975infrared}.

 This paper is organized as follows. In Section~\ref{data}, we
describe the observational data. In Sect.~\ref{meth}
details to the calculations of the ionic abundances from $T_{\rm e}$-method and infrared emission lines are presented. Descriptions of the calculation of the  total neon and oxygen abundances
are given in Sect.~\ref{icf}. The results and discussions are presented in Sect.~\ref{resdisc}  and Sect.~\ref{disc}, respectively. Finally, we summarize our conclusions in Sect.~\ref{conc}.

\section{Observational data}
\label{data}

In order to study the $\mathrm{Ne^{+2}/H^+}$ abundances we take into account the fact that \citet{dors2013optical}  found $\rm{ Ne^{2+}/H^{+}}$ abundance estimations in \ion{H}{ii} regions  using the $T_{\rm e}$-method are lower by a factor of $\sim 4$ than those obtained through infrared lines, which are less sensitive to electron temperature. Therefore, we consider AGN emission lines measured in both wavelength ranges in order to ascertain if similar discrepancy exists in AGNs. The caveat here is that it is unknown which among the $T_{\rm e}$- and IR-methods provides more accurate abundance values.

We limit the abundance determinations to the NLRs of Seyfert~2s because shocks with low velocity (lower than 400 km s$^{-1}$, \citealt{contini2017abundance, 2020MNRAS.tmp.3501D}) are expected in this type of object and the $T_{\rm e}$-method was adapted for this object type in a previous paper \citep{2020MNRAS.496.3209D}. The selection criteria for the objects are:
\begin{enumerate}
\item The objects must be classified as Seyfert 2 nuclei.
\item They must have the narrow optical [\ion{O}{ii}]$\lambda3726$ + $\lambda$3729, [\ion{Ne}{iii}]$\lambda$3869,  [\ion{O}{iii}]$\lambda$4363, H$\beta$, [\ion{O}{iii}]$\lambda$5007, H$\alpha$ and [\ion{S}{ii}]$\lambda$6716, $\lambda$6731 emission-line fluxes measured.
\item The  [\ion{Ne}{iii}]$\lambda$15.56 {\micron}
emission-line fluxes should also be measured. 
The flux of the [\ion{Ne}{ii}]12.81$\micron$ line is considered in the compiled data when it is available in the original work.
\end{enumerate}

The optical data consists of emission lines observed in the wavelength range of $3500 \: < \: \lambda $({\AA})$ \: < \: 8000$  obtained with  
  \begin{enumerate}
  \item low-dispersion spectra ($R\sim 5-10$ {\AA}) using telescopes at the Las Campanas, Anglo-Australian, Lick and European Southern observatories and
  \item Faint Object Spectrograph spectroscopy (FOS) on board the Hubble Space Telescope (HST) at  $3\,500 \: < \: \lambda$ ({\AA})$ \: < \: 7\,000$  ($R\sim 5$ {\AA}). 
\end{enumerate} 

The infrared observational data from near to mid infrared spectroscopic observations were obtained from the following:

 \begin{enumerate}
 \item Photodetector Array Camera and Spectrometer (PACS) instrument on board the European Space Agency (ESA)  Herschel Space Observatory in the short cross-dispersed mode ($R\sim 360$) covering the $JHK-$bands, together with an ancillary data,
 \item $Spitzer-$Infrared Spectrometer (IRS) spectroscopic survey consisting of the short wavelengths ranging from 9.9 to {19.6} {\micron} covered by the Short-High (SH) module  in the high spectral resolution  mode ($R \sim 600$) and from 8 to {2.4} {\micron},
 \item medium resolution ($R \sim 1\,500$) of Infrared Space Observatory Short Wavelength Spectrometer (ISO-SWS) $2.4 - 45$ {\micron} spectra, 
 
  \item the cooled grating spectrometer 4 (CGS4) on United Kingdom Infrared Telescope (UKIRT) for both high-resolution ($R = 1\,260$) and low-resolution ($R = 345$ and 425) $JHK-$band spectra of {4} {\micron} spectroscopy with ISAAC 
 at the European Southern Observatory Very Large Telescope array (ESO VLT),
 \item Infrared  array spectrometer - IRSPEC ($R \sim 1\,500$) at the ESO 3.6~m telescope,
 \item  Infrared Spectrometer And Array Camera Long Wavelength (ISAAC-LW) 
 medium resolution spectroscopy mode covering a range of 3.93 to {4.17} {\micron} at spectral resolving power $\sim2500$,  
 \item Anglo-Australian Telescope NIR integral field spectroscopy of moderate resolution ($R\sim 2\,100$) $KL-$bands spectra from 2.17 - 2.43{\micron}, and
 
 \item $H$ (1.5 - 1.8 {\micron}) and
$K$ (2.0 - 2.4 {\micron}) bands corresponding to  the spectral resolutions
$\lambda/\Delta\lambda = 1\,700$ and $\lambda/\Delta \lambda = 1\,570$, respectively, using the Keck NIR longslit spectrograph NIRSPEC.
  \end{enumerate}

  In Tables~${\color{blue}\text{A1}}$ and ${\color{blue}\text{A2}}$, available as supplementary material, the objects identifications,
the optical and infrared observational emission line fluxes and the bibliographic references to the origins of the data are listed.

The observational data considered in this work consist of a heterogeneous sample, thus, the data were obtained with different instrumentation and observational techniques with different apertures, which could potentially introduce some uncertainties in the derivation of physical properties for the objects under consideration. \citet{dors2013optical} analysed these 
effects on oxygen abundance determinations in star-forming regions and did not find 
 any bias in the physical conditions of the objects obtained by using a similarly heterogeneous samples. A  particular concern in AGN studies is the emission contribution from  \ion{H}{ii} regions to the measured AGN flux,
 which can be located at few parsecs away from the AGN (e.g. \citealt{1993A&A...277..397B, 2002AJ....123.1381E, 2007MNRAS.382..251D, 2008A&A...482...59D, 2008AJ....135..479B, 2009MNRAS.393..783R, 2013MNRAS.432..810H, 2015MNRAS.451.3173A, 2016MNRAS.461.4192R, 2019MNRAS.482.4437D}).
 In fact, \citet{thomas18}, who considered a large sample of AGNs data
 taken from the Sloan Digital Sky Survey (SDSS, \citealt{york00}), found that, even for strong AGNs $\mathrm{[with  \: log([\ion{O}{iii}]\lambda5007/H\beta) \: \ga \:0.9]}$,  $\sim30$ per cent 
of the  Balmer line fluxes, on average, is emitted by \ion{H}{ii} regions (see also \citealt{2014MNRAS.439.3835D,  2014MNRAS.444.3961D, 2018MNRAS.479.4907D, 2019MNRAS.487.4153D, 2018ApJ...861L...2T}).

 \citet{dors2020chemical} investigated the aperture effect on oxygen abundance and electron density estimates in a sample of local AGNs ($z \: \la \: 0.4)$ using SDSS spectra \citep{york00}, which were obtained with a fixed diameter of the fibres of $\sim3$ arcsec.  Since \ion{H}{ii} regions generally have lower O/H abundances (e.g. \citealt{2003ApJ...591..801K, groves2006emission}) and electron density values (e.g. \citealt{2000A&A...357..621C, dors2014metallicity})  than similar estimations in AGNs, 
if the emission from \ion{H}{ii} regions contributes significantly to the observed emission-line fluxes in AGNs, a decrease in O/H and $N_{\rm e}$ with increasing redshift (a greater number of \ion{H}{ii} regions were included within the fiber at larger distances) would be expected. However, no correlation between O/H or $N_{\rm e}$ with the redshift was derived by these authors, indicating negligible  aperture effects on the AGN parameter estimations. 
Moreover, \citet{kewley05} found that the derived metallicity can vary by a factor of only 0.14 dex from the value obtained when the  fluxes are measured with the assumption of an aperture capturing less than 20 per cent of the total  emissions  from a galaxy. The object of our sample with the highest redshift is Cygnus~A ($z=0.05607$), where assuming a spatially flat  cosmology with the present-day
Hubble parameter being $H_{0}$\,=\,71 $ \rm km\:s^{-1} Mpc^{-1}$, the total present matter density $\Omega_{m}=0.270$, the total present vacuum density $\Omega_{\rm vac}=0.730$  \citep{2006PASP..118.1711W} and a typical aperture of 2 arcsec,  corresponds to   a  physical scale   in the center  of  this galaxy of about 2 kpc, i.e. the emissions are mainly from an AGN.

\begin{figure}
\includegraphics[width=\columnwidth]{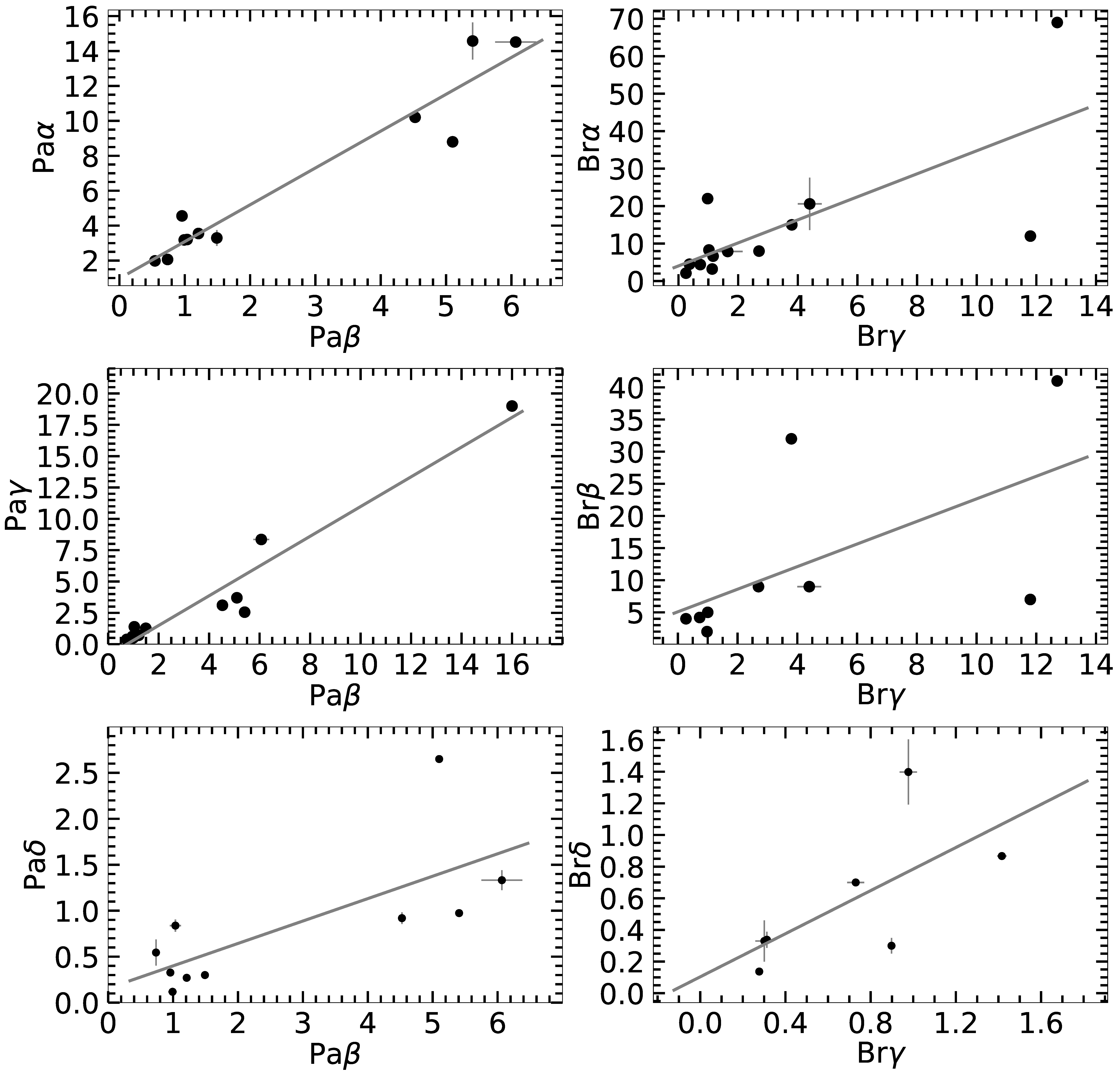}
\caption{ Strong IR emission line (in units of 10$^{-14}$ erg cm$^{-2}$ s$^{-1}$) ratios for each spectra in our sample for which the Paschen and Brackett series  were detected. Left-column: plots for the measured Pa$\beta$ line flux versus measured Pa$\alpha$, Pa$\gamma$ and Pa$\delta$. Right-column: plots for the measured Br$\gamma$ line flux versus other measured Brackett series (Br$\alpha$, Br$\beta$ and Br$\delta$).}
\label{series}
\end{figure}

 In Figure~\ref{series} we show plots for Pa$\beta$ and Br$\gamma$ versus all other strong Paschen and Brackett line series samples of our IR observational  data compiled from the literature (i.e. taken from \citealt{moorwood1988, 1989ApJ...337..230K, oliva1994, 1995ApJ...444...97G, 1997ApJS..108..449G, veilleux1997infrared, 1999ApJ...522..139V, 1999MNRAS.308..431B, gilli2000, 2000MNRAS.316....1W, lutz2002, 2002MNRAS.331..154R, sturm2002, 2005MNRAS.364.1041R, riffel2006spectral, 2009ApJ...694.1379R, 2017MNRAS.464.1783O}) which have strong Paschen and Brackett line series corresponding to Pa$\beta$ and Br$\gamma$.
It can be seen from Fig.~\ref{series}   that the IR \ion{H}{i} line fluxes have a clear linear correlation for all the strong emission lines with somewhat scattering in the points, which is probably due to the heterogeneity of the sample.
  However, we show (see below) that abundance estimates assuming different IR \ion{H}{i} lines have a very good agreement with each other, therefore, 
 this observed scatter in  Fig.~\ref{series} has no effect on our abundance
 results. Since most of the observed hydrogen recombination lines have been reddening-corrected by the original authors and the infrared line series show little deviations with Pa$\beta$ and Br$\gamma$, we considered them in our abundance estimations without further consideration for extinction correction.

 \subsection{Diagnostic diagrams}

\begin{figure*}
\includegraphics[width=2.1\columnwidth]{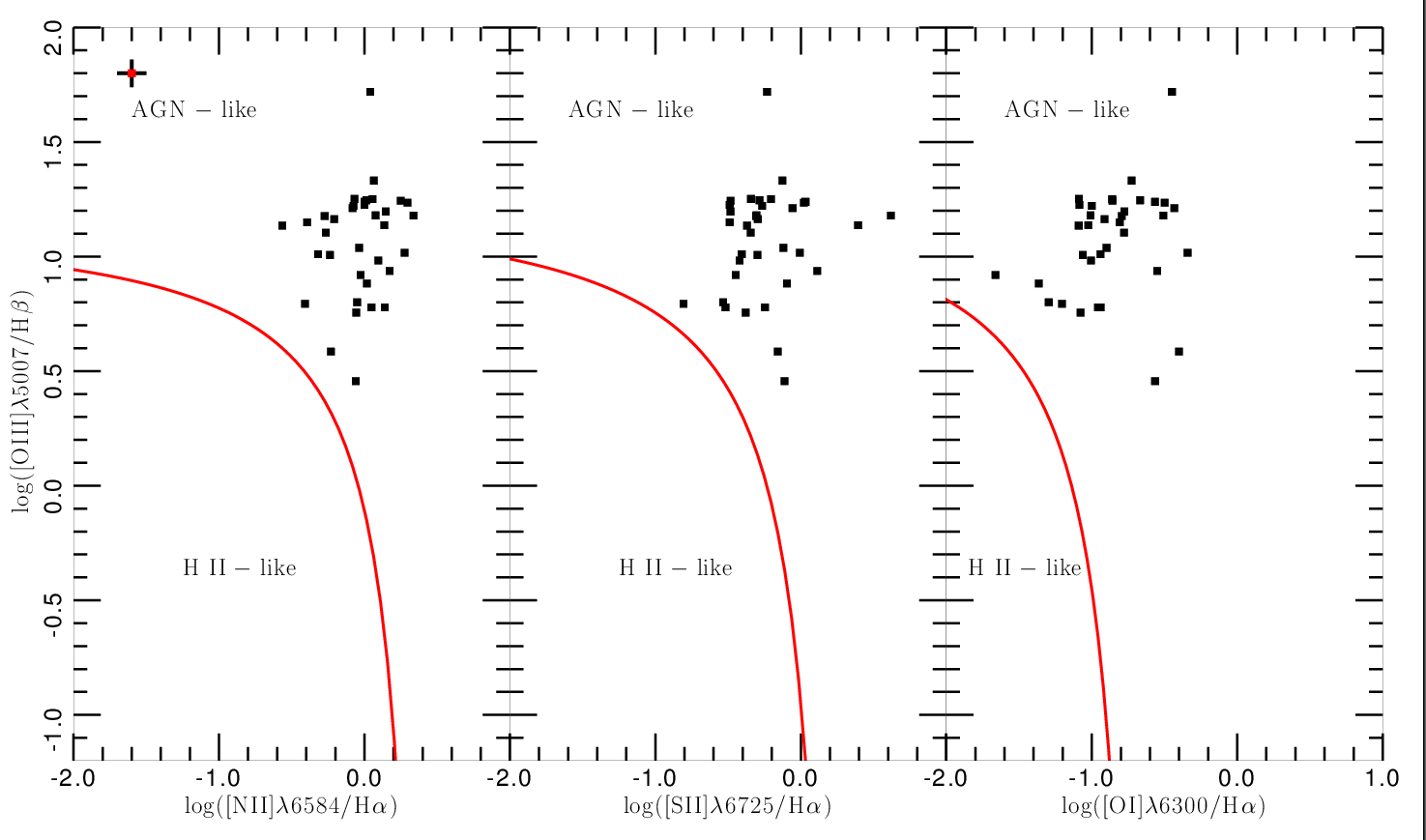}
\caption{Diagnostic diagrams for emission-line ratios of log([\ion{O}{iii}]$\lambda5007$/H$\beta$) versus log([\ion{N}{ii}]$\lambda6584$/H$\alpha$),   log([\ion{S}{ii}]($\lambda$6725)/H$\alpha$),  
and log([\ion{O}{i}]($\lambda$6300)/H$\alpha$).
[\ion{S}{ii}]$\lambda$6725 represents the
sum of the lines [\ion{S}{ii}]$\lambda$6717 and [\ion{S}{ii}]$\lambda$6731.
Points represent objects of our sample (see Sect.~\ref{data}).
Red lines, by  \citet{kewley2001theoretical} and represented by Eqs.~\ref{eqn20}, \ref{eqn21} and \ref{eqn21k}, separate objects ionized by massive stars   from those ionized by gas shocks and/or AGN-like objects, as indicated.
Error bar, in left panel, represents the typical uncertainty (0.1 dex) in emission-line ratio measurements (e.g. \citealt{kraemer1994spectra}).}
\label{bpt_figure}
\end{figure*}

Although the objects in our sample have been classified as AGNs by the authors from which the data were compiled, we produced an additional
test based on standard Baldwin-Phillips-Terlevich (BPT) diagrams \citep{baldwin1981, 1987ApJS...63..295V}. These diagnostic diagrams,
based on optical emission-line ratios,  have been used to distinguish objects whose main ionization mechanisms are massive stars  from those that are mainly ionized by AGNs and/or gas shocks 
(see also \citealt{kewley2001theoretical, 2013ApJ...774L..10K, 2003MNRAS.341...33K, 2013A&A...549A..25P, 2020MNRAS.tmp.3047J}). We adopted  the criteria proposed by \citet{kewley2001theoretical} where all objects with 
 \begin {equation}{\label{eqn20}}
\mathrm{\log([\ion{O}{iii}]\lambda5007/H\beta}) >  \frac{0.61}{[\mathrm{\log([\ion{N}{ii}]\lambda6584/H\alpha)-0.47}]} + 1.19, 
\end {equation}
\begin {equation}{\label{eqn21}}
\mathrm{\log([\ion{O}{iii}]\lambda5007/H\beta}) >  \frac{0.72}{[\mathrm{\log([\ion{S}{ii}]\lambda6725/H\alpha)  - 0.32}]} + 1.30
\end {equation} 
and
\begin {equation}{\label{eqn21k}}
\mathrm{\log([\ion{O}{iii}]\lambda5007/H\beta}) >  \frac{0.73}{[\mathrm{\log([\ion{O}{i}]\lambda6300/H\alpha)  + 0.59}]} + 1.33
\end {equation}

\noindent have AGNs as their main ionization mechanism. The [\ion{S}{ii}]$\lambda$6725 line above represents the sum 
of the  [\ion{S}{ii}]$\lambda$6717 and [\ion{S}{ii}]$\lambda$6731 lines.
Fig.~\ref{bpt_figure} further confirms that the ionizing sources of the objects in our sample are indeed AGNs.  Additionally, it can be seen
that the objects cover a large range of ionization degree and
metallicity since a wide range of [\ion{O}{iii}]/H$\beta$ 
and [\ion{N}{ii}]/H$\alpha$ are observed   (e.g. \citealt{feltre2016nuclear, 2021arXiv210807812A}).

\subsection {Reddening correction} \label {Redness correction}

We performed the reddening correction to the optical emission lines by considering the expression
\begin {equation}{\label{eqn2.1}}
\frac {I (\lambda)} {I\mathrm{(H\beta)}} = \frac{F(\lambda)} {F\mathrm{(H\beta)}} \times 10^{c\mathrm{(H\beta)} [f(\lambda) -f\mathrm{(H\beta)}]},
\end {equation}

\noindent where $I(\lambda)$ is the intensity (reddening corrected) of the emission line at a given wavelength $\lambda$, $F(\lambda)$ is the observed flux of the emission line, $ f (\lambda)$ is the adopted reddening curve normalized to H$\beta$ and $c$(H$\beta$) is the interstellar extinction coefficient.
The extinction coefficient of interest is normally calculated using the H$\alpha$/H$\beta$ line ratio and comparing it with its theoretical value. For instance, the estimation by \citet{hummer1987recombination} for a temperature of 10\,000 K and an electron density of 100 cm$^{- 3}$ produces $I$(H$\alpha$/H$\beta$) = 2.86. Following the parameterization by \citet{whitford1958law}, adopting the reddening curve  by \citet{miller1972recombination} and using a consensual assumed value of the ratio of total to selective absorption in the optical $V$ band, with $R_V = 3.1$, for the diffuse interstellar medium \citep[see][and references therein]{cardelli1989relationship, 1994ApJ...422..158O, 1999PASP..111...63F}, we deduce the logarithmic extinction at H$\beta$ expressed as

\begin {equation}{\label{eqn2.2}}
c(\mathrm{H\beta}) = 3.10 \times \left [\log \left(  \frac{F\mathrm{(H\alpha)}} {F\mathrm{(H\beta)}}\right) - \log \left(  \frac {I \mathrm{(H\alpha)}} {I\mathrm{(H\beta)}} \right) \right].
\end {equation}


\noindent The optical extinction curves in the extragalactic environment are closely parallel to those of the Milky Way in all related extinction studies, with $R_V$ values comparable to the canonical value of 3.1 (e.g. \citealt{2004AJ....128.2144M}; \citealt{2008MNRAS.390..969F}).

 In comparison with the Case B recombination value of 2.86, \citet{halpern1982xray}
 and  \citet{halpern1983ionization}, adopting photoionization models, found that $I$(H$\alpha$/H$\beta$) is close to 3.10 in AGNs with high and low ionization degree.
This contradicts \citet{heckman1980optical} preposition of an anomalously high Balmer decrement in these objects. Therefore, $I$(H$\alpha$/H$\beta$) = 2.86  and   3.10 intrinsic ratios are usually considered to be estimations for \ion{H}{ii} regions and AGNs, respectively \citep{ferland1983shock,  1982PASP...94..891G, gaskell1984reddening, gaskell1984theoretical, 1987ApJS...63..295V, 1988LNP...307...79W}. Particularly, in AGNs, there is a large transition zone, or partly ionized region, in which H$^0$ coexists with H$^+$ and free electrons. In this zone, collisional excitation is also important in addition to recombination  \citep{ferland1983shock, halpern1983ionization}. The main effect of collisional excitation is to enhance H$\alpha$. The higher Balmer lines are less affected because of their large re-excitation energies and smaller excitation cross-sections.

 In order to check the H$\alpha$/H$\beta$ value
assumed in our reddening correction, we consider results from AGNs photoionization models  built with the {\sc Cloudy} code \citep{2013RMxAA..49..137F}
by \citet{2020MNRAS.492.5675C}.  This grid of models assume
a wide range of nebular parameters, i.e. a Spectral Energy Distribution
with power law $\alpha_{ox}=-0.8, -1.1, -1.4$, oxygen abundances in the range
of $\rm 8.0 \: \lid \: 12+\log(O/H) \: \lid \: 9.0$, logarithm of the ionization parameter ($U)$
in the range of $ -4.0 \: \lid \: \log U \: \lid \: -0.5$, and 
electron density $N_{\rm e}$= 100, 500 and 3000 $\rm cm^{-3}$. 
The AGN parameters considered in the models built by \citet{2020MNRAS.492.5675C}
cover practically all the range of physical properties  of a large sample of  Seyfert~2 nuclei. We excluded models with $\alpha_{ox}=-1.4$ and $\log U=-4.0$ because they predicted
emission lines which are not consistent with observational data (see \citealt{2019MNRAS.489.2652P, 2020MNRAS.492.5675C}). 
The \citet{2020MNRAS.492.5675C} models assume constant electron density along the
nebular radius while spatially resolved studies of AGNs have found
$N_{\rm e}$ variations from $\sim100$ to $\sim3000 \: \rm cm^{-3}$ along the NLRs of some AGNs (e.g.  \citealt{2018MNRAS.476.2760F, 2018A&A...618A...6K, 2019A&A...622A.146M, 2021MNRAS.501L..54R}). 
However, to provide a simple test for this problem, \citet{2019MNRAS.486.5853D} built AGN photoionization models assuming a profile density similar to observational estimations  by \citet{2018ApJ...856...46R} in the Seyfert~2 Mrk~573, i.e.
with a central electron density peak at $\rm \sim 3000 \: cm^{-3}$ and a decrease in this value following a shallow power law.  \citet{2019MNRAS.486.5853D} found that predicted emission lines assuming this density profile are very similar to those considering a constant electron density along the AGN radius. Therefore, $N_{\rm e}$
variations have almost a negligible effect on  emission lines and abundances
predicted by photoionization models, at least for the low electron density limit ($\la 10^{4} \: \rm cm^{-3}$).

\begin{figure}
\includegraphics[angle=-90, width=1.0\columnwidth]{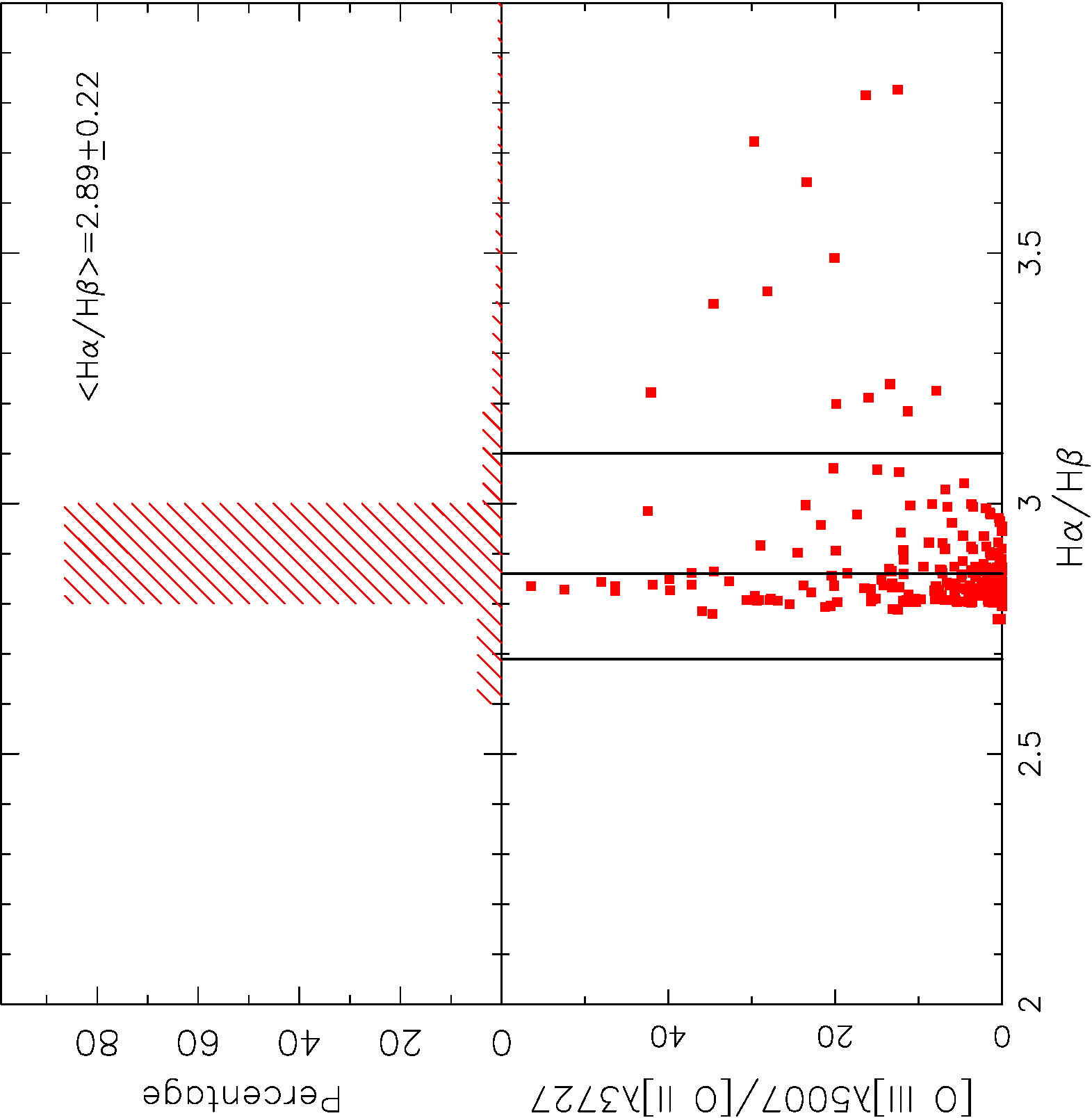}
\caption{ Bottom panel:  [\ion{O}{iii}]$\lambda5007$/[\ion{O}{ii}]$\lambda5007$
versus  H$\alpha$/H$\beta$. 
The red points represent AGN photoionization model predictions 
taken from \citet{2020MNRAS.492.5675C}.
The black lines represent the theoretical values from \citet{1995MNRAS.272...41S} for temperatures of 5000 K (3.10), 10\,000 K (2.86) and 20\,000 K (2.69). Top panel: Distribution
of H$\alpha$/H$\beta$ values.
The average for the H$\alpha$/H$\beta$  values is indicated.}
\label{fhahb}
\end{figure}

In Fig.~\ref{fhahb}, bottom panel, we show the  model predictions of  the gas ionization degree parameterized by the [\ion{O}{iii}]$\lambda5007$/[\ion{O}{ii}]$\lambda3727$
line ratio versus H$\alpha$/H$\beta$ ratio.
In this figure, the expected
values for the H$\alpha$/H$\beta$  ratio, considering the theoretical values by \citet{1995MNRAS.272...41S} for different temperature values of  
 5000 K, 10\,000 K and 20\,000 K are indicated by the solid black lines representing 3.10, 2.86 and 2.69, respectively. We notice that most of the models  ($\sim$95\,\%) predict H$\alpha$/H$\beta$ values in the range from 2.69 to 3.10. In  Fig.~\ref{fhahb}, top panel, the distribution
of H$\alpha$/H$\beta$ values predicted by the models is shown, where it can be seen that,
 the most representative value is
around (H$\alpha$/H$\beta$)=2.90 with an average value 
of $2.89\pm0.22$.
Therefore, for the intrinsic ratio of Seyfert 2 nuclei, we adopted the theoretical value given by  (H$\alpha$/H$\beta$)=2.86.

 The wavelength dependence in the optical domain, $f(\lambda)$, is the reddening value for the line derived from the curve given by \citet{whitford1958law}, which is defined such that $f(\infty) = -1 $ and $f({\rm H}\beta) = 0 $. An analytical expression for the estimation of $f(\lambda)$ following the proposal by
 \citet{1976ApJS...31..517K} was used in the derivation of the extinction curve,  which is given by:
\begin {equation}{\label{eqn19}}
f(\lambda) = 2.5659 \lambda^2 - 4.8545 \lambda + 1.7545,
\end {equation}

\noindent  with $\lambda$ in units of micrometers within the range $\mathrm{0.35 \: \la \: \lambda (\mu m)  \: \la \: 0.70}$. 
 We adopted negligible intrinsic reddening when the apparent Balmer decrement from the original work is $\la 2.86$ and the extinction correction constant indicates a value of zero as shown in  Table~${\color{blue}\text{A3}}$, thus, $c$(H$\beta$) = 0.0

Since several measurements for the  emission lines compiled from the literature do not have their errors  listed in the original papers where the data were compiled, we adopted a typical  error of  10\% for strong emission-lines (e.g. [\ion{O}{iii}]$\lambda$5007) and 20\,\% for weak emission lines, in the case of
[\ion{O}{iii}]$\lambda$4363 (see, for instance, \citealt{kraemer1994spectra, 2008MNRAS.383..209H}). These errors were used to calculate the uncertainties
in the derived values of electron temperatures (in order of 800 K) and abundances
(in order of 0.1 dex).

\section{Ionic abundance determinations} 
\label{meth}

The main goal of this work is to estimate the total abundance of neon in relation with hydrogen (Ne/H) for the NLRs of a sample of Seyfert 2 objects.
This can be carried out by using optical and infrared emission lines. 
In view of this, for optical lines, we adopted the $T_{\rm e}$-method used by \citet{2020MNRAS.496.3209D} to be 
applied in the studies of Seyfert~2 nuclei. Regarding abundances obtained through infrared lines, the methodology
proposed by \citet{dors2013optical} was adopted in this work, which is based on
\citet{1970ApJ...159..833P}, \citet{simpson1975infrared}, \citet{forster2001near}, and \citet{2002A&A...391.1081V}.

The observational  optical data 
 compiled from the literature make it possible to estimate only the 
$\rm Ne^{2+}/H^{+}$ ionic abundance. Therefore, to obtain Ne/H, Ionization
Correction Factors (ICFs) based on the neon infrared lines
and photoionization models were employed. The O/H abundance for our sample
was calculated by using only the $T_{\rm e}$-method,  since there are few emission lines of this element measured  in the infrared wavelength range under consideration 
 (e.g. \citealt{2010AJ....139.1553V}). In the succeeding subsections, each of the adopted methodology in the estimations of the Ne and O abundances is succinctly described.

 \subsection{\texorpdfstring{$T_{\rm e}$}--method}
 \label{meth1}
 
In determining ionic abundances  using the $T_{\rm e}$-method, basically,
 it is necessary  to obtain  measurements of the intensity of the emission lines emitted by the ions under consideration 
 and the representative values of the electron temperature ($T_{\rm e}$) and electron density ($N_{\rm e}$) of the gas region where these ions are located \citep{1989agna.book.....O}.

 \citet{2008MNRAS.383..209H} obtained, from the task {\scshape temden} of {\scshape iraf}\footnote{Image Reduction and   Analysis Facility ({\scshape iraf}) is distributed by the National Optical Astronomy Observatories,
which are operated by the Association of Universities for Research in
Astronomy, Inc., under cooperative agreement with the National Science
Foundation.} \citep{1987JRASC..81..195D, 1995PASP..107..896S},  functions to determine electron temperatures. It is considered that  $t_3$  and  $t_2$ are the electron temperatures (in units of $10^{4}$ K) for the electrons that are exciting the
 O$^{2+}$ and O$^{+}$ ions in the high and low ionization zones, respectively.
The expressions obtained by  \citet{2008MNRAS.383..209H} were 
assumed to calculate
$t_3$, $\rm O^{2+}/H^{+}$, $\rm O^{+}/H^{+}$ and $\rm Ne^{2+}/H^{+}$.
 
 First, for each object in our sample, the   electron temperature in the high-ionization zone ($t_3$) was obtained by using the expression 
\begin{equation}{\label{eqn3.5}}
t_3 = 0.8254 - 0.0002415\times {R_{\rm O3}} + \frac{47.77}{{R_{\rm O3}}}, 
\end{equation}

\noindent where ${R_{\rm O3}}$ = [\ion{O}{iii}]($\lambda4959${\AA} + $\lambda5007${\AA})/$\lambda4363${\AA}. This relation is valid for a range ${30\la R_{\rm O3}\la 700}$, corresponding to a temperature range of $0.7 \: \la \: t_{3} \: \la \: 2.3$. 
Only objects with $t_{3}$ in this range of values were
considered in the present analysis.

Consequently, it is not possible to explicitly estimate $t_2$ in the AGN spectra of our sample where the [\ion{O}{ii}]$\lambda$3727{\AA}/$\lambda$7325{\AA} emission line ratio can not be measured. Thus, we assumed the $t_{2}$-$t_{3}$ relation derived by 
\citet{2020MNRAS.496.3209D}  from a grid of photoionization models  built using the {\sc Cloudy} code \citep{2013RMxAA..49..137F}.
The theoretical resulting  relation is given by
\begin{equation}
\label{t2t3new}
t_{2}=({\rm a} \times t_{3}^{3})+({\rm b} \times t_{3}^{2})+({\rm c} \times t_{3})+{\rm d},
\end{equation}
where $\rm a=0.17$, $\rm b=-1.07$, $\rm c=2.07$ and $\rm d=-0.33$.\\
 
 The $t_{3}$-$t_{2}$ relation for SFs has issue of some uncertainties due to 
the large scatter between these temperatures, around 900 K (e.g. \citealt{2020ApJ...893...96B})
 and this problem has been addressed in
several chemical abundance studies. For example, \citet{2008MNRAS.383..209H}
pointed out that the scatter in the $t_{3}$-$t_{2}$ relation can be due to electron density effects because the [\ion{O}{ii}] temperature is somewhat dependent on the density. \citet{2017MNRAS.465.1384C} pointed out that the [\ion{O}{iii}]$\lambda$4363 can be contaminated by the neighboring [\ion{Fe}{ii}]$\lambda$4360 line, mainly for objects
with high metallicity ($\rm 12+\log(O/H) \: \ga \:8.4$).
Recently, \citet{2020MNRAS.497..672A} showed that the $t_{3}$ and $t_{\rm e}$(\ion{N}{ii}) ($\approx t_{2})$ relation depends on  the  ionization degree of the gas phase in SFs. In fact, the model results adopted to derive Eq.~\ref{t2t3new}  by \citet{2020MNRAS.496.3209D} also present a large scatter, which is not explained by electron density effects.
However, \citet{2021MNRAS.501L..54R} showed that the relation
given by Eq.~\ref{t2t3new} is in consonance with direct estimates of
temperature for AGNs when no clear gas outflows are present in these objects. Unfortunately, direct estimates of  $t_{2}$ for AGNs are rare in the literature and we stress that the use of Eq.~\ref{t2t3new} can yield somewhat
uncertain in $\rm O^{+}$ temperature estimates.

We make use of the relations to estimate ionic abundance of the singly and doubly ionized oxygen   originally derived by \citet{1992MNRAS.255..325P} and in its current form given by \citet{2008MNRAS.383..209H} as:

\begin{equation}{\label{eqtempo}}
\begin{split}
    12 + \log \left(\frac{\mathrm{O^{2+}}}{\mathrm{H^+}}\right) = & \log \left(\frac{I(4959\:{\angstrom})+I(5007\:{\angstrom})}{I\mathrm{(H\beta)}}\right) + 6.144  \\
    \\
    & + \frac{1.251}{t_3} - 0.55 \times \log t_3
\end{split}
\end{equation}

\noindent and

\begin{equation}
\begin{split}
    12 + \log \left(\frac{\mathrm{O^{+}}}{\mathrm{H^+}}\right) = &\log \left(\frac{I(3727\:{\angstrom})}{I\mathrm{(H\beta)}}\right) + 5.992 + \frac{1.583}{t_2}\\
    \\
    & - 0.681 \times \log t_2 + \log [1 + 2.3 n_{\mathrm e}] 
\end{split}
\end{equation}
where $n_{\rm e} = 10^{-4} \: \times \: N_{\rm e}$.

The electron density $N_{\rm e}$ for each object was derived 
through the relation of this parameter with the line ratio
[\ion{S}{ii}]$\lambda6717$/[\ion{S}{ii}]$\lambda6731$  by using the {\sc iraf} code \citep{1986SPIE..627..733T, 1987JRASC..81..195D, 1995PASP..107..896S} and assuming the $t_{2}$ value obtained for each object.  We derived electron density values in the
range of $ 300 \: \la \: N_{\rm e} \rm (cm^{-3}) \: \la 3\,500$, with an average value
of $\rm \sim 1\,000 \: cm^{-3}$. 
In \citet{2020MNRAS.496.3209D}, a detailed analysis of the effect of the electron density
on the direct abundance determination was presented and it is not repeated here.
We only point out to the fact that, despite high $N_{\rm e}$ values in order of 13\,000 - 80\,000 $ {\rm cm^{-3}}$ derived when optical lines emitted by ions with higher ionization potential than
the ${\rm S^{+}}$ are used to derive the electron density, e.g. [\ion{Ar}{iv}]$\lambda4711$/$\lambda4740$  line ratio 
(see \citealt{2017MNRAS.471..562C, 2021MNRAS.501L..54R}),
these values are much lower than the  critical densities (e.g. see \citealt{2012MNRAS.427.1266V}) for the optical lines used here. Additionally, the electron density determined from the line ratio 
[\ion{S}{ii}]$\lambda6717$/[\ion{S}{ii}]$\lambda6731$  is much lower than that obtained using auroral and transauroral lines, as well as ionization parameter based approach \citep{2020MNRAS.498.4150D}.

\begin{figure}
\centering
\includegraphics[angle=-90,width=1\columnwidth]{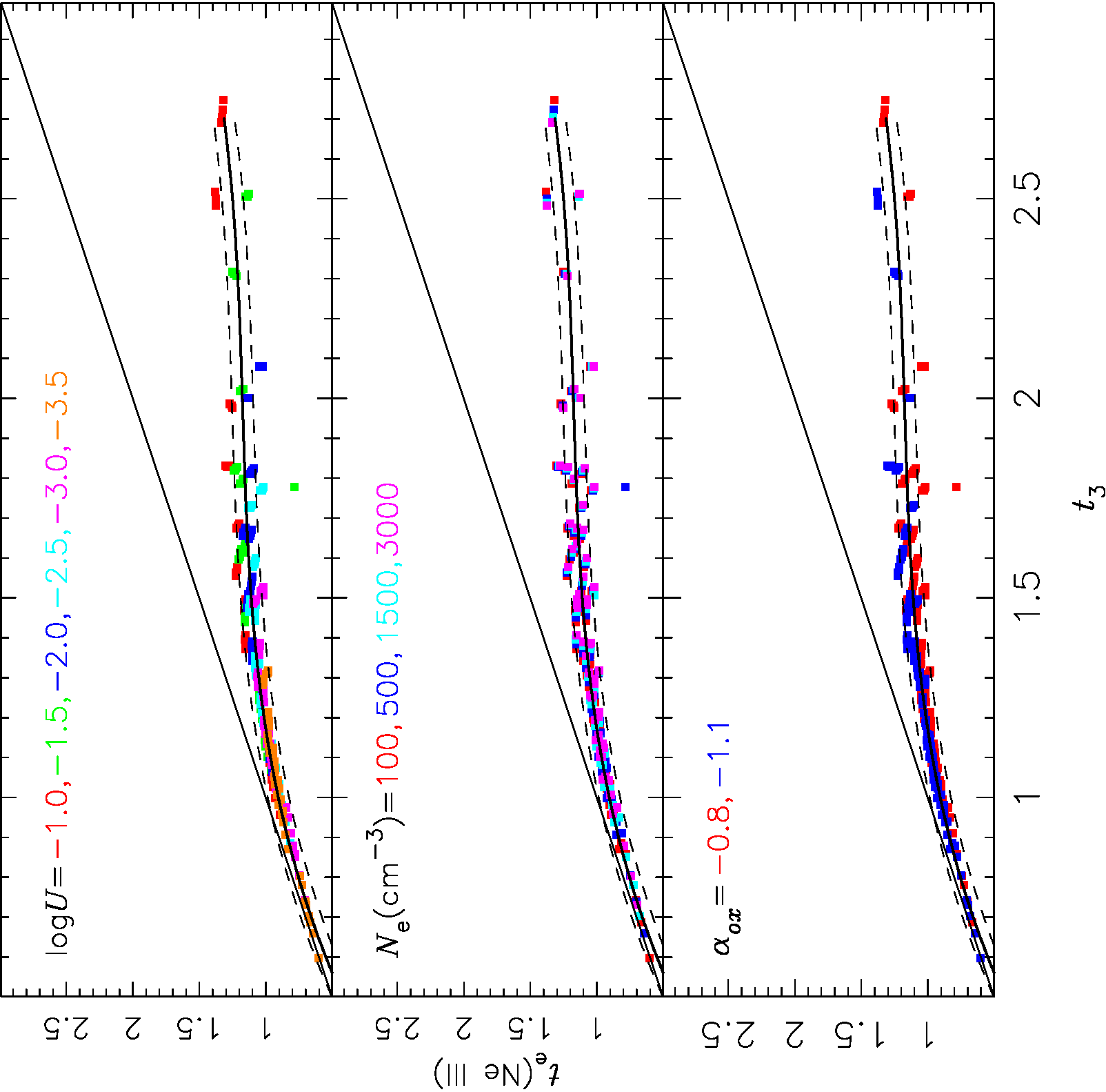}
\caption{Temperature values for $t_{\rm e}({\rm \ion{Ne}{iii}})$ versus  $t_{3}$
predicted by the photoionization models
built by \citet{2020MNRAS.492.5675C}. Temperatures are in units of $10^{4}$ K.
The solid line represents the equality between the
estimates while the continuum curve represents the fitting
to the points given by Eq.~\ref{eqtnet3}. Dashed curves represent the deviations
of  Eq.~\ref{eqtnet3} by $\pm 800$ K, i.e. typical  uncertainties  derived in direct electron temperature estimations
(e.g. \citealt{2003ApJ...591..801K, 2008MNRAS.383..209H}). 
In each panel, points with different colours represent photoionization models assuming different nebular parameters, as indicated.}
\label{figt3ne3}
\end{figure}

Generally, in \ion{H}{ii} regions studies, the same temperature $t_{3}$ is used to estimate the ${\rm O^{2+}}$ and ${\rm Ne^{2+}}$ ionic abundances.
 This approach is based on
the similarity of ${\rm O^{+}}$ and ${\rm Ne^{+}}$ ionization potentials, i.e. 35.12 and 40.96 eV, respectively,
which indicates that both ions coexist in  similar nebular regions.
The same assumption is considered for ${\rm O^{+}}$ and ${\rm N^{+}}$, which is to assume $t_{2}$ for both cases whenever it is not possible to directly  derive the $T_{\rm e}$ from the 
[\ion{N}{ii}]$\mathrm{\lambda6584/\lambda5755}$. However,  \citet{2020MNRAS.496.3209D}
found for AGNs (see also \citealt{2021MNRAS.501L..54R})
a slight deviation from the equality of  the temperature for the $\rm O^{+}$ ($t_{2}$) and $\rm N^{+}$ [$t_{\rm e}({\rm \ion{N}{ii}})$]. Therefore, 
in order to test if the temperature for ${\rm Ne^{2+}}$, defined for $t_{\rm e}({\rm \ion{Ne}{iii}})$,
can be considered to be the same as $t_{3}$, we used results from the grid of AGN photoionization models
built with the {\sc Cloudy} code by \citet{2020MNRAS.492.5675C}. 
 In Fig.~\ref{figt3ne3}, the model predicted values for $t_{\rm e}({\rm \ion{Ne}{iii}})$ versus $t_{3}$ are shown. In each panel of
Fig.~\ref{figt3ne3}, the model results are discriminated in accordance with the parameters $\alpha_{ox}$ (botton panel), $N_{\rm e}$ (midle panel) and $\log U$ (top panel) assumed in the models.
 It can be seen that for $t_{3} \: \ga \:1.0$, the models predict $t_{\rm e}({\rm\ion{Ne}{iii}})$ lower than  $t_{3}$ and the outlier of a point can not be explained by the variation in the nebular parameter assumed in the models. It is worth mentioning that the variations in the nebular parameters produce temperatures, in most part, within the uncertainty of
 $\pm 800$ K derived in direct estimates (e.g. \citealt{2003ApJ...591..801K, 2008MNRAS.383..209H}).
In the top panel of Fig.~\ref{figt3ne3}, we can note that for objects with higher ionization parameter a high difference between the temperatures closer to the 1 to 1 relation is derived.
 The fit to the estimations considering  all the points 
 in Fig.~\ref{figt3ne3} produces the relation

\begin{equation}
\label{eqtnet3}
t_{\rm e}({\rm \ion{Ne}{iii}})=0.1914 \times t_3^3 - 1.1344 \times  t_{3}^{2}+ 2.334 \times t_{3} - 0.4854.
\end{equation}

\begin{figure}
\includegraphics[angle=-90,width=1\columnwidth]{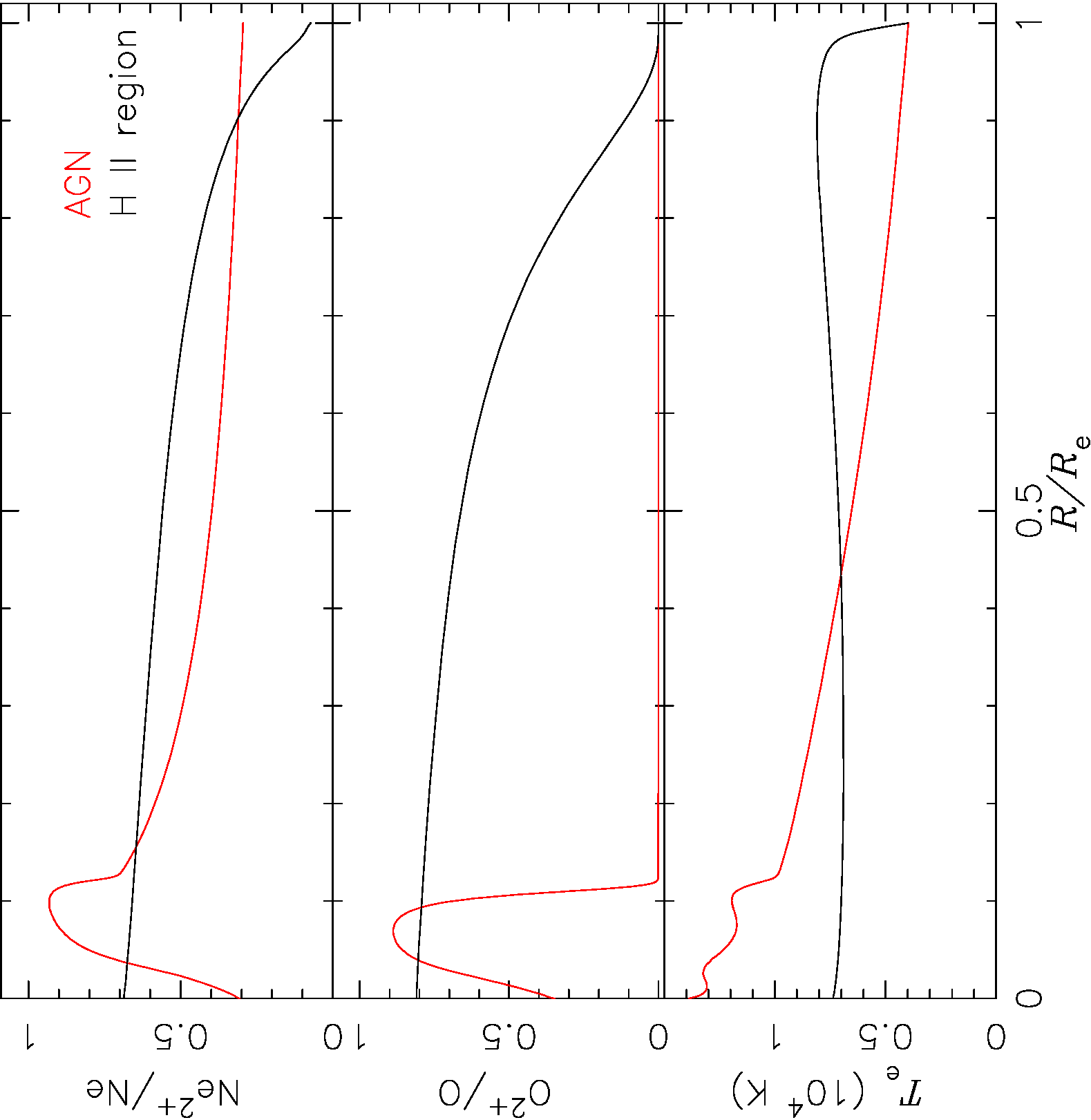}
\caption{ Bottom panel:
Profiles for the   electron temperature ($T_{\rm e}$,  in units of $10^{4}$ K) over the nebular radius  predicted by  AGN and \ion{H}{ii} region photoionization models built with the {\sc Cloudy} code \citep{2013RMxAA..49..137F}
versus the distance $R$ from the innermost gas region   normalized by the outermost radius $R_{\rm e}$ of each model. Different colours represent predictions for AGN and \ion{H}{ii}  region models, as indicated.
Middle and top panels: Same as bottom panel but for  
predictions of the fractional abundances
 $\rm O^{2+}/O$ and $\rm Ne^{2+}/Ne$ as indicated.
The same nebular parameters
are assumed in both AGN and \ion{H}{ii} region models: electron density $N_{\rm e}=500 \:\rm  cm^{-3}$, solar metallicity $(Z/\rm Z_{\odot})=1.0$, and logarithm of the ionization parameter $\log U = -2.5$.
The AGN SED was considered as a power law described by the slope $\alpha_{ox}=-1.1$. The \ion{H}{ii} region SED was assumed to be a stellar cluster  formed instantaneously with the age of 2.5 Myr taken from the {\sc starburst99} code  \citep{1999ApJS..123....3L}.}
\label{fstruc}
\end{figure}

In order to produce an additional test for ascertaining the $t_{\rm e}({\rm \ion{Ne}{iii}})$ and  $t_{3}$ relations between AGNs and \ion{H}{ii}
regions, we analyse the electron temperature ($T_{\rm e}$)  as well as the $\rm O^{2+}/O$ and $\rm Ne^{2+}/Ne$ ionic abundance structures along the nebular radius. In view of this, we consider photoionization models built
with the {\sc Cloudy} code in order to represent both kind of objects.
For both models we adopt the same nebular parameters, i.e.
electron density $N_{\rm e}=500 \:\rm  cm^{-3}$, solar metallicity
$(Z/\rm Z_{\odot})=1.0$, and logarithm of the ionization parameter $\log U = -2.5$.  The outer radius in both AGN and \ion{H}{ii} region models was considered to be the radius at which the electron temperature of the gas reaches 4\,000 K, i.e. the default lowest allowed kinetic temperature by the {\sc Cloudy} code. It is worth noting that gases  cooler than $\sim$4\,000 K practically do not emit the optical and infrared emission lines considered in this work. Despite the fact that AGNs have slightly larger $N_{\rm e}$ values (by a factor of $\sim2$) than \ion{H}{ii} regions (see, e.g. \citealt{2000A&A...357..621C, 2012MNRAS.427.1266V}),  the same value for this parameter was used in both models in order to maintain consistency.
For the lower electron density regime, the
$N_{\rm e}$ value  does not change the temperature
and ionization structure predicted by the photoionization models.
For the AGN model, we
adopt the SED as being a power law described by the slope $\alpha_{ox}=-1.1$
(for a detailed description of this SED see \citealt{2021MNRAS.505.2087K}).
The SED for the \ion{H}{ii} region model was taken from
{\sc starburst99} code \citep{1999ApJS..123....3L} and it assumes  a 
stellar cluster  formed instantaneously with the age of 2.5 Myr, which is a typical age of normal star-forming regions (e.g. \citealt{2008A&A...482...59D}).
For detailed  description of the AGN and \ion{H}{ii} region models see \citet{dors2018hemical}
 and  \citet{2020MNRAS.492.5675C}.
The model results from AGN and \ion{H}{ii} region are compared with each other in Fig.~\ref{fstruc}.
In the bottom panel of this figure, it can be seen that the AGN model
presents a very distinct temperature distribution over
the nebular radius as compared to the \ion{H}{ii} region one,
implying that the former has a stronger decrease with the
radius than the latter. Also, the $\rm O^{2+}/O$ and $\rm Ne^{2+}/Ne$
ionization structures are very distinct for both kind of objects.  Similar ionic abundance distributions for both ionic ratios are derived for the \ion{H}{ii} region, confirming the assumption of $T_{\rm e}(\ion{O}{iii}) \approx T_{\rm e}(\ion{Ne}{iii})$.
However, for the AGN model, the $\rm Ne^{2+}/Ne$ ionic abundance extends to an outer nebular radius (lower temperature)  in comparison with
$\rm O^{2+}/O$, implying that the approach $T_{\rm e}(\ion{O}{iii}) \approx T_{\rm e}(\ion{Ne}{iii})$ is not valid for this object class. Moreover,
the   neon and oxygen ionic abundance structures 
for the AGN clearly indicate that
the supposition $\rm (Ne^{2+}/O^{2+})=(Ne/O)$, usually assumed to derive
the total neon abundance in \ion{H}{ii} region studies
(e.g. \citealt{2003ApJ...591..801K}), can not be applied to AGNs.
The result shown in Fig~\ref{figt3ne3} is further supported by this
simulation.

To estimate the ${\rm Ne^{2+}/H^{+}}$ abundances  we use the relation given by \citet{2008MNRAS.383..209H}:
\begin{equation}{\label{ne}}
\begin{split}
    12 + \log \left (\mathrm{ \frac{Ne^{2+}}{H^{+}}} \right)_{\mathrm{Op.}} =  & \log \left[\frac{I(3869\:{\angstrom})}{I\mathrm{(H\beta)}} \right] + 6.486 + \frac{1.558}{t_{\rm e}} \\
    \\
    & - 0.504 \times \log t_{\rm e},
    \end{split}
\end{equation}
where $t_{\rm e}$ is the electron temperature.  We
considered both  $t_{3}$ and $t_{\rm e}({\rm \ion{Ne}{iii}})$ in the estimations for ${\rm Ne^{2+}/H^{+}}$.

\subsection{Infrared – lines method}
\label{infm}

The infrared${-}$lines method (hereafter, IR-method) is based on determining the abundance of a given element using intensities of emission lines in the infrared spectral region
 (for a review, see \citealt{2017PASA...34...53F}). Infrared emission lines have the advantage over  optical lines for being less
 dependent on the electron temperature and on reddening correction, however, they have  lower critical density ($10^{4-6}~{\rm cm^{-3}}$; e.g. 
 \citealt{forster2001near}) than the others  ($10^{4-8}~ {\rm cm^{-3}}$; e.g. 
 \citealt{2012MNRAS.427.1266V}).  
 
 Regarding the IR lines involved in our study, the critical electron density $N_{\rm c}$ for the [\ion{Ne}{ii}]$12.81\micron$ and [\ion{Ne}{iii}]$15.56\micron$
 emission lines are $7.1\times 10^{5}$ and $2.1\times 10^{5}$
 cm$^{-3}$ \citep{osterbrock2006astro}, respectively.  The electron densities for the NLRs of our sample (see Sect.~\ref{meth1}) derived from the $\rm S^{+}$ line ratio
 $(300 \: \la \: N_{\rm e}(\rm cm^{-3}) \: \la \: 3500)$
 are much lower than the $N_{\rm c}$ values. 
   However, $N_{\rm e}$ derived from line ratios
 emitted by ions with  different ionization potential (IP) other
 than the $\rm S^{+}$ (IP=10.36 eV) can reveal  gas regions with higher $N_{\rm e}$ values than the ones derived for our objects and, consequently, indicate an influence on physical properties based on IR lines. In fact, $N_{\rm e}$ estimates from the   [\ion{O}{iii}]52$\micron$/88$\micron$ line ratio
 [PI($\rm O^{2+})=35.12$ eV]
 carried out by \citet{2002A&A...391.1081V} for \ion{H}{ii} regions showed  electron densities lower than 2\,000 $\rm cm^{-3}$. Also, \citet{2009MNRAS.394.1148S}, who built  $N_{\rm e}$ map
 based on the [\ion{Fe}{ii}]1.533$\micron$/1.644$\micron$ [PI($\rm Fe^{+})=7.90$ eV] for the NLR of NGC\,4151, found values between 1\,000 and 10\,000 $\rm cm^{-3}$. Finally, $N_{\rm e}$ determinations based on [\ion{S}{ii}]$\lambda6716/\lambda6731$ and [\ion{Ar}{iv}$\lambda4711/\lambda4740$ [IP($\rm A^{3+})=40.74$ eV] 
line ratios in two Seyfert 2 (IC\,5063 and NGC\,7212)
by  \citet{2017MNRAS.471..562C} show $N_{\rm e}$ values ranging from $\sim$200   to  $\sim$13 000 $\rm cm^{-3}$.
Although studies indicate the existence of an electron density stratification  in NLRs of AGNs with values higher than the ones for our sample (see also, e.g. \citealt{2018A&A...618A...6K, 2018MNRAS.476.2760F}), effects of collisional de-excitation are probably negligible in our IR abundance estimates. 
 Furthermore, we selected emissivity ratio values considering lower electron density compared to the aforementioned $N_{\rm c}$ values (see Table~\ref{tab01}).

 The Ne$^{+}$ and Ne$^{2+}$ ionic abundances can be determined using the intensities of the [\ion{Ne}{ii}]12.81$\micron$ and [\ion{Ne}{iii}]15.56$\micron$ emission lines
following a similar methodology presented by \citet{dors2013optical}. 
 Considering two ions X$^{\rm i+}$ and H$^+$, the ratio of their ionic abundances is determined by

\begin{equation}
   \frac{{\rm N({X^{\rm i+}}})}{{\rm N}(\mathrm{{H^+}})} = \frac{I_{\lambda}\mathrm{(X^{\rm i+})}N_{e}j_{{\lambda}\mathrm{(H^+)}}}{I_{\lambda}\mathrm{(H^{+})} N_{e}j_{\mathrm{{\lambda}(X^{\rm i+})}}},
\end{equation}

\noindent where, N(${\mathrm{X^{\rm i+}}}$)  and N(${\mathrm{H^+ }}$) are the abundances of the X$^{\rm i+}$ and H$^+$ ions, $I_{\lambda}$(X$^{i+})$ is the intensity of a given emission line emitted by X$^{\rm i+}$, $I_{\lambda}$(H$^+)$ is the intensity of a reference hydrogen line, while $j_{{\lambda}\mathrm{(H^{\rm i+})}}$ and $j_{{\lambda}\mathrm{(X^+)}}$  are the emissivity values. In \citet{dors2013optical}
the emissivity values were obtained from the {\sc ionic} routine of the nebular package of {\sc iraf}, which uses the Ne atomic parameters from \citet{mendoza1983recent}, \citet{saraph1994atomic}, \citet{galavis1997atomic}, \citet{badnell2006}, \citet{griffin2001electron}, \citet{kaufman1986forbidden}, \citet{butler1994atomic} and \citet{mcLaughlin2000electron}. In all abundance determinations, these emissivity values are believed to be constant as they differ by less than $5\,\%$ over a wide temperature range  \citep{simpson1975infrared}.

Using this method, any error in the determination of these emissivities directly translates into a systematic shift to the derived  neon ionic abundance. To obtain the  Ne$^{+}$ and Ne$^{2+}$ ionic abundances with respect to hydrogen (H$^+$), near to
 mid-infrared \ion{H}{i} recombination lines must be preferably used as reference line, 
such as P$\alpha$, P$\beta$, P$\gamma$, P$\delta$, Br$\alpha$, Br$\beta$, Br$\gamma$, Br$\delta$ and Br11, which are detected in most of the sources under consideration. 

 The emission coefficient  for lines in the infrared has a weak dependence on the electronic temperature, which, in general, is disregarded. Therefore, abundance of a given ion can be obtained directly from the ratio between an emission line observed in the infrared  and a hydrogen reference line. The calculation of Ne$^{+}$ and Ne$^{2+}$ ionic abundances can be obtained by the general relations with their emission lines:

\begin{equation}{\label{eq15}}
\frac{\mathrm{Ne^{+}}}{\mathrm{H^+}} = \frac{I({12.81}{\micron})}{I\mathrm{(H\beta)}} \times 1.322 \times 10^{-4}
\end{equation}

\noindent and

\begin{equation}{\label{eq16}}
\frac{\mathrm{Ne^{2+}}}{\mathrm{H^+}} = \frac{I({15.56}{\micron})}{I\mathrm{(H\beta)}} \times 6.323 \times 10^{-5},
\end{equation}

\noindent respectively.

 Based on the above assumptions together with the values of the hydrogen line emissivities relative to $\mathrm{H\beta}$ listed in Table \ref{tab01} and Eqs.~\ref{eq15} and \ref{eq16}, we deduce the following relations:

\begin{equation}
\label{ifq1}
\frac{\mathrm{Ne^{+}}}{\mathrm{H^+}} =\frac{I({12.81}~{\micron})}{I\mathrm{(Paschen)}} \times k\mathrm{_1(H\beta)},
\end{equation}

\begin{equation}
\label{ifq2}
\frac{\mathrm{Ne^{+}}}{\mathrm{H^+}} =\frac{I({12.81}~{\micron})}{I\mathrm{(Brackett)}} \times k\mathrm{_1(H\beta)},
\end{equation}

\begin{equation}
\label{ifq3}
\frac{\mathrm{Ne^{2+}}}{\mathrm{H^+}} =\frac{I({15.56}~{\micron})}{I\mathrm{(Paschen)}} \times k\mathrm{_2(H\beta)}
\end{equation}

\noindent and

\begin{equation}
\label{ifq4}
\frac{\mathrm{Ne^{2+}}}{\mathrm{H^+}} =\frac{I({15.56}~{\micron})}{I\mathrm{(Brackett)}} \times k\mathrm{_2(H\beta)},
\end{equation}

\noindent where $k\mathrm{_1(H\beta)}$ and $k\mathrm{_2(H\beta)}$ are the constants derived from the emissivity ratio values presented in Table~\ref{tab01}.

\begin{table}
\centering
\caption{\ion{H}{i}  emissivity ratio values assuming the  Case B taken
from \citet{osterbrock2006astro}  for electron
density $N_{\rm e}=10^4 \: \rm cm^{-3}$ and  electron temperature
$T_{\rm e}=10^4$ K. $k\mathrm{_{i}(H\beta)}$, where i = 1 and 2, represent $\mathrm{Ne^{+}/H^+}$ and $\mathrm{Ne^{2+}/H^+}$ ionic abundance constants after the emissivity ratio values have been applied to Eqs.~\ref{eq15} and ~\ref{eq16}, respectively.}
\label{tab01}
\begin{tabular}{lcccc}
\hline

    $j_{\mathrm{\lambda}}/j_{\mathrm{H\beta}}$                &   Value         &  $k\mathrm{_1(H\beta)}$    &  $k\mathrm{_2(H\beta)}$ \\  
\hline
\multicolumn{4}{c}{Paschen series} \\

$j_{\mathrm{P\alpha}}/j_{\mathrm{H\beta}}$ & 0.33200 & $4.389\times 10^{-5}$  & $2.099 \times 10^{-5}$\\

$j_{\mathrm{P\beta}}/j_{\mathrm{H\beta}}$ & 0.16200 & $2.141 \times 10^{-5}$ & $1.024\times 10^{-5}$ \\

$j_{\mathrm{P\gamma}}/j_{\mathrm{H\beta}}$ & 0.09010 & $1.191 \times 10^{-5}$& $5.697\times 10^{-6}$\\

$j_{\mathrm{P\delta}}/j_{\mathrm{H\beta}}$ & 0.05540 & $7.323\times 10^{-6}$ & $3.502\times 10^{-6}$ \\

$j_{\mathrm{P_{8}}}/j_{\mathrm{H\beta}}$ & 0.03740 & $4.944 \times 10^{-6}$ & $2.365\times 10^{-6}$ \\
\hline

\multicolumn{4}{c}{Bracket series} \\

$j_{\mathrm{Br\alpha}}/j_{\mathrm{H\beta}}$ & 0.07780 & $1.028\times 10^{-5}$ & $4.919\times 10^{-6}$\\

$j_{\mathrm{Br\beta}}/j_{\mathrm{H\beta}}$ & 0.04470 & $5.909\times 10^{-6}$ & $2.826\times 10^{-6}$\\

$j_{\mathrm{Br\gamma}}/j_{\mathrm{H\beta}}$ & 0.02750 & $3.635\times 10^{-6}$& $1.738\times 10^{-6}$ \\

$j_{\mathrm{Br\delta}}/j_{\mathrm{H\beta}}$ & 0.01810 & $2.392\times 10^{-6}$& $1.144\times 10^{-6}$\\

$j_{\mathrm{Br_{10}}}/j_{\mathrm{H\beta}}$ & 0.00910 & $1.203\times 10^{-6}$& $5.753\times 10^{-7}$ \\

$j_{\mathrm{Br_{11}}}/j_{\mathrm{H\beta}}$ & 0.00695 & $9.181\times 10^{-7}$& $4.391\times 10^{-7}$\\

$j_{\mathrm{Br_{13}}}/j_{\mathrm{H\beta}}$ & 0.00425 & $5.613 \times 10^{-7}$ & $2.684\times 10^{-7}$\\

\hline
\end{tabular}
\end{table}

The Case B was assumed to derive the above equations because, as opposed to the broad-line region gas,  much of the narrow-line region  is believed to be optically thick to the ionizing radiation, even though studies of the \ion{He}{ii} $\lambda4686$ {\AA}/H$\beta$ ratio in AGNs indicates the presence of some optically thin gas \citep{2003eaa..book.....M}. The [\ion{Ne}{iii}] {$\lambda15.56$} {\micron} line is always chosen over the [\ion{Ne}{iii}] {$\lambda36.0$} {\micron} when both are measured, because  the  spectrum is  noisier  at  the  long  wavelength end of the long high-resolution (LH) module in the Infrared  Spectrograph (IRS). Therefore, we preferred to use the [\ion{Ne}{iii}] {$\lambda15.56$} {\micron} line flux for the abundance determination of this ion, which also has larger transition probability and critical density.


\section{Total abundance determinations}
\label{icf}

\subsection{Oxygen}

In general, the total abundance of an element relative to hydrogen abundance is difficult
to be calculated because
not all emission line intensities emitted by the ions of this element are measured in the same spectral range.
This fact, in principle, is circumvent by the use of 
ionization correction factor (ICF) proposed by \citet{peimbert1969chemical}. The ICF for the unobserved ionization stages of an element X  is defined as
\begin{equation}{\label{eq19}}
{\mathrm{ICF(X^{i+})}}  =  \frac{\mathrm{N(X/H})}{\mathrm{N(X^{i+}/H^+)}},
\end{equation}
being N the abundance and $\rm X^{i+}$ the ion whose ionic abundance can be calculated from
its observed emission lines.
\noindent For instance, considering optical emission lines of oxygen, it is relatively easy to derive the
$\rm O^{+}/H^{+}$ and $\rm O^{2+}/H^{+}$ abundances when $T_{\rm e}$ and $N_{\rm e}$ are derived. However, emission lines of oxygen ions with higher
ionization states are observed in other spectral bands as, for instance, in
X-rays (e.g. \citealt{2009A&A...505..541C, 2010MNRAS.405..553B, 2017ApJ...848...61B}).
Recent studies \citep{2020MNRAS.496.2191F, 2020MNRAS.496.3209D} 
indicate that the  contribution of ions with ionization stage  higher than 
$\rm O^{2+}$ in AGNs is in order of 20 per cent of the total O/H abundance.
A smaller contribution of these ions, at least, in poor metal star-forming
regions, is in order of  only  1-5 per cent  \citep{1993ApJ...411..655S, 2004ApJ...614..698L}.

To calculate the total oxygen abundance N(O/H) for our sample, we  assumed 
\begin{equation}
\label{eqt6}
{\rm
N\left(\frac{O}{H}\right)=ICF(O)\: \times \: N\left(\frac{O^{2+}}{H^{+}}+\frac{O^{+}}{H^{+}}\right),} 
\end{equation}
where  ICF(O) is the Ionization Correction Factor for oxygen. We consider the ICF(O) expression proposed  by \citet{1977RMxAA...2..181T}
\begin{equation}
\label{icfox}
\rm
ICF(O)=\frac{N(He^{+})+N(He^{2+})}{N(He^{+})},
\end{equation}
 (see also \citealt{izotov2006chemical, 2020MNRAS.496.2191F}).
This ICF expression is based on the similarity between
the $\rm He^{+}$ and $\rm O^{2+}$ ionization potential (about 54 eV).

To calculate the ionic helium abundance for each object taking into account the assumption that $t=t_{3}$,  we use the relations
proposed by \citet{izotov1994primordial} expressed as,

\begin{equation}
\label{abundhe}
 {\rm \frac{N(He^{+})}{N(H^{+})}}=0.738 \: t^{0.23} \: \frac{I(\lambda 5876)}{I(\rm H\beta)}
\end{equation}
and
\begin{equation}
 {\rm \frac{N(He^{2+})}{N(H^{+})}}=0.084 \: t^{0.14} \: \frac{I(\lambda 4686)}{I(\rm H\beta)}.
\end{equation}

\subsection{Neon}
\label{tneon}

 The total neon abundance   determination in Seyfert 2 nuclei from either $T_{\rm e}$ or IR method can be realised by using 
 an ICF  taking into account the unobserved ionization stages   ions of this element, such as
 Ne$^{3+}$, whose emission lines are observed at 12$\mu$m and 24$\mu$m (e.g. \citealt{2007ApJ...664...71D}).
\citet{peimbert1969chemical} and \citet{peimbert2009chemical}, based on 
the similarity between the ionization structures of neon and oxygen [$\mathrm{(Ne^{2+}/Ne) \approx (O^{2+}/O)}$], 
proposed

\begin{equation}
\label{ICFNe}
\mathrm{ICF(Ne^{2+})} = {\rm
N\left(\frac{O}{O^{2+}}\right) \approx N\left(\frac{O^+ + O^{2+}}{O^{2+}}\right)
}.
\end{equation}

However, this approach does not seem to be valid for AGNs. For example, 
\citet{1997A&A...323...31K}, by using multi-component photoionization models which permitted a successful match of a large set of line intensities from the UV to the NIR for Seyfert~2 nuclei,  showed that  the Ne$^{2+}$ ion extends to a larger (where a lower temperature is expected) radius of the AGN than O$^{2+}$. Similar result was found 
by \citet{1997AJ....114..713A} for Planetary Nebulae (PNs), which also exhibited gas with high ionization.
In fact, it can be seen from Fig.~\ref{figt3ne3} that, generally, $t_{3}$ is higher than $t_{\rm e}({\rm \ion{Ne}{iii}})$. Therefore, 
 based on  the results shown in Fig.~\ref{fstruc},  
it is necessary to produce a new formalism to replace Eq.~\ref{ICFNe} for AGNs. We developed a  semi-empirical neon ICF 
following a  similar methodology assumed by \citet{dors2013optical} for SFs.

\begin{figure}
\includegraphics[angle=-90,width=0.9\columnwidth]{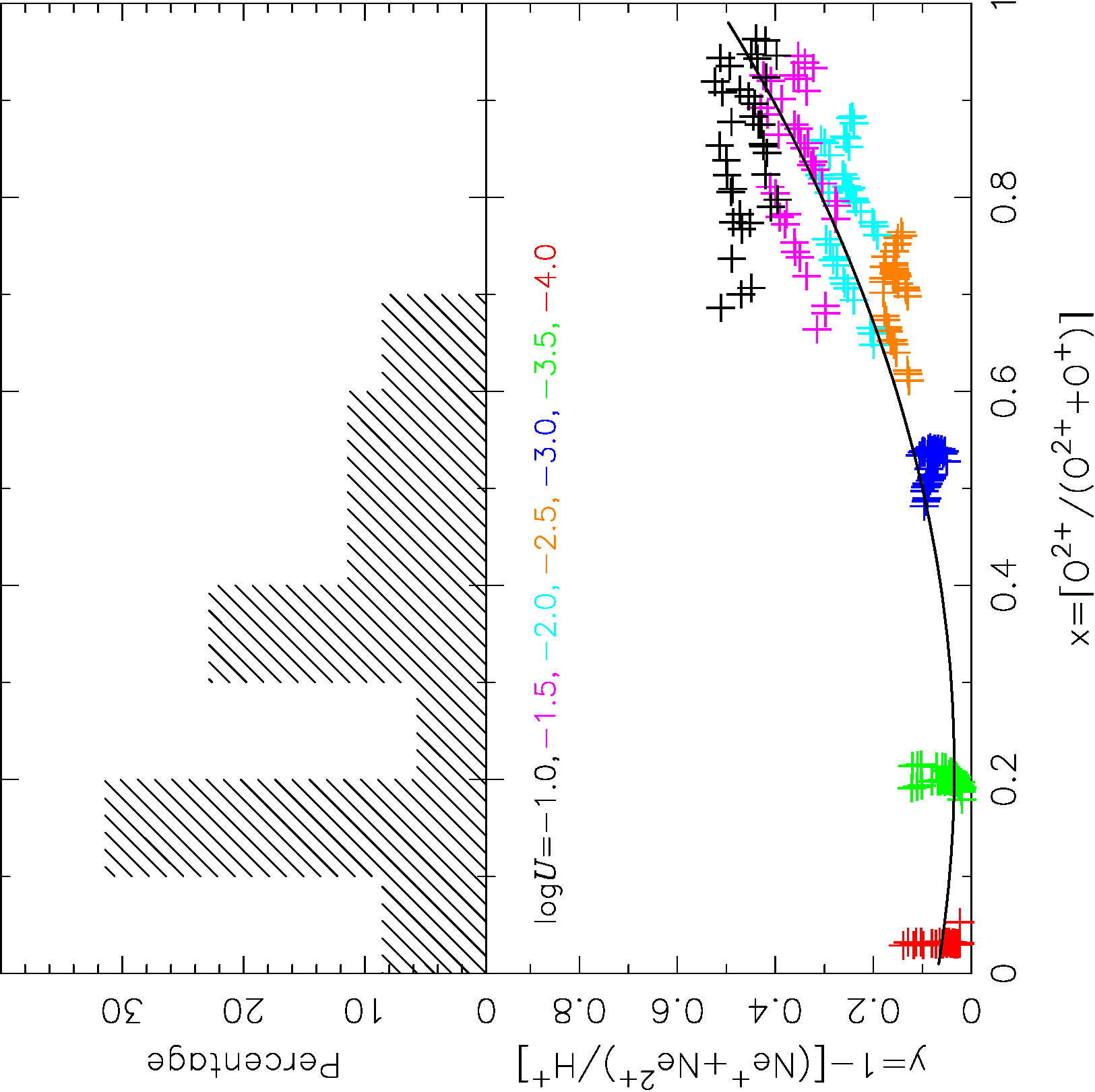}
\caption{ Bottom panel: Neon ionic abundance ratio $\rm y=1-[\rm (Ne^{+}+Ne^{2+})/H^{+}]$ versus oxygen ionic abundance ratio
$\rm x=[O^{2+}/(O^{+}+O^{2+})]$ predicted 
by photoionization model built by \citet{2020MNRAS.492.5675C}.
Results from photoionization models assuming distinct ionization
parameter ($U$)  values are indicated by different colours. 
The black solid line represents a fit to the points represented
by Eq.~\ref{fitnei}. Top panel: The distribution of oxygen ionic abundance ratios for our sample of objects (see Sect.~\ref{meth}) calculated by using the $T_{\rm e}-$method.}
\label{ananet1}
\end{figure}

The total neon abundance in relation  to hydrogen is usually assumed to be
\begin{equation}
\label{eq26}
\rm
\frac{Ne}{H}\approx\frac{Ne^{+}}{H^{+}}+\frac{Ne^{2+}}{H^{+}}.
\end{equation}
This approximation can be more reliable for SFs than AGNs,
because it considers  a null abundances of  $\rm Ne^{i\:  > \: 2+}$. We use the photoionization model results by \citet{2020MNRAS.492.5675C}  to ascertain the validity of Eq.~\ref{eq26} for AGNs. In the bottom panel of Fig.~\ref{ananet1},  the model results for
 $\rm y=1-[\rm (Ne^{+}+Ne^{2+})/H^{+}]$ versus
$\rm x=[O^{2+}/(O^{+}+O^{2+})]$ is shown. It can be seen that, for
$\rm x \: \la \: 0.7$ or for $\log U \: \la \: -2.5$
the abundance sum  $(\rm Ne^{+}+Ne^{2+})$ 
represents more than 80\,\% of the total Ne abundance.
In the top panel of Fig.~\ref{ananet1}, a distribution of x values
for the objects in our sample, calculated by using the $T_{\rm e}-$method
(see Sect.~\ref{meth1}), is shown. It can be seen that most
of the objects ($\sim$90\,\%) have $\rm x \: \la \: 0.6$ within the range  $\rm 0.02 \: \lid \: x \lid \: 0.72$.  Thus, a small correction factor  is  necessary in 
Eq.~\ref{eq26}. A fit to the
points in Fig.~\ref{ananet1} produces
\begin{equation}
\label{fitnei}
\rm y=(0.78\pm0.06) x^{2} - (0.33\pm0.06) x +  (0.07\pm 0.01)   
\end{equation}
and  Eq.~\ref{eq26}  can be rewritten in the form

\begin{equation}
\label{eq26aa}
{\rm
\frac{Ne}{H}}= f \times {\rm \left(\frac{Ne^{+}}{H^{+}}+\frac{Ne^{2+}}{H^{+}} \right)},
\end{equation}
where 
\begin{equation}
\label{feqc}
f=\frac{1}{1-y}.    
\end{equation}

For the infrared abundance determinations, Eq.~\ref{eq26aa} was applied,
where the $\rm Ne^{+}$ and $\rm Ne^{2+}$ estimates were based on Eqs.~\ref{ifq1}, \ref{ifq2}, \ref{ifq3} and \ref{ifq4} and the $f$ factor was calculated from Eqs.~\ref{fitnei} and \ref{feqc} with x estimates obtained by using the
$T_{\rm e}$-method (see Sect.~\ref{meth1}).

For Ne/H estimates based on optical lines, it is only possible to calculate the $\rm Ne^{2+}/H^{+}$ abundance based on the Eq.~\ref{ne} and 
assuming $t_{3}$ and $t_{\rm e}({\rm \ion{Ne}{iii}})$. For that,
the total neon abundance estimates via optical lines must be assumed
\begin{equation}
\rm 
\frac{Ne}{H} =  ICF(Ne^{2+}) \times \frac{Ne^{2+}}{H^{+}}.  
\end{equation}

Using Eqs.~\ref{eq15}, \ref{eq16} and \ref{eq26aa}, we derive a semi-empirical
neon ICF given by 

\begin{equation}
\label{eqicfnew}
{\rm ICF(Ne^{2+})}=2.10 \: f \times \left[\frac{I({12.81}~{\micron})}{I({15.56}~{\micron})}+0.48 \right].
\end{equation}

The photoionization
models and expressions employed to derive the ionic abundances, previously presented, probably use different set of atomic parameters which could introduce a small systematic
uncertainty in the resulting abundances. However, \citet{2017MNRAS.469.1036J}
found that atomic data variations introduce differences in the derived abundance ratios as low as $\sim$0.15 dex at low density ($N_{\rm e} \: \la \: 10^{3} \: \rm  cm^{-3}$). Since most NLRs of Seyfert~2 present $N_{\rm e}$ values lower than
$10^{3} \: \rm  cm^{-3}$ (e.g. \citealt{2012MNRAS.427.1266V, 
dors2014metallicity, dors2020chemical, 2018MNRAS.476.2760F, 2018ApJ...867...88R, 2018A&A...618A...6K, 2021ApJ...910..139R})
the consideration of distinct atomic parameters in our calculations is expected to have a
small effect in our abundance results.

\section{Results}
\label{resdisc}

\begin{figure}
\includegraphics[width=\columnwidth]{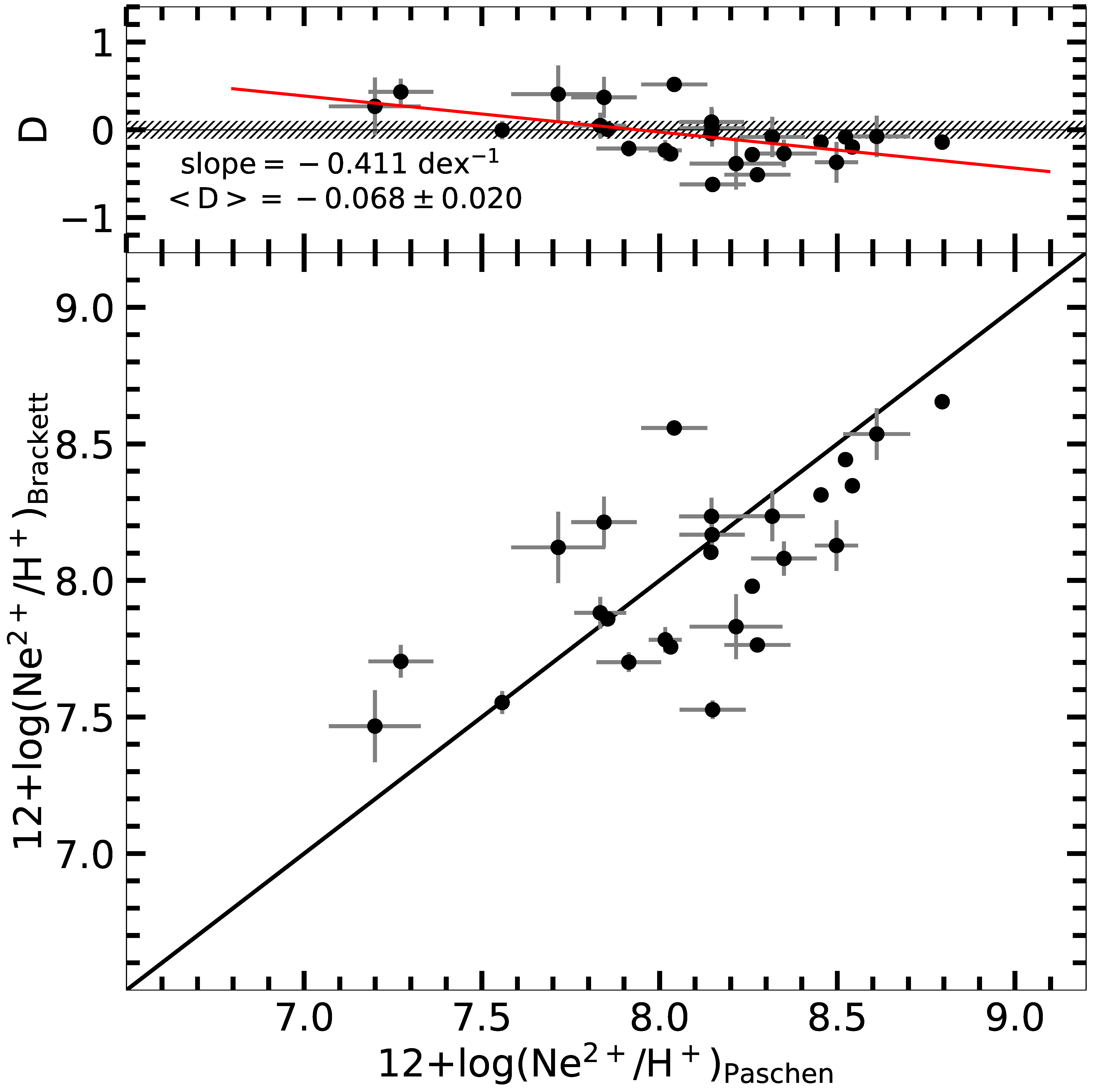}
\caption{Comparison between the ionic abundance of 
12+log($\mathrm{Ne^{2+}/H^+}$) derived via IR$-$lines by using Brackett and Paschen (see Sect.~\ref{infm}). The points represent estimations for the objects presented in Table~${\color{blue}\text{A4}}$.
The solid line represents the equality of the two estimates.
Top panel: difference ($\rm D= ordinate-abscissa$) between the estimations. The black line represents the null difference, while the red line represents a linear regression to these differences whose slope is indicated. The average difference ($\rm < D >$) is also shown. The hatched area indicates the uncertainty of $\pm 0.1$ derived in the abundance estimations.}
\label{comb_figure}
\end{figure}

The bottom panel of Fig.~\ref{comb_figure} shows a comparison between the ionic abundance 12+log($\mathrm{Ne^{2+}/H^+}$) obtained using the infrared lines method (see Sect.~\ref{infm}) considering  Paschen and Brackett
emission lines, i.e. calculated from Eqs.~\ref{ifq3} and \ref{ifq4}.
The average and standard deviation are derived for $\mathrm{Ne^{2+}/H^+}$
estimation of each object  assuming  different Paschen  
and Brackett lines which are in order of 0.02 dex. 
In the top panel of Fig.~\ref{comb_figure}, the mean differences between the estimations  versus the estimations via Paschen lines 
are shown. The average difference is about 0.1 dex, similar to error  derived in ionic abundance estimates by \citet{2003ApJ...591..801K}.
We notice a slight trend (see  Fig.~\ref{comb_figure} top panel) of 
$\mathrm{Ne^{2+}/H^+}$ abundances via Paschen lines to be lower than those via Brackett lines,  reaching up to $\sim0.4$ dex for the lowest values of Paschen neon ionic
determinations. This discrepancy could suggest either some uncertainties in the physical constants of the H line ratios (probably for high temperature) or in the line measurements (e.g. aperture corrections and/or distinct calibrations in the
data compiled from the literature).
In any case, this result is marginal because  only two  objects present 12+log($\mathrm{Ne^{2+}/H^+}$) lower than $\sim7.3$ dex.
If this objects are not considered, we derive about a null difference 
among the estimates.
Thus, it is shown that the $\mathrm{Ne^{2+}/H^+}$  estimates based on any IR hydrogen reference line of a particular series are in agreement with each other taken into account a discrepancy of $\sim 0.1$ dex.
It was possible to calculate  $\mathrm{Ne^{2+}/H^+}$ by using Paschen lines for 27 objects of our sample and Brackett lines for 34 objects with 26 correspondingly Paschen and Bracket emission line series.

 In Table~${\color{blue}\text{A4}}$, the $\mathrm{Ne^{+}/H^+}$, $\mathrm{Ne^{2+}/H^+}$, $f$ factor and the total neon abundance [12+log(Ne/H)]
values for our sample obtained using IR-method (see Sect.~\ref{infm} and \ref{tneon})
are listed. The infrared  $\mathrm{Ne^{2+}/H^+}$ ionic abundance values listed in Table~${\color{blue}\text{A4}}$ represent the mean values from the estimates
based on Paschen and  Brackett lines.
 It was not possible to calculate the  $\mathrm{Ne^{+}/H^+}$ abundance,
and consequently the Ne/H, for three objects from our sample
(i.e. NGC\,1320, NGC\,3393 and ESO428$-$G014) due to the absence of the 
[\ion{Ne}{ii}]12.81$\micron$ emission lines in the original works where the data were
compiled. In Table~${\color{blue}\text{A5}}$, the  12+(Ne$^{2+}$/H$^{+}$) values calculated 
via $T_{\rm e}$-method assuming $t_{3}$ and $t_{\rm e}({\rm \ion{Ne}{iii}})$
(see Sect.~\ref{meth1}), the ICF(Ne$^{2+}$) obtained from Eq.~\ref{eqicfnew}
and the total neon abundance for our sample are listed.

\begin{figure*}
\centering
\includegraphics[width=2.1\columnwidth]{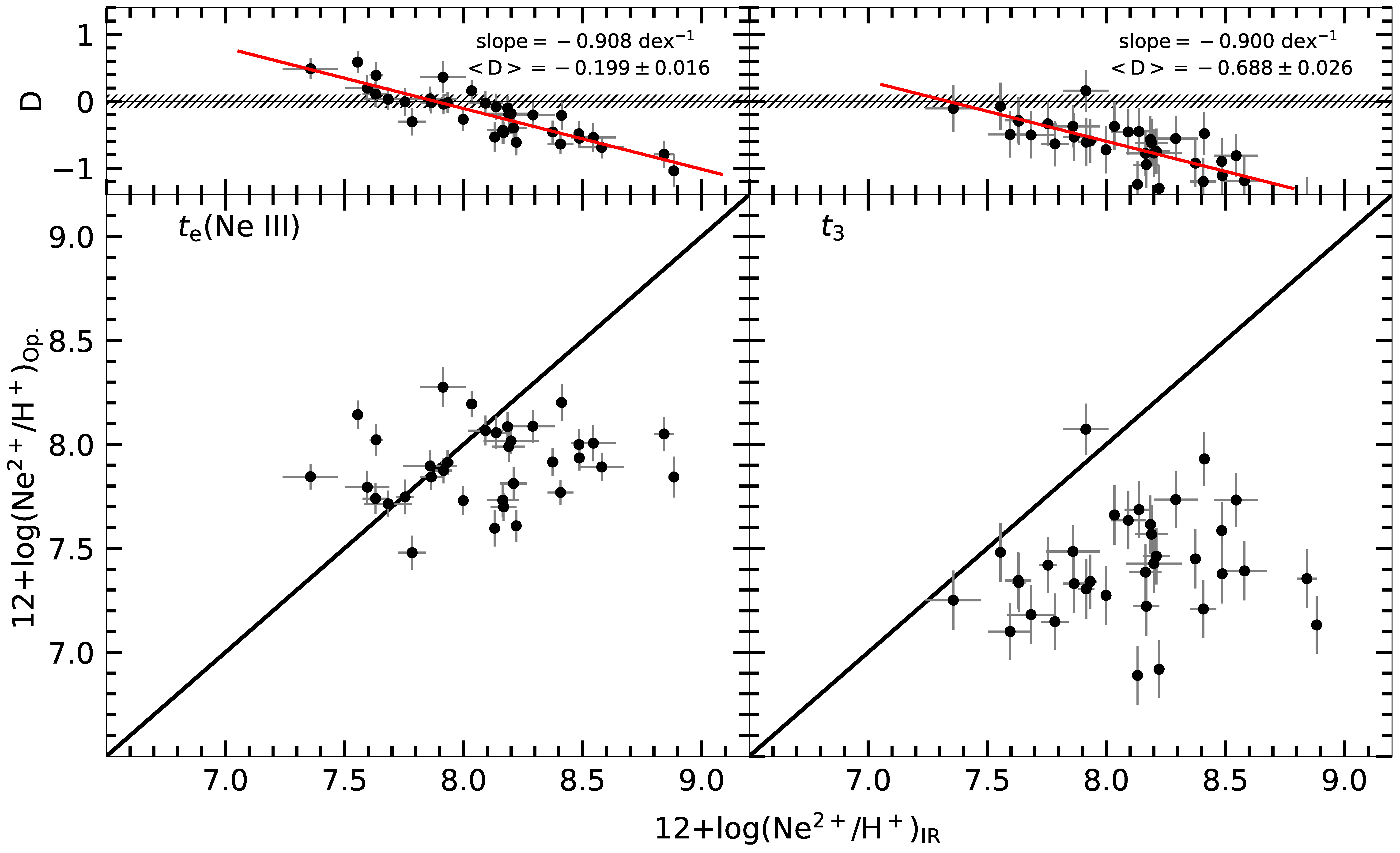}
\caption{Same as Fig.~\ref{comb_figure} but for estimations
 derived using the $T_{\rm e}$-method versus IR-method  (see Sect.~\ref{meth}). In left panel $T_{\rm e}$-method estimates are based on $t_{\rm e}({\rm \ion{Ne}{iii}})$ and in the right panel on $t_{3}$, as indicated.}
\label{figcompne}
\end{figure*}

In the bottom panel of Fig.~\ref{figcompne}, the $\mathrm{Ne^{2+}/H^+}$ values estimated 
using the $T_{\rm e}$-method, assuming $t_{\rm e}({\rm \ion{Ne}{iii}})$ and $t_{3}$, versus estimations obtained from IR-method
are shown. In the top panel of this figure, the
differences between these estimations versus the  IR ionic estimates
are  shown. It can been seen that, for most of the objects, the IR estimations are higher than those obtained via the $T_{\rm e}$-method by using  both $t_{\rm e}({\rm \ion{Ne}{iii}})$ and $t_{3}$ electron temperatures. An average difference value from the comparison between the IR-method and the $t_{\rm e}({\rm \ion{Ne}{iii}})$ estimations is $\sim -$0.20 dex. However, when $t_{3}$ is considered, an average difference between the estimates of $\sim-$0.69 dex is found. The differences between the IR-method and the $T_{\rm e}$-method estimates imply  systematic differences in both cases, i.e. they increase with $\mathrm{Ne^{2+}/H^+}$, until $\sim 1$ dex and $\sim 2$ dex, 
for $t_{\rm e}({\rm \ion{Ne}{iii}})$ and $t_{3}$, respectively.
 The difference between $\mathrm{Ne^{2+}/H^+}$ found in 
Fig.~\ref{figcompne} is due to systematic  derivation (from Eq.~\ref{eqtnet3}) of lower
 $t_{\rm e}({\rm \ion{Ne}{iii}})$ values in comparison with   $t_{3}$, which
 translate into higher ionic abundances when $t_{\rm e}({\rm \ion{Ne}{iii}})$ is assumed. In other words, according to our photoionization model results (see Fig.~\ref{figt3ne3}), the [\ion{O}{iii}] temperature is likely an overestimation of the [\ion{Ne}{iii}] temperature.  

 Neon ICFs for AGNs
are still not available in the literature, however, we can compare
the values obtained for our sample with those derived for \ion{H}{ii} regions by \citet{dors2013optical}. These authors derived the ICF($\rm Ne^{+2}$) directly (from neon IR lines)  for 23 \ion{H}{ii} regions with  oxygen abundance in the range 
of $\rm 7.1 \: \la \: 12+\log(O/H) \: \la \: 8.5$ and ionization degree 
$\rm 0.4 \: \la \: [O^{2+}/(O^{+}+O^{2+})] \: \la \: 1.0$.
Our AGN sample is based on more metallic objects
$\rm 8.0 \: \la \: 12+\log(O/H) \: \la \: 9.2$ and similar ionization degree 
$\rm 0.2 \: \la \: [O^{2+}/(O^{+}+O^{2+}) \: \la \: 0.7$. From Table~${\color{blue}\text{A5}}$,
we notice that the  ICF($ \rm Ne^{+2}$) values for the AGN sample range
from $\sim$1.5 to $\sim$12, where the highest value (11.83) is derived
for NGC\,5953. Even not considering this high value, we find an ICF($\rm Ne^{+2}$) 
range of 1.5-6.5, a wider range of values than those derived for 
\ion{H}{ii} regions by \citet{dors2013optical}, i.e. from $\sim 1$ to $\sim 2$.

 In Table~${\color{blue}\text{A6}}$, the $12+(\log\mathrm{O^{+}/H^+})$,
$12+(\log\mathrm{O^{2+}/H^+})$, the ICF(O) (by using
Eq.~\ref{eqicfnew}), the total oxygen abundance [12+log(O/H)] as well as the log(Ne/O) values, assuming neon abundance derivations via $t_{3}$ 
and $t_{\rm e}({\rm \ion{Ne}{iii}})$, are listed.
With regard to the oxygen ICFs for the 35 objects where the
values ($\sim 80\,\%$ of the sample) could be estimated, we derived values ranging from  $\sim1.1$ to $2.2$, with an average
value of $\sim1.30$, which indicates a correction in the total oxygen
abundance in order of only $\sim$0.1 dex (see also \citealt{2020MNRAS.496.2191F, 2020MNRAS.496.3209D}).

\begin{figure}
\includegraphics[angle=-90,width=\columnwidth]{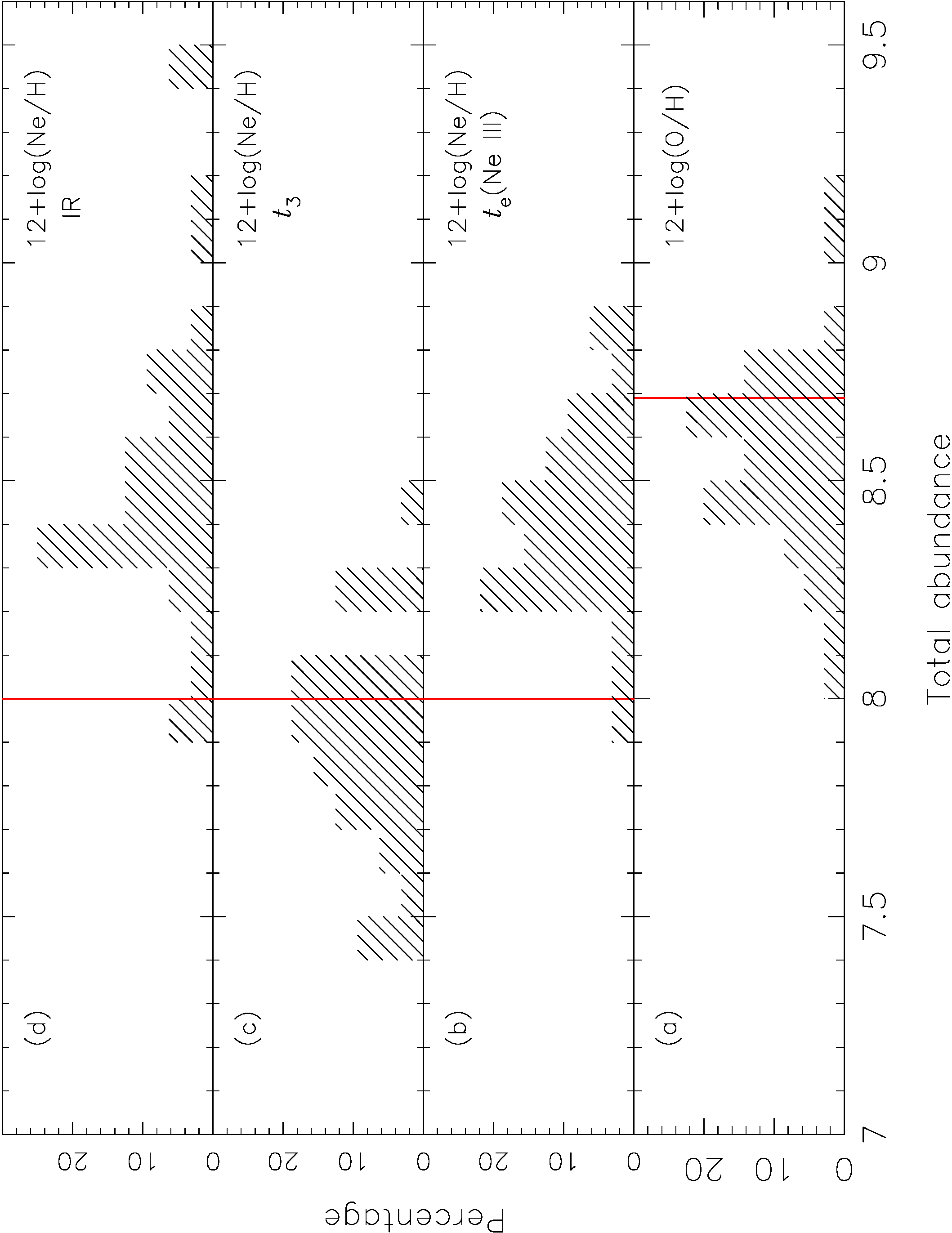}
\caption{Histograms containing the total abundance distributions 
for our sample of objects (see Sect.~\ref{data}).
Panel (a): Distribution of 12+log(O/H) calculated from $T_{\rm e}$-method (see Sects.~\ref{meth1}
and \ref{icf}). Panels (b) and (c):  Distribution of 12+log(Ne/H) calculated from $T_{\rm e}$-method assuming $t_{\rm e}({\rm \ion{Ne}{iii}})$
and  $t_{3}$, as indicated. (d) 
Distribution of 12+log(Ne/H) calculated from IR-method (see Sects.~\ref{infm}
and \ref{icf}).
 Red lines indicate the 12+log(O/H)$_{\odot}$=8.69 and the 12+log(Ne/H)$_{\odot}$=8.0 solar values
derived by \citet{2001ApJ...556L..63A} and \citet{2001AIPC..598...23H}, respectively.}
\label{redtemp1}
\end{figure}

In Fig.~\ref{redtemp1},  histograms showing the distributions of total oxygen abundance (O/H) and the total neon abundances for our sample, calculated from $T_{\rm e}$-method assuming $t_{3}$ and $t_{\rm e}({\rm \ion{Ne}{iii}})$, as well as Ne/H via IR-method, are shown. The solar values $\rm 12+\log(O/H)_{\odot}=8.69$
and $\rm 12+\log(Ne/H)_{\odot}=8.0$, obtained by \citet{2001ApJ...556L..63A}
and \citet{2001AIPC..598...23H}, respectively, are indicated in Fig.~\ref{redtemp1}.
In Table~\ref{tab2}, the minimum, maximum and average values of
the distributions of O/H, Ne/H and Ne/O derived using the distinct methods are listed. It can be observed that, in Fig.~\ref{redtemp1}, most ($\sim64\,\%$) of the objects for the sample have oxygen abundance in the range $\rm 8.4 \:\la \: 12+\log(O/H) \: \la 8.8$ or $0.50 \:\la \: (Z/{\rm Z_{\odot}}) \: \la 1.3$,
which implies that only $8\,\%$ of oxygen abundance values
are found in the low metallicity regime (i.e. $\rm  12+\log(O/H) \: \la \:8.2$).
\citet{groves2006emission}, who considered a photoionization model sequence to
reproduce the optical emission line intensities of AGNs, found a similar result, i.e. low metallicity AGNs are rarely found
in the local universe. The maximum O/H value derived
for our sample $\mathrm{(  12+\log(O/H)\approx 9.2)}$ is about 0.2 dex higher than 
the maximum value derived for star-forming galaxies by \citet{2007MNRAS.376..353P},
who adopted the $P$-method \citep{2000A&A...362..325P, 2001A&A...369..594P}.

 In the case of the Ne/H abundance in Fig.~\ref{redtemp1},  the estimates based on $t_{3}$ indicate that most objects
($\sim65\,\%$) present lower values than the solar value. The abundance estimates via 
$t_{\rm e}({\rm \ion{Ne}{iii}})$ and IR-method indicate that majority ($\ga 90\,\%$) of the objects
have higher Ne/H abundances than the  solar value. For  some few
objects ($\sim10\,\%$), IR estimates indicate values  rising up to 10 times the solar value.

\begin{figure}
\includegraphics[angle=-90,width=\columnwidth]{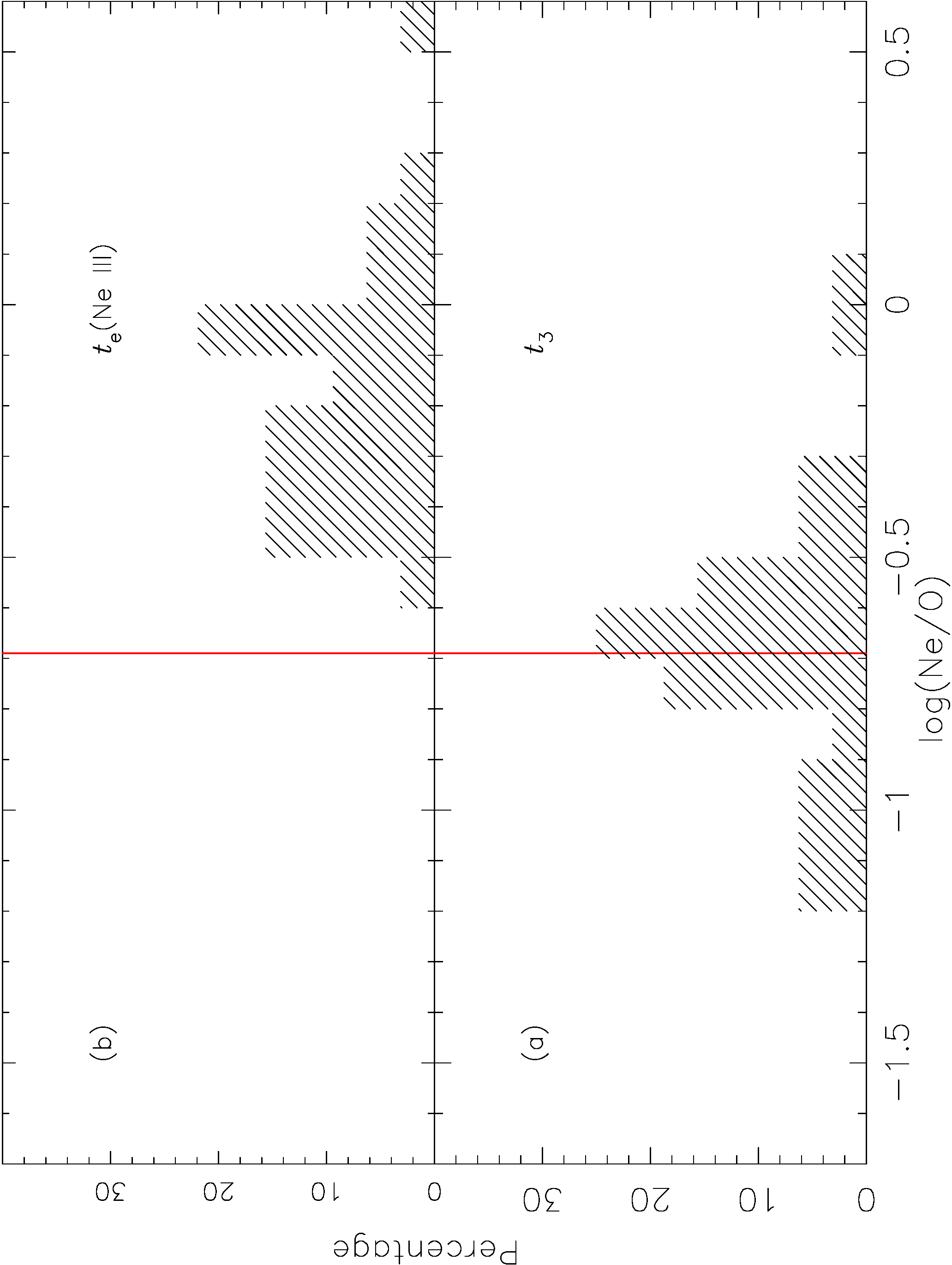}
\caption{Same as Fig.~\ref{redtemp1} but for log(Ne/O).
Panels (a) and (b) show distributions obtained with
Ne abundances calculated from $T_{\rm e}$-method assuming $t_{3}$   and $t_{\rm e}({\rm \ion{Ne}{iii}})$, as indicated. Red line indicates log(Ne/O)$_{\odot}$=$-0.69$ solar value \citep{2001ApJ...556L..63A, 2001AIPC..598...23H}.}
\label{redtemp2}
\end{figure}

In Fig.~\ref{redtemp2}, histograms showing the Ne/O abundance ratios 
distribution for our sample, whose values were calculated via $T_{\rm e}$-method by assuming $t_{3}$ and $t_{\rm e}({\rm \ion{Ne}{iii}})$,
are presented. No Ne/H values derived via IR-lines are considered in
Fig.~\ref{redtemp2} because the O/H values are based on a distinct method, i.e. the $T_{\rm e}$-method. The line indicating the Ne/O solar value is also depicted
in this figure.  We can see that the majority ($\sim60\,\%$) of the $t_{3}$ estimates
are higher than the solar ratio and all  values based on $t_{\rm e}({\rm \ion{Ne}{iii}})$ lead to oversolar Ne/O abundances.

\begin{table}
\caption{Minimum, maximum and the mean  abundance ratio values 
for our sample (see Sect.~\ref{data})
derived by  the use of the  distinct methods (see Sects.~\ref{meth} and \ref{icf}).
The values obtained from the abundance distributions are presented in
Figs.~\ref{redtemp1} and \ref{redtemp2}.}
\label{tab2}
\begin{tabular}{lcccc}
\hline
Abundance ratio                                &        Min.          &       Max.          &  Mean            \\
\hline              
12+log(O/H)                                    & $  8.03 \pm0.05$     & $ 9.17  \pm 0.06$   & $ 8.55 \pm 0.22$ \\
12+log(Ne/H)-${t_{3}}$                         & $  7.44 \pm 0.12$    & $ 8.48  \pm 0.06$   & $ 7.90 \pm 0.24$ \\ 
12+log(Ne/H)-$t_{\rm e}({\rm \ion{Ne}{iii}})$  & $  7.90 \pm 0.05$    & $ 8.88  \pm 0.21$   & $ 8.39 \pm 0.22$ \\
12+log(Ne/H)-IR                                & $  7.99 \pm 0.01$    & $ 9.47  \pm 0.11$   & $ 8.54 \pm 0.36$ \\
log(Ne/O)-${t_{3}}$                            & $ -1.21 \pm 0.01$    & $+0.03  \pm 0.03$   & $-0.66 \pm 0.27$ \\ 
log(Ne/O)-$t_{\rm e}({\rm \ion{Ne}{iii}})$     & $ -0.51 \pm 0.01$    & $+0.58  \pm 0.03$   & $-0.17 \pm 0.24$ \\
\hline
\end{tabular}
\end{table}

\section{Discussion}
\label{disc}

Along decades, several studies have been carried out  to address the determination of chemical abundances of AGNs at both low and high redshift, mainly based on comparing photoionization model
results with observational data. However, these studies have been primarily
focused on the determination of the metallicity and in some few instances on the determination of oxygen and nitrogen abundances. Pertaining to the low redshift objects,
where optical emission lines are easily observed, for instance, \citet{thaisa90}, who compared
Seyfert and LINERs observational
and photoionization model line predictions in the diagram [\ion{N}{ii}]($\lambda6548$ +$\lambda6584$)/H$\alpha$ versus 
[\ion{S}{ii}]($\lambda6716$+$\lambda6731$)/H$\alpha$, found sulfur and nitrogen abundances ranging from one-half-solar to five times the solar values.
After this pioneering work, \citet{dors17} built detailed photoionization models
to reproduce narrow optical emission lines for a sample consisting of 44 local ($z\: < \: 0.1$) 
Seyfert~2 nuclei and found nitrogen abundances ranging from $\sim$0.3 to $\sim$ 7.5 times the solar value. 

Direct elemental abundance of AGNs, based on
the $T_{\rm e}$-method, are rare in the literature.
Probably, the first $T_{\rm e}$-method estimation in AGN was undertaken by \citet{1975ApJ...197..535O} for Cygnus~A, in the derivation of 12+log(O/H)$\sim8.60$, 
12+log(Ne/H)$\sim8.0$ and other elemental abundances. After this pioneering work, other authors also applied the $T_{\rm e}$-method to AGNs (e.g. \citealt{1992A&A...266..117A, 2008ApJ...687..133I, dors2015central, dors2020chemical, 2020MNRAS.496.3209D}) but focused
mainly on O/H abundance. Recently,  \citet{2020MNRAS.496.2191F}, by assuming an approach for estimating abundances of heavy elements which involves a reverse-engineering of the
$T_{\rm e}$-method, derived the first (N/O)-(O/H) relation for AGNs based on the direct method.
On the other hand, for the elemental abundances in high redshift AGNs, oversolar nitrogen
abundance have been derived for the most part of the objects (see \citealt{2019MNRAS.486.5853D} and reference therein). In summary, hitherto, the unique neon abundance
in AGNs appears to have been the derivation obtained by \citet{1975ApJ...197..535O}, who estimated a value
approximately equal to the solar abundance. In subsequent sections, we discuss the neon abundance results derived for our sample.

\subsection{\texorpdfstring{${\rm Ne^{2}/H^{+}}$}~~abundance}

\citet{2002A&A...391.1081V} obtained optical (by using the Boller \& Chivens spectrograph on the ESO 1.52 meter telescope) and  infrared (by using Short Wavelength Spectrometer - SWS and Long Wavelength Spectrometer- LWS on board the Infrared Space Observatory - ISO) spectra for 15  \ion{H}{ii} regions located in the Magellanic Clouds.
From these objects, it was possible to derive the Ne$^{2+}$ ionic abundances via both IR and $T_{\mathrm{e}}{-}$method for 13 out of the 15  \ion{H}{ii} regions.
The differences (D) between these estimations  ranges from  $- 0.6$ to +0.6  dex, thus, for some objects the $T_{\mathrm{e}}{-}$method resulted in higher abundances. The averaged value of D was about zero. The result obtained by \citet{2002A&A...391.1081V} is in disagreement with the findings by \citet{dors2013optical}, who found that the abundances obtained via infrared emission lines  are higher than those obtained via optical lines in \ion{H}{ii} regions, by a factor of $\sim 0.60$ dex.

In Fig.~\ref{figcompne}, 12 + log($\mathrm{Ne^{2+}/H^+}$) abundances via 
$T_{\rm e}$-method assuming $t_{\rm e}$(\ion{Ne}{iii}) (left panel) and $t_{3}$ (right panel)  are compared with the results via IR${-}$method for our sample. In the top panels of Fig.~\ref{figcompne}, the differences between both estimates are plotted versus
the IR estimates. As noted earlier in Fig.~\ref{figcompne}, the difference (D) is systematic
in both cases, where D increases with $\rm Ne^{2+}/H^{+}$ from IR-method estimations. 
The average difference ($\rm <D>$) between  ionic abundances values via $T_{\rm e}$-method assuming
$t_{3}$ and IR estimates is obviously the same value  ($\sim 0.60$ dex) as the average value found for \ion{H}{ii} regions by \citet{dors2013optical}.  Therefore, probably, any artificial effects attributed to the use of heterogeneous sample of data sets,
aperture effects,  different regions in the objects which are considered in optical and IR 
observations, can have influence on our results.

The origin of D was discussed in details by \citet{dors2013optical} for \ion{H}{ii}
regions and it was attributed to be mainly the presence of abundance and/or electron 
temperature variations across the nebula rather than extinction effects in the area of the sky covered by the IR and optical observations, as proposed by 
\citet{2002A&A...391.1081V}. An overview of the discrepancy derived from this work will be presented in a subsequent paper, even though we refer to few possible scenarios here. Recently, \citet{2020MNRAS.tmp.3501D} by using the
{\sc suma} code \citep{1989ApJ...339..689V}, which assumed that the  gas ionization/heating is
due to photoionization and shocks, found that Seyfert 2 nuclei have gas shock velocities in the range of 50-300
$\mathrm{ km \: s^{-1}}$. These shocks can produce an extra gas heating source in the NLRs,
which translates into underestimation of the elemental abundances via $T_{\rm e}$-method in relation with abundances  derived from IR lines (less dependent
on temperature). As an addition  support to the presence of
electron temperature fluctuations in AGNs, \citet{2021MNRAS.501L..54R}
 presented 2D electron temperature maps, based on  Gemini GMOS-IFU observations at
spatial resolutions ranging from   110 to 280 pc, in the central region of three luminous Seyfert galaxies,
 where a large variation of temperatures  (from $\sim 8000 $ to $\ga 30\:000$ K) were derived.
 This result indicates a large fluctuation of $t_{3}$.

 The Pa$\alpha$ to Pa$\delta$ and Br$\alpha$ to Br$\delta$ emission lines are not only the strongest emission lines found in the NIR and MIR, they are also relatively free from blending features and dust attenuation. This makes them valuable tools to derive the chemical abundances of AGNs. The optical Balmer emission lines, although stronger, can suffer blending with other lines (i.e. H$\alpha$ normally blends with [\ion{N}{ii}] $\lambda6548$, $\lambda6584$ {\AA}), and at least to
some degree, are expected to be more affected by dust absorption than Paschen and Brackett emission
lines. Independent measurements of the narrow component fluxes can yield important constraints on the presence of dust within the line of sight
which could also affect the emitter regions of IR lines. In fact, effects of dust on hydrogen emission line measurements are clearly observable only in the Balmer emission line ratios but they can not be detected at a significant level using the Paschen and Brackett emission lines alone \citep{landt2008near}.

For the NIR broad emission line region (BLR) of AGNs,
\citet{landt2008near} obtained the dust extinction in the order of $A_V\sim 1$ to $\sim 2$ mag in consonance with other studies (e.g. \citealt{cohen1983narrow, crenshaw2001bsorption, crenshaw2002reddening, 2009MNRAS.394.1148S}). From these results, the effect of the dust causing the observed extinction of the narrow emission line region depends on the location of the dust, thus, being internal dust if it is mixed  with the gas phase or if it is located outside the NLR, for instance, in the host galaxy. Since the covering factor of the narrow emission line clouds is assumed to be only a few per cent, the line of sight towards the BLR will not necessarily intercept the dust. However, dust external to the NLR will act as a screen to affect the smallest scale components such as the BLR and the continuum emitted by the accretion disk. However, 
reddening in Seyfert galaxies by means of NIR line ratios performed by \citet{riffel2006spectral}, led to the fact that Sy2s tend to lie close to the locus of points of the reddening curve, with $E(B-V)$ in the interval $0.25 {-} 1.00$ mag.

Despite these drawbacks, IR transitions offer the opportunity to examine the metallicity of galaxies almost without being affected by dust extinction, therefore, it is worthwhile to be explored and used whenever possible (\citealt{moorwood1980spectroscopy, moorwood1980line, lester1987infrared, rubin1988, tsamis2003}).
For instance, the metallicities of the central and obscured regions of starburst galaxies can only be accessed via far-infrared (FIR) lines, while metallicities derived via optical lines are likely related to only the outer, less dust-extincted part of these galaxies \citep{puglisi2017, calabro2018}. Considering non-consensus on dust extinction in the NIR coupled with the fact that little is known about the shapes of the NIR
extinction curves of the Small and Large Magellanic Cloud  (for a review see, for instance, \citealt{2020ARA&A..58..529S}), it will probably take observations from FIR to settle IR dust extinction and its effects on metallicities in AGNs. As a result, we chose the approach to extinction correction problems to be most relevant to optical line fluxes, while we considered extinction to be essentially negligible for our infrared data.

\begin{figure*}
\centering
\includegraphics[width=2.1\columnwidth]{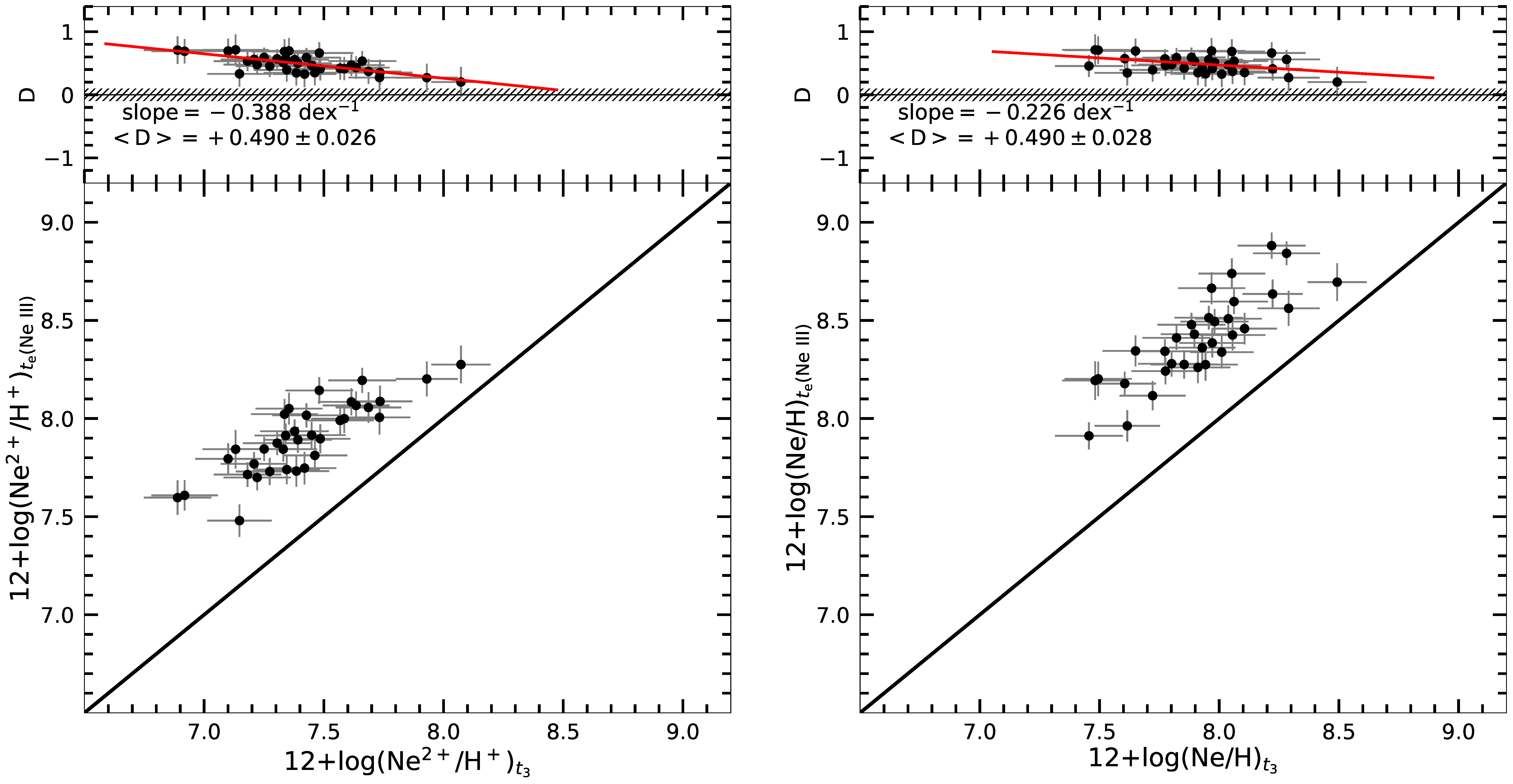}
\caption{Same as Fig.~\ref{comb_figure} but for the ionic abundance of 
12+log($\mathrm{Ne^{2+}/H^+}$) and the total abundance of 
12+log(Ne/H) derived using the $T_{\rm e}$-methods. In both panels $T_{\rm e}$-method estimates are based on $t_{\rm e}({\rm \ion{Ne}{iii}})$ versus $t_3$, as indicated.}
\label{figneh}
\end{figure*}

 We notice that the comparison between the ionic abundance of 12+log($\mathrm{Ne^{2+}/H^+}$) derived via IR$-$lines by using Brackett and Paschen series presents a linear correlation with a Pearson correlation coefficient of $R = 0.70$ (see Fig.~\ref{comb_figure}). Also, the twice neon ionic abundance estimations
 derived using the $T_{\rm e}$-method based on $t_{\rm e}({\rm \ion{Ne}{iii}})$ and $t_{3}$ have a positive linear correlation with a Pearson correlation coefficient of $R = 0.84$  (see Figs.~\ref{figcompne} and \ref{figneh}). However, we find no correlation between estimations
 derived using the $T_{\rm e}$- and IR-methods.   Separating the Paschen and Brackett series ionic abundance estimations with or without discriminating against the outliers, we do not find any significant change in the disparity of the doubly ionized neon ionic abundance trend. Consequently, we find the use of either only Paschen or only Bracket series or both to be reliable  estimations of neon ionic abundance in Seyfert~2 nuclei. Comparison of values estimated from Eqs.~\ref{eqn3.5} and \ref{eqtnet3} clearly shows a high disparity between $t_{3}$ and $t_{\rm e}({\rm \ion{Ne}{iii}})$. This discrepancy  translate into underestimations of 12 + log($\mathrm{Ne^{2+}/H^+}$) abundances by $t_{3}$ as compared to $t_{\rm e}({\rm \ion{Ne}{iii}})$ estimates. Despite this difference and the positive correlation between the $T_{\rm e}$-methods, there is no correlation between the $T_{\mathrm{e}}$- and IR-methods. Therefore, it is worthwhile investigating the non-existence of mutual relation between the $T_{\mathrm{e}}$- and IR-methods. 
 
The temperature problem in AGNs, thus, the cause of higher electron temperature values usually derived from observational $R_{\mathrm{O3}}$ ratio other than  predictions by photoionization models is a potential cause of the neon ionic abundance discrepancy. 
 It is important to highlight that the origin of the electron temperature fluctuation is an open problem in nebular astrophysics. A $t^2$ value of  $\sim 0.04$ typically results in an underestimation of C/H, O/H and Ne/H by about 0.2 to 0.3 dex (\citealt{1967ApJ...150..825P, peimbert1969chemical}). Therefore, it is extremely important to ascertain whether the fluctuations in temperature exist or whether there are inherent potential errors from the adopted methodology. If temperature variations exist, it is imperative to better understand their nature and possibly derive some methodology to reconcile them in chemical abundance derivations. It is worth noting that, hitherto, the $t^2$ values available in the literature are, in most part, indirectly based on the comparison of different methods to the estimation of  $T_{\mathrm{e}}$ and the majority of the studied objects are \ion{H}{ii} regions
 and Planetary Nebulae (PN). Only mapping the AGNs with appropriate sensitivity and spatial resolution in the temperature diagnostic lines could conceivably provide direct evidence of small or large scale fluctuations. Recently,
 \citet{2021MNRAS.506L..11R}, who used the  Gemini GMOS-IFU observations of three luminous nearby Seyfert galaxies 
 (Mrk\,79, Mrk\,348 and Mrk\,607), found electron temperature fluctuations in these objects in the same order or larger than the maximum values reported in star-forming regions and Planetary Nebulae. Thus, the discrepancy derived from optical and IR abundance estimates can be due to the presence of electron temperature fluctuations in AGNs.  Moreover, another potential source of temperature fluctuations could be the presence of density variations in the gas but we did not observe high-scale of density fluctuations in our selected sample.

Furthermore, as previously stated in this paper, aperture effect is not the primary cause of the neon ionic abundance discrepancy (e.g. \citealt{dors2013optical, dors2020chemical}). Following from the foregoing, we point out here two potential key reasons for the absence of connection between the $T_{\mathrm{e}}$- and IR-methods.
 It is worth stating from the onset that only IR tracers can explore the gas-phase elemental abundances in the interstellar medium of dusty galaxies because the IR emission lines are insensitive to interstellar reddening.  Internal dust extinction could have a significant impact on the comparison of abundances obtained from IR and optical emission lines. The blue optical [\ion{Ne}{iii}]$\lambda$3869 emission line suffers more dust absorption than the IR emission lines. Secondly, unlike optical emission lines, the emissivity of IR lines has weak dependence on electron temperature, because the atomic levels involved in the transitions are much closer to the ground state as compaered to the optical.

\subsection{Ne/H abundance}

In this work we determine for the first time the neon abundances for a large sample of local AGNs. These abundance determinations have deep implications in the studies of the chemical evolution of galaxies and stellar nucleosynthesys, mainly because, due to their localization in the disk and according to
scenario inside-out of galaxy formation (e.g. \citealt{2005MNRAS.358..521M}), 
it is expected a high metallicity in AGNs  in comparison to disk \ion{H}{ii} regions. 

\begin{table}
\centering
\caption{Parameters of the Ne/H abundance gradients in a sample of spiral galaxies.
$N$ represents the number of \ion{H}{ii} regions considered in the estimations of the
gradients. $Y{_0}$,  $grad\,Y$ and $W_{0}$ are defined in Eqs.~\ref{eqY2}
and \ref{neonsolar}. In the last column, the original works from which the Ne/H abundance values were compiled are listed.} 
\label{tabneong}
\begin{tabular}{lccccc}
\hline
 Galaxy   & $N$     & $Y{_0}$           & $grad\,Y$                   &$W_{0}$     & Reference \\
M\,33     &  6      &  $-4.23\pm0.25$   &  $-0.057\pm 0.005$          & 0.58       &  1 \\
M\,33     & 16      &  $-4.07\pm0.04$   &  $-0.058\pm 0.014$          & 0.85       &  2  \\
NGC\,2403 &  6      &  $-4.40\pm0.03$   &  $-0.008\pm 0.005$          & 0.40       &  3 \\
NGC\,3184 &  29     &  $-3.57\pm0.21$   &  $-0.080\pm 0.029$          & 2.70       &  4 \\
NGC\,628  &  35     &  $-4.23\pm0.08$   &  $-0.004\pm 0.013$          & 0.60       &  4 \\
NGC\,5194 &  8      &  $-4.01\pm0.20$   &  $-0.028\pm 0.037$          & 0.97       &  4 \\
NGC\,5457 & 70      &  $-4.05\pm0.05$   &  $-0.021\pm 0.003$          & 0.89       &  4 \\
NGC\,925  & 23      &  $-3.67\pm0.18$   &  $-0.059\pm 0.021$          & 2.13       &  5 \\
NGC\,2805 & 8       &  $-3.39\pm0.19$   &  $-0.050\pm 0.015$          & 4.07       &  5  \\
NGC\,4395 & 8       &  $-4.13\pm0.20$   &  $-0.056\pm 0.038$          & 0.74       &  5  \\
NGC\,300 & 27       &  $-4.33\pm0.04$   &  $-0.057\pm 0.016$          & 0.46       &  6  \\
\hline
\end{tabular}
\begin{minipage}{1.0\linewidth}
 {References: (1) \citet{2006ApJ...637..741C}, (2) \citet{2008MNRAS.387...45R}
 (3) \citet{2013ApJ...775..128B}, (4)\citet{2020ApJ...893...96B},
 (5) \citet{van1998spectroscopy}, (6) \citet{2009ApJ...700..309B}.}
\end{minipage}
\end{table}

IR spectra of AGNs have been obtained in many studies and certain properties have been extensively derived from them. For example, \citet{1998ApJ...498..579G},
by using ISO observations from the Infrared Astronomical Satellite (IRAS) ultraluminous galaxies, proposed a
methodology to separate the relative contribution of  AGNs and star-forming regions (see also \citealt{2007ApJ...667..149F, 2010ApJ...716.1151W, 2014MNRAS.443.1358M, 2017ApJ...838...26H}, among others).
Also, theoretical calibrations between metallicity, ionization parameter and IR emission lines 
have been proposed in the literature (e.g. \citealt{2011A&A...526A.149N, 2017MNRAS.470.1218P}). However, for the most part, these studies have not derived the elemental abundance of heavy metals (e.g. Ne, Ar, S).  

Measurements  of emission lines for the most abundant neon lines  have been undertaken
by several authors (e.g. \citealt{2011ApJ...740...94D, 2016ApJS..226...19F}) but no direct determination
of the neon abundance has been obtained either in AGNs or star-forming regions,  mainly due to difficulty in the observation of the hydrogen reference lines
and the metal lines within the same  spectral range. However, using the ISO Short Wavelength Spectrometer, where recombination hydrogen and metal lines were measured, \citet{2003A&A...403..829V} obtained IR data ($\rm 2.3 \:\la \:\lambda(\mu m) \:\la \: 45$) for 12 starburst galaxies. These authors found that Ne abundances span approximately over one up to three  times order of magnitude the solar value ($ \mathrm{1 \: \la \: (Ne/Ne_{\odot}) \: \la \: 3}$).
\citet{2009ApJS..184..230B} obtained IR observational data (from 10 to 37 $\mu$m) for 24 starburst by using the $Spitzer$ telescope and derived the Ne/H abundances ranging from $\sim$0.60 to $\sim$2 times the solar value.  Finally, \citet{2008MNRAS.389L..33W} obtained
the neon and oxygen abundances for a large sample of Planetary Nebulae and \ion{H}{ii} regions,
whose the observational data were compiled from the literature. Taking into account the findings of these aforementioned authors, we 
 can assumed for SFs   Ne/H values  ranging from $\sim 0.6$ to $\sim 3$  times the solar value.
Our Ne/H results based on $T_{\rm e}$-method  indicate a wider range of Ne/H abundances than those derived for star-forming objects, with the maximum values (see Table~\ref{tab2}) ranging from $\sim 7$ to $\sim 30$ times
the solar value when $t_{3}$ and $t_{\rm e}({\rm \ion{Ne}{iii}})$ are considered, respectively. Similarly, we find a very high maximum value considering the Ne/H estimates based on IR lines, i.e. $\sim 30$ times the solar value.
Thus, it appears the Ne/H abundances in AGNs reach higher values than Ne/H estimations in star-forming regions.

As an additional test, in order to verify the higher Ne/H abundance in AGNs
in comparison with values derived in star-forming regions,
we estimate the total neon abundance (Ne/H) in the central parts of galaxies based on the extrapolation of the radial abundance gradients of this element, which is generally found in spiral galaxies (e.g. \citealt{2002ApJ...568..679W, 2006ApJ...637..741C, 2008ApJ...675.1213R, 2009ApJ...696..729M, 2010A&A...521A...3S}). This procedure helps us to infer indirect and independent values of abundances in the nuclei of spiral galaxies (e.g. \citealt{1992MNRAS.259..121V, 1998AJ....116.2805V, 2004A&A...425..849P, zinchenko2019effective}). As usual, we assume that the Ne/H abundance 
gradient is represented by 
\begin{equation}
\label{eqY2}
Y  = Y{_0} + grad\,Y \times R (\rm kpc),
\end{equation} 
where $Y=\rm \log(Ne/H)$, $Y_0$ is the extrapolated value from the Ne/H abundance to the galactic center, i.e. at radial
distance $R=0$,
and  $grad\,Y$ is the slope of the distribution expressed in $Y$ units of $\rm dex \: kpc^{-1}$.
As pointed out by \citet{2004A&A...425..849P}, the reliability of  radial  abundance  gradient determinations is defined not only by the large number of objects considered
but also by the distribution  of these objects along the galactic radius.  Under this supposition, we take into consideration published data from the literature for Ne/H abundance values of  \ion{H}{ii} regions derived by using the $T_{\rm e}$-method and located at galactocentric distances in spiral galaxies within the range 
$0.2 \:  \:  \la \: (R/R_{25}) \: \la \: 1$, where $R$ is the galactocentric distance and $R_{25}$ is the $B$-band isophote at a surface brightness of 25 mag arcsec$^{-2}$.  In addition, Ne/H estimations in the M\,33 galaxy obtained through IR lines by \citet{2008MNRAS.387...45R} using  $Spitzer$ Space Telescope are considered.  It was possible to obtain the Ne/H abundance gradients
in 10 spiral galaxies. In Table~\ref{tabneong}, the identification of each galaxy,
the number ($N$) of \ion{H}{ii} regions considered in deriving  the Ne/H gradient,
the $Y{_0}$ and $grad\,Y$ values as well as references to the original works from which the data were obtained are listed. Also in Table~\ref{tabneong}, the extrapolation to the central
part of each galaxy of the Ne/H abundance in relation with the solar value, defined as 
\begin{equation}
\label{neonsolar}
 W_{0}=\rm (Ne/H)_{0}/(Ne/H)_{\odot}    
\end{equation}
is listed. It can be seen that the extrapolated values of $W_{0}$ range from $0.40$ to $\sim 4.0$ in Table~\ref{tabneong}, while our results indicate that AGNs have abundances of Ne/H in the range 0.30-3.00, 0.80-7.60 and 0.90-30 times the solar value, depending on the method considered (see Table~\ref{tab2}). Also, the average value of $W_{0}$ obtained in Table~\ref{tabneong} indicates that  Ne/H abundance of $\sim$1.30 times the solar value in the central parts of spiral galaxies, while  our results indicate  twice the average value of $W_{0}$ for AGNs ($\sim$2.24 times the solar value).  Therefore, we certainly find that the total neon abundances from both optical and IR-lines determinations in AGNs are  higher in comparison with  estimations from \ion{H}{ii} regions.

\subsection{Neon ICF}

The total neon abundances based on IR lines combined with the ionic   oxygen
abundance estimates present a good opportunity to obtain an expression for
the neon ICF to be applied in AGN abundance studies \citep{2003ApJ...591..801K, dors2013optical}. In most part of cases, in the optical
spectra of AGN and SFs only the [\ion{Ne}{iii}]$\lambda$3869 {\AA} line is measured, which makes the use of ICFs necessary to calculate the total neon
abundance, as suggested by \citet{peimbert1969chemical}.

\begin{figure}
\includegraphics[angle=0.0,width=1.15\columnwidth]{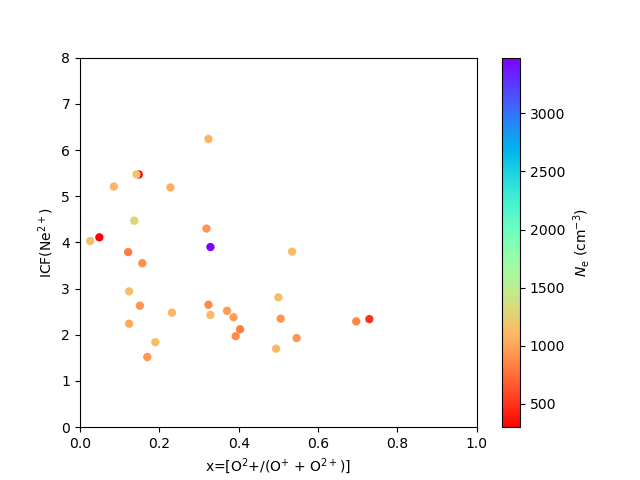}
\includegraphics[angle=0.0,width=1.15\columnwidth]{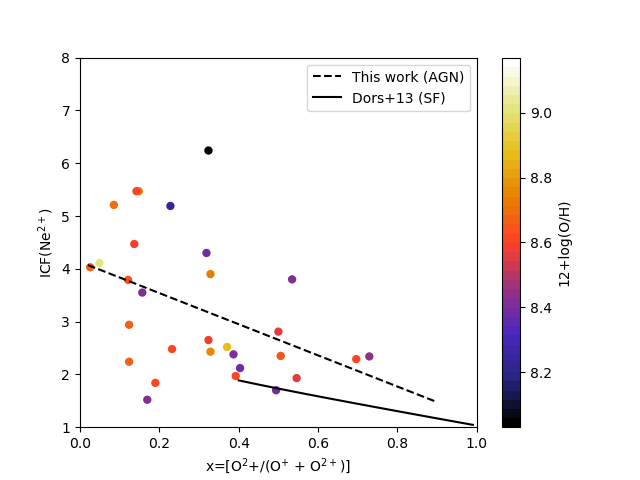}
\caption{ Relation between $\rm ICF(Ne^{2+})$ and 
and x=[$\rm O^{2+}/(O^{+}+O^{2+})$] ionic
abundance ratio. Points represent direct estimates
for our sample (see Sect.~\ref{data}) whose
$\rm ICF(Ne^{2+})$ and x are calculated by using  Eq.~\ref{eqicfnew} and the $T_{\rm e}$-method
(see Sect.~\ref{meth1}). 
Bottom panel: Red line represents a fitting to the
points obtained by using Eq.~\ref{icffit1}. Black line represent the relation for SFs derived by \citet{dors2013optical} assuming the
same methodology and given by Eq.~\ref{icfdors13}.
Colour bars indicate the 12+log(O/H) value for each object.
Top panel. As bottom panel but the colour bars  indicate the electron density ($N_{\rm e})$ for each object calculated through the [\ion{S}{ii}]$\lambda$6716/$\lambda$6731 (see Sect.\ref{meth1}).}
\label{figicfn}
\end{figure}

Neon ICFs for SFs have been proposed by several authors and, in most part,
based on photoionization models (e.g. \citealt{izotov2006chemical, 2007MNRAS.381..125P, 2021MNRAS.505.2361A}). \citet{dors2013optical}  proposed  an empirical ICF for the neon based on only infrared neon lines measurements,
i.e. free from the photoionization uncertainties. Unfortunately, no neon ICF expression has been proposed for AGN studies. 
In view of this, and following the  method proposed by  \citet{dors2013optical}, in Fig.~\ref{figicfn},
the neon ICF values for our sample  obtained from Eq.~\ref{eqicfnew} versus 
the $\rm O^{2+}/(O^{+}+O^{2+})$ abundance ratio 
are shown. Inspection of ICF(Ne$^{2+}$) values from Table~${\color{blue}\text{A5}}$
reveals a very discrepant and suspicious high ICF  value 
for NGC\,5953 in comparison with other objects, therefore, it was excluded from our analysis.
Despite the scattering, a clear relation between the 
estimates can be noted. A linear fit to the points in Fig.~\ref{figicfn}  produces
\begin{equation}
\label{icffit1}
\rm ICF(Ne^{2+})_{IR}= -2.95(\pm 1.17) \times x + 4.13(\pm 0.41),   
\end{equation}
where x=[$\rm O^{2+}/(O^{+}+O^{2+})$].  This expression is valid for  $\rm 0 \: < \: x \: < \: 0.8$, i.e. the range of values covered by our sample of objects.

Also in Fig.~\ref{figicfn},
the ICF derived for SFs by \citet{dors2013optical} given by   
\begin{equation}
\label{icfdors13}
\rm ICF(Ne^{2+})_{IR} = 2.382 - 1.301x + \frac{0.05}{x} 
\end{equation}
is shown. It can be noted in Fig.~\ref{figicfn} that AGNs present higher neon ICF values than  those of SFs for a fixed value of x, which is expected given their higher ionization degree.

 We investigate the scattering in the points observed in Fig.~\ref{figicfn}, taking into account the dependent of the ICF(Ne$^{2+}$)-x relation on some nebular parameters.
 \citet{izotov2006chemical} found, for some elements, a dependence between  ICF-x relations and the metallicity, moreover, other authors have been investigated the ICF-x dependence with other nebular parameters (e.g. \citealt{2021MNRAS.505.2361A, 2020MNRAS.492..950A, 2014MNRAS.440..536D}). 
 In order to ascertain if the dispersion in our estimations is due to a reliance on metallicity (measured by O/H), as found by \citet{izotov2006chemical}, the points in Fig.~\ref{figicfn} (bottom panel) are indicated in accordance with their oxygen abundances.
Also in Fig.~\ref{figicfn} (top panel), the scattering of the ICF-x relation due to the electron density is considered. Since infrared emission-line intensities are involved in the ICF determinations
and these present some dependence on the electron density,
some effects from this parameter on the ICF could be derived.
It can be observed from Fig.~\ref{figicfn} that, the point positions are independent from O/H abundance and $N_{\rm e}$ values. Probably, a larger sample of
objects from both infrared and optical emission lines measured
with high signal-to-noise ratio, which makes it possible to derive
reliable physical properties could  help to improve our understanding of the source of this scattering.

\subsection{Ne/O versus O/H}

The primary origin of neon is derived from the stellar nucleosynthesis theory, which predicts that neon and oxygen are formed by stars of similar masses  (e.g. \citealt{1995ApJS..101..181W}).
Thus, if  stars are formed  following an universal Initial Mass Function
\footnote{For a discussion on the universality of the IMF see, for example, \citet{2010ARA&A..48..339B}.} (IMF, \citealt{1955ApJ...121..161S}), 
the Ne/O abundance ratio must not be dependent on 
O/H abundance (or on metallicity). However, several studies on this subject have yielded conflicting results. On the one hand,  \citet{2008MNRAS.389L..33W} used direct abundance values
from PN and \ion{H}{ii} regions, leading to the findings which suggested that the Ne/O ratio increases with O/H in both types of nebulae. Additionally, 
\citet{2011A&A...529A.149G} also used a large sample of SFs and
found a slight increase in  Ne/O with O/H, which was interpreted by these authors as if this small increment would be likely due to a stronger depletion of oxygen onto dust grains in higher metallicity objects. On the other hand, 
several authors have derived a constant relation between
Ne/O and O/H based on independent sample of data and
ICFs (e.g. \citealt{2003ApJ...591..801K,  dors2013optical, 2016ApJ...830....4C, 2020MNRAS.496.1051A}). 

\begin{figure*}
\centering
\includegraphics[angle=0,width=2.1\columnwidth]{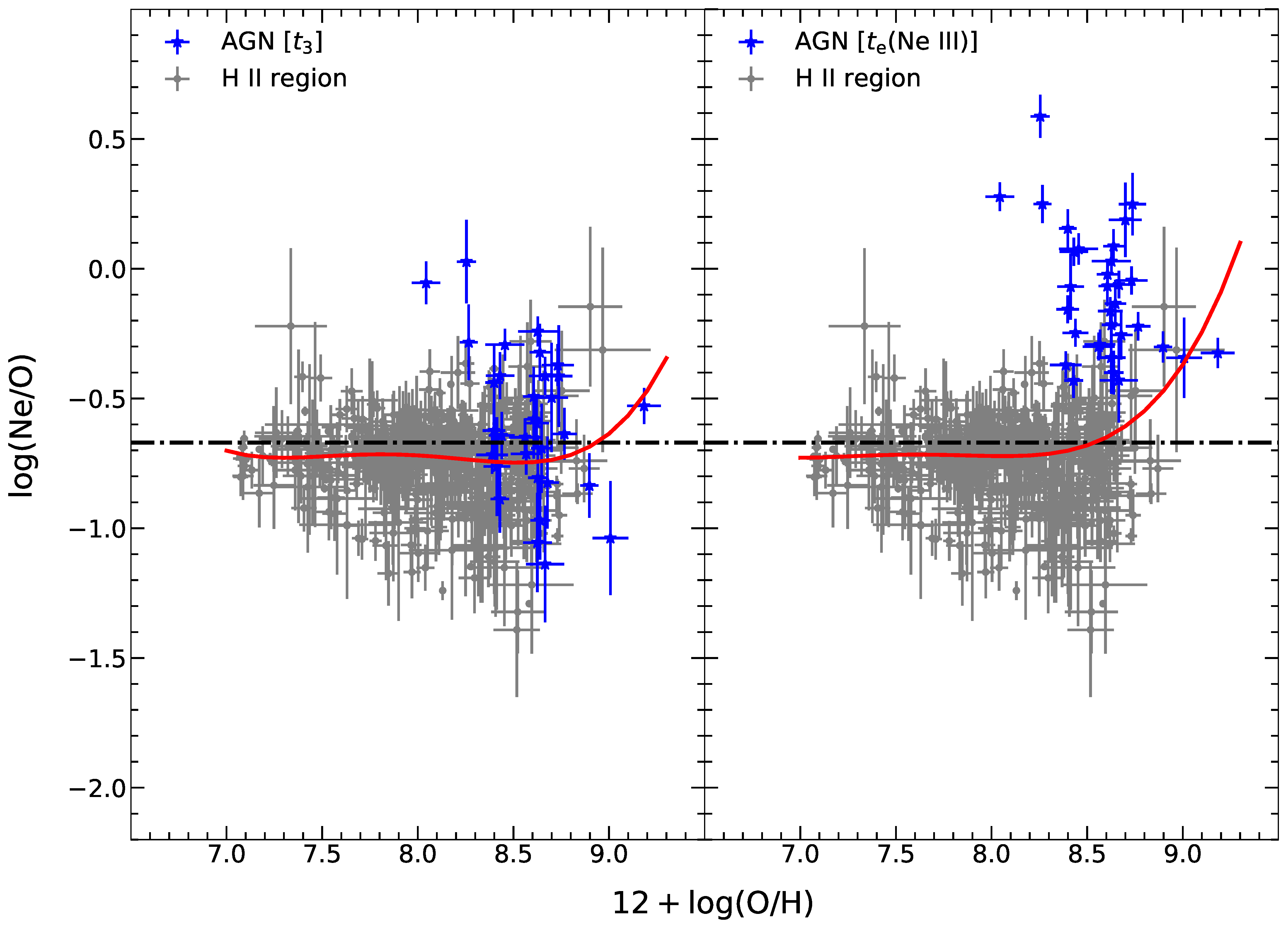}
\caption{Relation between log(Ne/O) and 12+log(O/H). 
Blue points represent  estimations for our sample of Seyfert~2 obtained by using $T_{\rm e}$-method where Ne abundances
are calculated assuming $t_{3}$ (left panel) and $t_{\rm e}({\rm \ion{Ne}{iii}})$
(right panel). Grey points represent estimations obtained by using $T_{\rm e}$-method
for star-forming regions (\ion{H}{ii} regions and \ion{H}{ii} galaxies) taken from CHAOS  project \citep{2015ApJ...806...16B, 2015ApJ...808...42C, 2016ApJ...830....4C, 2020ApJ...893...96B, 2020ApJ...894..138S},    \citet{2006MNRAS.372..293H,  2008MNRAS.383..209H}  and \citet{2007MNRAS.381..125P} . The red solid lines represent the polynomial fits to the points (Eqs.~\ref{eqneo1} and \ref{eqneo2}, respectively). Dashed black  line represents log(Ne/O) = $- 0.67$ \citep{2012A&A...539A.143N}.}
\label{fig10}
\end{figure*}

In the advent of the CHemical Abundances of Spirals (CHAOS)  project, thousands of direct abundances for the heavy elements have been possible in  \ion{H}{ii} regions located in spiral disks \citep{2015ApJ...806...16B, 2015ApJ...808...42C, 2016ApJ...830....4C, 2020ApJ...893...96B, 2020ApJ...894..138S}. These \ion{H}{ii} regions present  a wide range of metallicities [$\rm 7.8 \: \la \: 12+\log(O/H) \: \la \: 9.0$ or $ 0.10 \: \la \: (Z/{\rm Z_{\odot}}) \: \la \: 2$] and play an important role in the chemical abundance studies. 
This homogeneous sample combined with star-forming data from the literature and our abundance results
 expand  direct abundance determination in the emitting line  objects at $(Z/{\rm Z_{\odot}}) \: \ga \:3$,
providing a unique opportunity to analyse the neon  nucleosynthesis  in
the widest range of metallicity than previous studies.
 In Fig.~\ref{fig10},  we show the Ne/O versus O/H results for our AGN sample, considering neon estimations based on $T_{\mathrm{e}}{-}$ method assuming  $t_{3}$ (left panel) and  $t_{\rm e}$(\ion{Ne}{iii}) (right panel).
Estimates from the CHAOS project and abundance results of star-forming regions (\ion{H}{ii} regions and \ion{H}{ii} galaxies) taken from the literature, as well as polynomial fits to these estimations, are also shown in Fig.~\ref{fig10}.  Considering all the estimates
(SFs and AGNs) we found

\begin{equation}
\label{eqneo1}
{\rm \log(Ne/O)}_{t_{3}}=  ~{a_1}x^4 +{b_1} x^3 + {c_1}x^2 + {d_1} x  + {e_1} 
\end{equation}
\noindent and
\begin{equation}
\label{eqneo2}
{\rm \log(Ne/O)}_{t_{\rm e}(\ion{Ne}{iii})}=   ~{a_2}x^4 +{b_2} x^3 + {c_2}x^2 + {d_2} x  + {e_2}
\end{equation}
where $a_1 = 0.153$, $b_1 = -4.825$, $c_1 = 5.689 \times 10^{+1}$, $d_1 = -2.975\times 10^{+2}$, $e_1 = 5.816 \times 10^{+2}$, $a_2 = 1.084 \times 10^{-1}$, $b_2 = -3.279$, $c_2 = 3.713 \times 10^{+1}$, $d_2 = -1.865 \times 10^{+2}$, $e_2 = 3.500 \times 10^{+2}$ and $x$ = 12 + log(O/H).

 In Fig.~\ref{fig10},  we observe a  better agreement between SF estimates and those for AGNs when $t_3$ is assumed (left panel) instead of $t_{\rm e}$(\ion{Ne}{iii}) (rigth panel). For the very high metallicity regime [$\rm 12+\log(O/H)\:\ga \: 8.80$ or $(Z/\rm Z_{\odot}) \: \ga \: 1.3$] an oversolar Ne/O abundance is derived, which is more conspicuous in the estimations via $t_{\rm e}$(\ion{Ne}{iii}).
 \citet{2020MNRAS.496.3209D}, by using photoionization model
results, found that theoretical relations between temperatures
derived for AGNs differ considerably from those for \ion{H}{ii}
regions. This is due to the fact that AGNs present a very different ionization structure caused by, for instance, gas outflows
(e.g. \citealt{riffel2018outflows}) and gas shocks in the ionized-neutral region transition \citep{2020MNRAS.tmp.3501D}.
In fact, recently, \citet{2021MNRAS.501L..54R} obtained from Gemini Multi-Object Spectrograph-integral field unit observations at spatial resolutions of 110--280 pc  of three  luminous Seyfert galaxies: Mrk 79, Mrk 348, and Mrk 607. These authors found shocks due to gas outflows play an important role in the observed temperature distributions, which can produce very different electron temperature
distribution than those in \ion{H}{ii} regions (see, for instance,
\citealt{2021MNRAS.506L..11R}). Based on these results, we suggest that
$T_{\rm e}(\ion{Ne}{iii})$ must be used in the derivation of
$\rm Ne^{2+}$ ionic abundance, instead of $t_{3}$.

The observed increase in Ne/O can be attributed to two factors. First, it can be explained by the fact that   higher dust depletion of oxygen occurs in the NLRs than in SFs.  Some fraction of the oxygen, in order of 0.1-0.2 dex,  is expected to be trapped in dust grains in SFs  \citep{1998MNRAS.295..401E} and in the Interstellar Medium  (ISM) along the Galactic disk \citep{2006ApJ...641..327C, 2009ApJ...700.1299J}. While AGNs may have a higher rate of oxygen depletion onto dust in molecular clouds, it is unlikely that their abundance values vary significantly from SFs abundance estimations
(e.g. \citealt{1994ApJ...436L.131S}). Moreover, \citet{1997ApJS..110..287F} and \citet{2003AJ....125.1729N} concluded that refractory elements are not depleted in the coronal line region of  NLRs, indicating a low dust abundance in AGNs, probably due to the destruction of grains
by the hard radiation from the supermassive black hole accretion disk.
Therefore, in principle, we can exclude the oxygen depletion as the origin for high Ne/O values in AGNs.

In Fig.~\ref{fig10}, we also notice that a value of 0.5 dex oxygen depletion in NLRs is necessary to conciliate the high Ne/O abundance values with those derived for the majority of the objects. However, such level of depletion is not observed in SFs and in the ISM. Additionally, the Ne/O increase with O/H  is noted in both AGNs
and SFs.  Furthermore, the Ne/O deviation from applying $t_{\rm e}({\rm \ion{Ne}{iii}})$ is not due to the ICF, because the ICF was applied to both $t_3$ and $t_{\rm e}({\rm \ion{Ne}{iii}})$ estimates, and the Ne/O from $t_3$ still agrees with SFs estimations.  The total neon abundance estimations
 derived using the $T_{\rm e}$-method based on $t_{\rm e}({\rm \ion{Ne}{iii}})$ and $t_{3}$ have a positive linear correlation with a Pearson correlation coefficient of $R = 0.83$ (see Fig.~\ref{figneh}).  Therefore, it is unlikely that the offset in Fig.~\ref{fig10} is due to oxygen depletion.
Another plausible  explanation for the Ne/O increase with O/H at
high metallicity is that neon, in a similar way as nitrogen and carbon, may have a secondary origin in stellar nucleosynthesis, but at an oversolar metallicity. The stellar nucleosynthesis  studies by \citet{1995ApJS..101..181W} and even more recent
studies (e.g. \citealt{1999ApJS..125..439I, 2006ApJ...653.1145K, 2011MNRAS.414.3231K, 2018MNRAS.480..538R}) did not investigate star formation in environments with metallicities
higher than the solar value, despite the fact that 
$Z$ appears to have an impact on the stellar product
(e.g. \citealt{2021arXiv210314050G}).

\section{Conclusions}
\label{conc}

We compiled infrared and optical emission line fluxes from the literature for 35 Seyfert 2 galaxies in the local universe
($0\: \la  z \: \la \: 0.06$) and these emission lines were used to derive the ionic $\mathrm{Ne^{2+}/H^+}$  and elemental Ne/H  abundances through the $T_{\mathrm{e}}$-method and the IR-method. Also, O/H abundances
were derived by using the $T_{\mathrm{e}}$-method for our sample.
We obtained the following conclusions:

\begin{enumerate}

\item  We derived   $\mathrm{Ne^{2+}/H^+}$ ionic abundances  using  optical and IR emission lines. We found that the ionic abundance
ratio derived via IR emission lines are higher than those
calculated from optical lines by the factors  of $\mathrm{0.69 \pm 0.03~dex}$  and $\mathrm{ 0.20 \pm 0.02~dex}$ when
 $t_3$ and $t_{\rm e}({\rm \ion{Ne}{iii}})$
are assumed  in the determinations relying on the $T_{\rm e}$-method, respectively.

\item The $\mathrm{Ne^{2+}/H^+}$ abundance differences derived from the comparison between the $T_{\mathrm{e}}{-}$method (assuming $t_{3}$ to derive $\mathrm{Ne^{2+}/H^+}$) and the IR$-$method estimations are similar to those derived in  nearby \ion{H}{ii} regions.

\item  We found no correlation between estimations derived
using the $T_{\mathrm{e}}{-}$ and IR$-$methods.

\item  We have demonstrated  from  photoionization model results that, the assumption  $T_{\rm e}(\ion{O}{iii}) \approx T_{\rm e}(\ion{Ne}{iii})$  which is valid in \ion{H}{ii} regions, is not applicable to AGNs. As a result, we proposed a new  relation between  electron temperature $T_{\rm e}({\rm \ion{Ne}{iii}})$ and $T_{\rm e}({\rm \ion{O}{iii}})$, i.e.
the temperatures in the gas phase where the $\rm Ne^{2+}$
and $\rm O^{2+}$ are located, respectively.

\item  We proposed a semi-empirical Ionization Correction Factor (ICF) for neon based on [\ion{Ne}{ii}]12.81$\micron$, [\ion{Ne}{iii}]15.56$\micron$ which is derived
from oxygen ionic abundance ratio x=[$\rm O^{2+}/(O^{+}+O^{2+})$].
The scattering in the ICF(Ne$^2$)-x relation does not
correlate with the O/H abundance as well as the electron density.

\item We found that the average Ne/H value in AGNs is a factor of 2 times higher than estimations for  star-forming  regions (SFs). The maximum Ne/H abundance derived for our sample  spans from 8 to 30 times the solar value, a factor of $\sim4$-$10$
times the maximum Ne/H value derived in SFs.

\item  An increase in Ne/O with O/H was observed for the very high metallicity regime [$\rm 12+log(O/H)\: \ga \: 8.80$] when estimates for SFs  are combined with the ones for AGNs. We suggest that this phenomenon is due to secondary stellar production of the neon at very high metallicity regime rather than oxygen depletion onto dust.

\end{enumerate}

\section*{Acknowledgements}

We appreciate the detailed revision by the referee, Dr. Brent Groves, which has considerably improved this work. MA gratefully acknowledges support from Coordenação de Aperfeiçoamento de Pessoal de Nível Superior (CAPES). OLD and ACK are grateful to 
to  Funda\c c\~ao de Amparo \`a Pesquisa do Estado de S\~ao Paulo (FAPESP) and Conselho Nacional
de Desenvolvimento Cient\'{\i}fico e Tecnol\'ogico (CNPq). CPA is grateful for the financial support from FAPESP. AF acknowledges support from grant PRIN MIUR 2017-20173ML3WW4-001. RF and RAR acknowledge financial support from CNPq (202582/2018-3, 304927/2017-1,  400352/2016-8 and 312036/2019-1) and FAPERGS (17/2551-0001144-9 and 16/2551-0000251-7).

\section{DATA AVAILABILITY}
The data underlying this article will be shared on reasonable request
with the corresponding author.




\bibliographystyle{mnras}
\bibliography{marc} 



 


\begin{table*}
\renewcommand\thetable{A1}
\renewcommand{\arraystretch}{1.8}
\setlength{\arrayrulewidth}{1.2pt}
\LARGE
\centering\footnotesize
\caption{Flux (in units of 10$^{-14}$ erg cm$^{-2}$ s$^{-1}$) of [\ion{Ne}{iii}]$\lambda$12.81$\micron$, [\ion{Ne}{iii}]$\lambda$15.56$\micron$, 
Paschen and Brackett series for selected Seyfert 2 nuclei. In last but one and last columns, the redshift ($z$) and
the original works where the data were compiled are listed, respectively.}
\label{tableA1}
\resizebox{\textwidth}{!}{%
\Large
\begin{tabular}{@{}lcccccccccccccc@{}}
\hline
Object & [\ion{Ne}{iii}]$\lambda$12.81$\micron$ & [\ion{Ne}{iii}]$\lambda$15.56$\micron$  & $\mathrm{Pa\delta}$~$\lambda 10052$ {\AA} & $\mathrm{Pa\gamma}$~$\lambda 10941$ {\AA} & $\mathrm{Pa\beta}$~$\lambda 12822$ {\AA} & $\mathrm{Pa\alpha}$~$\lambda 18756$ {\AA} &  $\mathrm{Br11}$~$\lambda 16811$ {\AA} &  $\mathrm{Br\delta}$~$\lambda 19451$  {\AA} &  $\mathrm{Br\gamma}$~$\lambda 21661$ {\AA} &  $\mathrm{Br\beta}$~$\lambda 26259$ {\AA} & $\mathrm{Br\alpha}$~$\lambda 40523$ {\AA}  &  Redshift ($z$)  & Ref. & \\

\hline

NGC\,3081 & $12.62 \pm 1.16$ & $36.46 \pm 1.25$ & --- & ---  & 5.46  & --- & ---&  --- & --- &    --- & 1.12 & 0.00798 & [1, 5, 6] & \\

 NGC\,4388 & $79.74 \pm 4.76$ &$108.18 \pm 1.56$ &  --- & --- & 8.09 &  --- & --- &     --- & 1.13 & & 3.22 & 0.00842 & [1, 5, 7] & \\

NGC\,4507 & $33.73 \pm 2.63$&  $28.78 \pm 0.63$ & --- &  --- &  --- &  --- &  --- &  --- &  --- &   --- & 3.65 & 0.01180 & [1, 6] & \\
 
 NGC\,5135  &$ 112.00 \pm 0.00$  & $58.00 \pm 0.00$ & --- & ---  & ---   & ---   & ---  &--- & $1.65 \pm  0.5$  & --- & 7.9 & 0.01369 & [4, 9, 28] & \\

NGC\,5643 & $38.00 \pm 0.00$ & $56.00 \pm 0.00$ & --- &--- & 3.5 &  --- &  --- &  --- & --- & 4.0 & 2.92 & 0.00400 & [2, 4, 6, 7, 10] & \\

NGC\,5728 & $30.44 \pm 1.81$ & $54.76 \pm 0.51$ &  $0.545 \pm 0.142$ & $0.380 \pm 0.105$ & $0.737 \pm 0.116$ & $2.063 \pm 0.192$ &  --- & --- &  $0.211 \pm 0.019$ & --- & --- & 0.00935 & [1, 11] & \\

IC\,5063 & $28.22 \pm 3.34$ & $73.67 \pm 4.61$ &  --- &  --- & ---  &  --- & --- &  --- & $1.00 \pm 0.03$ &  $<5.0$ & --- & 0.01135 & [1, 6, 17] & \\
 
IC\,5135 & $71.00 \pm 5.00$ & $37.00 \pm 2.00$ &  --- &  --- &  --- &  --- &  --- &  --- & $1.02 \pm 0.08$ & --- & 8.3 & 0.01615 & [5, 9, 16, 28] & \\
 
MRK\,3 &$86.00 \pm 12.00$ & $207.00 \pm 29.00$ &  --- & ---  & 11.5  & ---  & ---   &  --- & $6.20 \pm 0.40$ & --- & --- & 0.01351 & [5, 16, 25] & \\
 
MRK\,273 & $44.49 \pm 0.79$ &$33.81 \pm 0.25$ &  --- & ---  &   --- &  8.84 &  --- &   0.70
 & $0.73 \pm 0.04$ & 4.2 & 4.4 & 0.03778 & [8, 9, 12, 29] & \\

MRK\,348 & $15.34 \pm 0.74$  &  $20.60 \pm 0.79$&  $0.27 \pm 0.04$ & $0.71 \pm 0.12$ & $1.21 \pm 0.06$ & $3.55 \pm 0.15$  &  --- & $0.33 \pm 0.13$ & $0.301 \pm 0.042$ & --- & --- & 0.01503 & [1, 14] & \\
 
 MRK\,573 & $13.00 \pm 0.00$  &   $24.0 \pm 0.00$ &  $0.327 \pm 0.016$ & $0.611 \pm 0.04$ & $0.958 \pm 0.017$ & $4.557 \pm 0.028$  &   --- & $0.137 \pm 0.006$ & $0.277 \pm 0.009$ & --- & --- & 0.01718 & [10, 11] & \\
 
NGC\,1068 &  $538.34 \pm 37.3$ & $1432.20 \pm 76.87$&  --- & --- & --- &  --- &  --- &    --- & 12.7 & 41.0 & 69.0 & 0.00379  & [5, 10, 12, 15] & \\

NGC\,2992 & $53.65 \pm 3.66$  & $61.06 \pm 1.98$ & 2.65 & 3.7  & 5.1  & 8.8  & 0.56 &  --- & 1.16 & --- & 6.65 &  0.00771  &  [1, 6, 19]   & \\

NGC\,5506 & $91.75 \pm 3.31$  & $152.13 \pm 9.13$ &  --- & ---  & 85.1 &  ---  &  --- & ---  & 11.8 & 7.0 & 12.0 &  0.00618 & [1, 5, 10] & \\

NGC\,7674 &  $ 18.00 \pm 1.00$&  $46.00 \pm 2.00$ & $0.838 \pm 0.067$ & $1.387 \pm 0.131$ & $1.036 \pm 0.058$ & $3.206 \pm 1.009$  &    $2.566 \pm 0.452$ &   $0.338 \pm 0.051$ & $0.313 \pm 0.025$ & ---  &--- & 0.02892 & [11, 16] & \\

$\mathrm{I\:Zw\:92}$ & $24.00 \pm 0.00$ & $16.00 \pm 0.00$ & --- &  --- & 10.4 & ---  & ---  &  --- & 0.953 & --- & --- & 0.03780 & [5, 10] & \\

 NGC\,2110 & $60.19 \pm 5.34$  & $47.40 \pm 0.71$& $0.300 \pm 0.027$ & $1.266 \pm 0.103$ & $1.491 \pm 0.086$ & $3.295 \pm 0.450$ & ---  & ---  & $0.250 \pm 0.022$ & --- & 2.11 &  0.00779  & [1, 6, 11] & \\

NGC\,5929 &  $13.20 \pm 0.34$ & $9.83 \pm 0.31$ &  --- & $0.379 \pm 0.027$
 & $0.768 \pm 0.020$ &  --- &  --- & --- & $0.135 \pm 0.025$ & --- & --- & 0.00831 & [11, 13] & \\

 MRK\,463E & $10.82 \pm 0.35$ & $40.46 \pm 0.73$ & --- &  --- & 3.01 &  --- &  ---& --- &   0.272 & 4.0 & --- & 0.05035 & [5, 8] & \\
 
 MRK\,622 & $6.00 \pm 2.00$ & $8.00 \pm 2.00$ & --- &  --- & 1.02 &  --- &  --- &  --- &  --- &  --- & --- & 0.02323 & [5, 16] & \\
 
NGC\,1386 & $17.8 \pm 1.02$ & $36.6 \pm 0.72$ & --- & --- & 3.50  & --- & --- & --- & $0.176 \pm 0.014$ & --- & ---  & 0.00290  & [2, 3, 13] & \\

NGC\,7582 & $250.94 \pm 3.53$  & $105.00 \pm  2.05$ &  --- &  --- & 7.8 &  --- &  --- &   --- & $4.4 \pm 0.4$ & 9.0 & $20.6 \pm 7.0$ &  0.00525  & [2, 10, 13, 18] & \\

 NGC\,1275 &  $46.15 \pm 0.80$  &$22.37 \pm 0.56$ & $1.332 \pm  0.108$ & $8.353 \pm  0.425$  & $6.066 \pm 0.315$ & $14.514 \pm  0.252$ &   --- & $1.398 \pm  0.206$ & $0.977 \pm  0.041$  & 2 & 22 &  0.01756  & [8, 10, 11] & \\

Circinus & $453.6 \pm 14.5$ & $400.00 \pm 9.00$ & --- & 10.4 &  --- &  --- &  --- &  --- & 3.8 & 32 & 15.0 &   0.00145  & [8, 10, 20] & \\
 
Centaurus\,A & $221.00 \pm 4.50$ &$140.00 \pm 1.20$ & --- & 19 & 16 &  --- &  --- &  ---  & 2.7 & 9.0  & 8.0 &  0.00183  & [3, 8, 10, 26] & \\
 
Cygnus\,A & $26.7 \pm 0.3$ & $41.30 \pm 0.40$  & --- &  --- &  --- & $2.6 \pm 0.2$ &  --- &  --- & $0.26 \pm  0.08$ & --- & --- & 0.05607 & [21, 22] & \\
  
MRK\,266SW & $57.00 \pm 0.00$& $28.00 \pm 0.00$ & --- &  & 5.51 &  --- & --- &  --- & 0.367 & --- & 4.5 &  0.02760  & [4, 5, 9] & \\

MRK\,1066 & $10.94 \pm 0.21$ & $46.91 \pm 0.76$ & $0.974 \pm 0.030$ & $2.553 \pm 0.120$ & $5.407 \pm 0.024$ & $14.574 \pm 1.060$  & $0.398 \pm 0.055$ & $0.867 \pm 0.004$ & $1.416 \pm 0.022$&  ---  &  ---  & 0.01202 & [8, 11] & \\

NGC\,1320 & --- & $9.00 \pm 1.00$  & --- &  --- &  --- &  --- & ---  &  ---  & $0.094 \pm 0.01$ & --- & --- & 0.00888 & [16, 25] & \\

NGC\,1667 & $10.1 \pm 3.00$  & $7.23 \pm 3.00$ & --- &  --- &  --- &  --- &  --- &  --- & $0.018 \pm 0.004$ & --- & --- &  0.01517 & [16, 25] & \\

NGC\,3393 & --- & $95.00 \pm 0.00$ & --- &  --- &  --- &  --- &  --- &  --- & $0.46 \pm 0.005$ & --- & --- & 0.01251 & [24, 27] & \\

NGC\,5953 & $105.00 \pm 2.00$ &$21.00 \pm 1.00$ & --- &  --- & 0.544 & $1.982 \pm 0.083$ &  --- &  --- & 0.277 & --- & --- &  0.00656 & [11, 16, 23] & \\

NGC\,7682 & $5.46 \pm 0.25$ & $8.07  \pm 0.15$ & $0.118  \pm 0.018$ & $0.498  \pm 0.112$ & $0.992  \pm 0.065$ & $3.182  \pm 0.081$ &  --- &   --- & $0.194  \pm 0.011$ & --- & --- & 0.01714 & [1, 11] & \\  
 
$\mathrm{ESO428\:-\:G014}$ & --- & $168.01 \pm 0.00$  & $0.919 \pm 0.063$ & $3.104 \pm 0.115$  & $4.526 \pm 0.057$ & $10.205 \pm 0.078$ & $0.822 \pm 0.029$  & $0.300 \pm 0.049$ & $0.898 \pm 0.014$ &  --- &  ---  & 0.00566 & [11, 24] & \\

\hline
\end{tabular}%
}

\begin{minipage}[l]{17.5cm}

{References: (1) \citet{2010ApJ...716.1151W}, (2) \citet{2000MNRAS.316....1W}, (3) \citet{2002MNRAS.331..154R}, (4) \citet{pereira2010infrared}, (5) \citet{veilleux1997infrared}, (6) \citet{lutz2002}, (7) \citet{2017MNRAS.464.1783O}, (8) \citet{2011ApJ...740...94D}, (9) \citet{2010ApJ...721.1233I}, (10) \citet{sturm2002}, (11) \citet{riffel2006spectral}, (12) \citet{1995ApJ...444...97G}, (13) \citet{2010ApJ...709.1257T},  (14) \citet{2009ApJ...694.1379R}, (15) \citet{2009MNRAS.398.1165G}, (16) \citet{2007ApJ...671..124D}, (17) \citet{moorwood1988}, (18) \citet{1989ApJ...337..230K}, (19) \citet{gilli2000}, (20) \citet{oliva1994}, (21) \citet{2012ApJ...747...46P}, (22) \citet{1991ApJ...382..115W}, (23) \citet{2005MNRAS.364.1041R}, (24) \citet{2011ApJS..195...17W}, (25) \citet{vanderlaan2013}, (26) \citet{1999MNRAS.308..431B}, (27) \citet{2001ApJS..136...61S}, (28) \citet{1997ApJS..108..449G} and (29) \citet{1999ApJ...522..139V}.}

\end{minipage}

\end{table*}

\begin{table*}
\renewcommand\thetable{A2}
\renewcommand{\arraystretch}{1.8}
\setlength{\arrayrulewidth}{1.2pt}
\LARGE
\centering\footnotesize
 \caption{Observed reddening-uncorrected optical emission-line intensities of Seyfert 2 nuclei compiled from the literature.
 The last column is the list of references for the original works where the data were obtained.}
\label{tableA2}
\resizebox{\textwidth}{!}{%
\Large
\begin{tabular}{@{}lccccccccccccc@{}}

\hline

Object  & [\ion{O}{ii}] $\lambda3727$ {\AA} & [\ion{Ne}{iii}] $\lambda3869$ {\AA} & [\ion{O}{iii}] $\lambda4363$ {\AA} & [\ion{O}{iii}] $\lambda4959$ {\AA} & [\ion{O}{iii}] $\lambda5007$ {\AA}
& [\ion{O}{i}] $\lambda6300$ {\AA} & H$\alpha~\lambda6563$ {\AA} & H$\beta~\lambda4861$ {\AA} & [\ion{N}{ii}] $\lambda6584$  {\AA} & [\ion{S}{ii}] $\lambda6717$ {\AA} & [\ion{S}{ii}] $\lambda6731$ {\AA} & Ref. & \\

\hline

NGC\,3081  & 1.47 &	0.88 &	0.20 &	4.53 &	13.30 &	0.37 &	4.53 &	1.00 &	3.87 &	0.99 &	1.07  & 1 & \\

 NGC\,4388  & 1.72 &	0.48 &	0.13 &	3.83 &	11.20 &	0.78 &	4.86 &	1.00 &	2.59 &	1.27 &	1.12 & 1 & \\

NGC\,4507  & 1.64 &	0.71 &	0.27 &	3.17 &	9.53 & 0.86 &	5.16 &	1.00 &	2.80 &	1.10 &	1.23
 & 1 & \\
 
 NGC\,5135   & 1.06  &	0.42  &	0.08  &	1.49  &	4.82  &	0.31  &	6.12  &	1.00  &	5.45  &	0.92  &	0.87 & 1 & \\

NGC\,5643  & 2.68  &	0.89  &	0.32  &	4.85  &	16.60  & 1.16  & 6.17  &	1.00  &	7.17  &	2.40  &	2.21 & 1 & \\

NGC\,5728  & 1.84 &	0.75 &	0.34 &	3.92 &	11.80 &	1.00 &	5.97 &	1.00 &	8.36 &	0.99 &	0.97 & 1 & \\
 
IC\,5063  & 2.90  &	0.75  &	0.22  &	3.55  &	11.00  &	0.68  &	5.55  &	1.00  &	3.44  &	1.50  &	1.31  & 1 & \\
 
IC\,5135  & 2.15 &	1.04 &	0.19 &	2.20 &	7.41 &	0.60 &	6.07 &	1.00 &	7.56 &	1.19 &	1.11
 & 1 & \\

MRK\,3 & 2.21 &	0.94 &	0.19 &	4.16 &	13.46 &	1.14 &	5.31 &	1.00 &	5.48 &	1.30 &	1.46
 & 2 & \\
 
MRK\,273 & 3.05 &	0.71 &	0.13 &	5.39 &	17.96 &	1.22 &	28.20 &	3.06 &	29.30 &	17.50 & 5.15 & 2, 3 & \\
 
 MRK\,348 & 3.05 &	1.23 &	0.21 &	3.96 &	12.33 &	1.58 &	4.27 &	1.00 &	3.54 &	1.74 &	2.01 & 2 & \\

 MRK\,573  & 2.11  &	1.01  &	0.15  &	4.01  &	12.64  &	0.43  &	4.30  &	1.00  &	3.62  &	1.12  &	1.21 & 2 & \\

NGC\,1068  & 0.76  &	0.94  &	0.17  &	4.28  &	13.22  &	0.62  &	4.47  &	1.00  &	7.94  &	0.48  &	0.99   & 2 & \\

NGC\,2992 &	0.19 &	0.04 &	0.01 &	0.32 &	1.00 &	0.15 &	1.73 &	0.13 &	1.00 &	0.46 &	0.41
 & 4 & \\

NGC\,5506  & 0.14  &	0.04  &	0.01  &	0.31  &	1.00  &	0.11  &	0.87  &	0.12  &	0.80  &	0.32  &	0.34 & 4 & \\

NGC\,7674  & 1.08  &	0.98  &	0.11  &	3.99  &	12.82  &	0.38  &	4.62  &	1.00  &	4.62  &	0.69  &	0.81 & 5 & \\

$\mathrm{I\:Zw\:92}$  &	1.95  &	0.94  &	0.28  &	3.60  &	10.50  &	0.55  &	3.54  &	1.00  &	1.43  &	0.55  &	0.60 & 5 & \\

 NGC\,2110$^{a}$  &	21.10  &		4.20  &		0.73  &		11.17  &	33.50  &	8.20  &	18.00  &	4.30 &	34.00 &	7.90 &	9.80
 & 6 & \\

NGC\,5929$^{a}$  &	14.80  & 2.21  &	0.40  &	4.23  &	12.70  &	8.20  &	20.60  &	4.40  &	12.10  &	7.60  &	6.70 & 6 & \\
 
 MRK\,463E   &	0.21  &	0.07  &	0.013  &	0.33  &	1.00  &	0.055  &	0.48  &	0.13  &	0.23  &	0.10  &	0.09 & 7 & \\
 
 MRK\,622    &	0.49  &	0.06  &	0.004  &	0.33  &	1.00  &	0.041  &	1.88  &	0.16  &	1.77  &	0.32  &	0.35  & 7 & \\
 
 NGC\,1386$^a$   &	1.81  &	0.77  &	0.19  &	3.78  &	11.34  &	0.46  &	4.70  &	1.00  &	5.60  &	1.04  &	1.29 & 8 & \\
 
 NGC\,7582  & 124.10 & 32.80 & 2.90 & 71.60 & 214.70 & 8.70 & 286.00 & 100.00 & 186.90 & 41.80 & 38.80 & 9 & \\
 
 NGC\,1275  & 2.90  & 1.43  &	0.33  &	4.33  &	12.99  & 1.48 &	5.44 &	1.00 &	5.44 &	1.33
 &	4.52 & 10 & \\

 Circinus  &  78.00  &	41.00  &	16.00  & 317.00  & 1048.00  &	46.00  & 565.00  &	100.00  &	154.00  & 128.00  & 113.00 & 11 & \\
 
Centaurus\,A & 2.49 & 0.48 & 0.10 & 2.38 & 6.28 & 2.05 & 7.27 & 1.00 & 10.83 & 5.24 & 4.17 & 12 & \\
 
Cygnus\,A &	2.44 &	0.66 &	0.16 &	4.08 & 13.11 & 2.10 & 6.61 &	1.00 &	13.07 & 3.65 & 3.29 & 13 & \\
  
MRK\,266SW  &	5.20  &	0.90  &	0.08  &	1.50  &	4.50  &	0.38  &	3.30  &	1.00  &	3.68  &	0.54  &	0.46 & 14 & \\

MRK\,1066  & 0.32  & 0.08  & 0.01  & 0.31  &	1.00  &	0.15  &	1.80  &	0.23  &	1.58  &	0.36  &	0.39 & 15 & \\

NGC\,1320  & 0.38 &	0.49 &	0.29 &	3.57 &	9.86 & 0.38 & 4.86 &	1.00 &	3.36 &	0.93 & 1.07 & 16, 17 & \\

NGC\,1667   & 12.08  &	1.98  &	0.42  &	3.99  &	11.10  & 0.94  & 3.03  &	1.00  &	6.59  &	2.86
 &	9.72 & 18, 19 & \\

NGC\,3393 &	155.00 & 77.00 &	10.00 &	341.00 & 1030.00  &	34.00  &	359.00  &  100.00 &	492.00 &	202.00 & 686.80  & 20  & \\

NGC\,5953   & 2.60  &	0.90  &	0.12  &	1.70  &	4.30  &	0.32  &	2.90  &	1.00  &	4.00  &	0.80  &	0.84  & 21 & \\

NGC\,7682$^{a}$  &	575.00  &	158.00  &	77.40  &	1310.00  &	3930.00  &	167.00  &	470.00  &	100.00  &	515.00  &	134.00  &	141.00 & 22 & \\

$\mathrm{ESO428\:-\:G014}$  & 2.49  &	1.13  &	0.28  &	4.20  &	13.60  &	0.49  &	3.55  &	1.00  &	4.03  &	1.07  &	1.14  & 23 & \\

 \hline
\end{tabular}%
}

\begin{minipage}[l]{17.5cm}
{References: (1) \citet{phillips1983nearby}, (2) \citet{koski1978spectrophotometry}, (3) \citet{2017ApJ...846..102M}, (4) \citet{1980ApJ...240...32S}, (5) \citet{kraemer1994spectra}, (6) \citet{1999ApJ...523..147F}, (7) \citet{1981ApJ...250...55S}, (8)  \citet{bennert2006}, (9) \citet{dopita2015probing}, (10) \citet{1975PASP...87..879S}, (11) \citet{oliva1994}, (12) \citet{1981MNRAS.197..659P}, (13) \citet{1975ApJ...197..535O}, (14) \citet{1983ApJ...273..478O}, (15) \citet{goodrich1983galaxies}, (16) \citet{1986ApJ...301...98D}, (17) \citet{2017ApJS..232...11T}, (18) \citet{1993ApJ...417...63H}, (19) \citet{radovich1996}, (20) \citet{2000ApJS..129..517C}, (21) \citet{1996MNRAS.281..781G}, (22)  \citet{durret1994} and (23) \citet{bergvall1986}.}

{Note: $^{a}$Value of $I$([\ion{O}{iii}]$\lambda4959$) estimated from the theoretical relation $I$[\ion{O}{iii}]$\lambda4959$ = $I$[\ion{O}{iii}]$\lambda5007/3.0$ \citep{2000MNRAS.312..813S}.}

\end{minipage}

\end{table*}

\begin{table*}
\renewcommand\thetable{A3}
\renewcommand{\arraystretch}{1.5}
\centering
\caption{Reddening-corrected optical emission-line intensities  (relative to H$\beta$=1.0) of Seyfert 2 nuclei compiled from the literature. Original works which the data were obtained are presented in Table~\ref{tableA2}.}
\label{tableA3}
\resizebox{\textwidth}{!}{%
\Large
\begin{tabular}{@{}lcccccccccccc@{}}
\hline
Object  & [\ion{O}{ii}] $\lambda3727$ {\AA} & [\ion{Ne}{iii}] $\lambda3869$ {\AA} & [\ion{O}{iii}] $\lambda4363$ {\AA} & [\ion{O}{iii}] $\lambda4959$ {\AA} & [\ion{O}{iii}] $\lambda5007$ {\AA} & [\ion{O}{i}] $\lambda6300$ {\AA} & H$\alpha~\lambda6563$ {\AA}  & [\ion{N}{ii}] $\lambda6584$  {\AA} & [\ion{S}{ii}] $\lambda6717$ {\AA} & [\ion{S}{ii}] $\lambda6731$ {\AA} & c(H$\beta$) & \\

f($\lambda$)  & 0.302 &	0.260 &	0.125 &	 $- \:0.022$ & $- \:0.033$ &	 $- \:0.285$ &	$- \:0.326$ & $- \:0.329$ & $ - \:0.349$ & $- \: 0.350$ & \\

\hline

NGC\,3081  & 2.26 &	1.28 &	0.24 &	4.39 &	12.69 &	0.25 &	2.85 & 2.42 &	0.60 &	0.65 &	0.6192 & \\

 NGC\,4388  & 2.82 & 0.74 & 0.16 & 3.70 & 10.61 & 0.49 & 2.84 & 1.51 & 0.72 & 0.63 & 0.7139 & \\

NGC\,4507  & 2.85 &	1.14 &	0.34 &	3.05 &	8.97 &	0.51 &	2.84 &	1.53 &	0.58 &	0.65 &	0.7946 & \\
 
 NGC\,5135   &	2.16 &	0.78 &	0.11 &	1.42 &	4.46 &	0.16 &	2.84 &	2.51 &	0.40 &	0.38 &	1.0243  & \\

NGC\,5643 & 5.50 &	1.66 &	0.43 &	4.60 &	15.35 &	0.59 &	2.83 &	3.27 &	1.05 &	0.96 &	1.0352 &  \\

NGC\,5728  & 3.66 &	1.36 &	0.45 &	3.73 &	10.95 &	0.52 &	2.84 &	3.94 &	0.45 &	0.44 &	0.9909 & \\
 
IC\,5063  &	5.39 &	1.28 &	0.28 &	3.39 &	10.28 &	0.38 &	2.84 &	1.75 &	0.73 &	0.64 &	0.8927 & \\
 
IC\,5135  & 4.35 &	1.91 &	0.25 &	2.09 &	6.86 & 0.31	& 2.84 & 3.51 &	0.53 &	0.49 &	1.0132 & \\

MRK\,3  & 3.94 &	1.55 &	0.24 &	3.99 &	12.64 &	0.66 & 2.84 &	2.91 &	0.67 &	0.75 &	0.8331  & \\
 
MRK\,273  &	2.98 &	0.60 &	0.07 &	1.63 &	5.21 &	0.14 &	2.82 &	2.90 &	1.62 &	0.47 &	1.5755 & \\
 
 MRK\,348 & 4.44 &	1.70 &	0.25 &	3.85 &	11.84 &	1.11 &	2.85 &	2.35 &	1.13 &	1.30 &	0.5396 & \\

 MRK\,573 &	3.09 &	1.40 &	0.18 &	3.90 &	12.13 &	0.30 &	2.85 &	2.39 &	0.72 & 0.78 &	0.5491 & \\

NGC\,1068  & 1.15 &	1.35 &	0.20 &	4.15 &	12.63 &	0.42 &	2.85 &	5.03 &	0.30 &	0.61 &	0.6013 & \\

NGC\,2992 &	6.16 &	1.17 &	0.18 &	2.22 &	6.58 &	0.30 &	2.81 &	1.60 &	0.67 &	0.59 &	2.0702 & \\

NGC\,5506 & 2.78 &	0.69 &	0.13 &	2.43 &	7.58 &	0.40 &	2.83 &	2.58 &	0.98 &	1.03 &	1.2524 & \\

NGC\,7674  & 1.69 &	1.44 &	0.13 &	3.86 &	12.21 &	0.25 &	2.84 & 2.83 &	0.41 &	0.48 &	0.6457 & \\

$\mathrm{I\:Zw\:92}$  & 2.38 &	1.12 &	0.30 &	3.55 &	10.27 &	0.46 &	2.85 &		1.15 &	0.44 &	0.48 &	0.2872  & \\

 NGC\,2110 & 7.01 &	1.33 &	0.20 &	2.53 &	7.49 &	1.36 &	2.85 &	5.36 &	1.22 &	1.51 &	0.5129 & \\

NGC\,5929  & 5.33 &	0.75 &	0.11 &	0.93 &	2.75 &	1.21 &	2.84 &	1.66 &	1.01 &	0.89 &	0.6636  & \\
 
 MRK\,463E  & 2.05 &	0.61 &	0.11 &	2.50 &	7.49 &	0.34 &	2.85 &	2.30 &	0.57 &	0.52 &	0.3439 & \\
 
 MRK\,622  & 11.48 &	1.25 &	0.04 &	1.87 &	5.41 &	0.07 &	2.81 &	2.61 &	0.43 &	0.47 &	1.9026 & \\
 
 NGC\,1386  & 2.88 & 1.15 &	0.23 &	3.66 &	10.78 &	0.30 &	2.84 &	3.37 &	0.61 &	0.75 &	0.6688 & \\
 
 NGC\,7582 & 1.24 &	0.33 &	0.03 &	0.72 &	2.15 &	0.09 &	2.86 &	1.87 &	0.42 &	0.39 &	0.0000 & \\
 
 NGC\,1275  & 5.29	& 2.40 &	0.42 &	4.15 &	12.17 &	0.84 &	2.84 &	2.82 &	0.66 &	2.25 &	0.8657  & \\

Circinus  &	1.47 &	0.71 &	0.21 &	3.03 &	9.78 &	0.25 &	2.84 &	0.77 &	0.61 &	0.54 &	0.9167 & \\
 
Centaurus\,A   & 5.96 &	1.02 &	0.14 &	2.23 &	5.71 &	0.90 &	2.83 &	4.18 &	1.91 &	1.51 &	1.2561 & \\
 
Cygnus\,A   & 5.34 &	1.30 &	0.22 &	3.86 &	12.04 &	1.00 &	2.83 & 5.56 &	1.48 &	1.32 &	1.1280 & \\
  
MRK\,266SW & 5.94 &	1.01 &	0.08 &	1.49 &	4.43 &	0.33 &	2.86 &	3.18 &	0.46 &	0.39 &	0.1927 & \\

MRK\,1066 &	3.57 &	0.74 &	0.08 &	1.26 &	3.92 &	0.27 &	2.83 & 2.46 &	0.53 &	0.57 &	1.3554 & \\

NGC\,1320 & 0.62 &	0.74 &	0.36 &	3.44 &	9.34 &	0.24 &	2.84 &	1.96 &	0.52 &	0.60 &	0.7139 & \\

NGC\,1667 	& 12.75	& 2.07 &	0.43 &	3.97 &	11.03 &	0.89 &	2.86 &		6.21 &	2.69 &	9.13 &	0.0777 & \\

NGC\,3393 & 1.92 &	0.93 &	0.11 &	3.36 &	10.06 &	0.28 &	2.85 &	3.90 &	1.58 &	5.36 &	0.3061 & \\

NGC\,5953  & 2.63 &	0.91 &	0.12 &	1.70 &	4.29 &	0.32 &	2.86 &	3.94 &	0.79 &	0.83 &	0.0187 & \\
 
NGC\,7682  & 9.15 &	2.36 &	0.94 &	12.67 &	37.36 &	1.08 &	2.84 &	3.10 &	0.78 &	0.82 &	0.6688 & \\

$\mathrm{ESO428\:-\:G014}$  & 3.05 &	1.13 &	0.30 &	4.14 &	13.30 &	0.40 &	2.85 &	3.23 &	0.85 &	0.90 &	0.2910 & \\

\hline
\end{tabular}%

}

\end{table*}

\begin{table*}
\renewcommand\thetable{A4}
\renewcommand{\arraystretch}{1.0}
\setlength{\arrayrulewidth}{0.8pt}
\caption{Ionic and total neon abundances for the Seyfert~2 sample obtained through IR-method using the methodology described in Sect.~\ref{tneon}.
The abundances $\rm 12+\log(Ne^{+}/H^{+})$ and $\rm 12+\log(Ne^{2+}/H^{2+})$
are calculated by using the Eqs.~17 to 20.
The term $f$ represents the correction for the total neon abundance
$\rm 12+\log(Ne/H)$ due to the presence of ions with ionization
stages higher than Ne$^{2+}$ (see Eq.~\ref{eqicfnew}) which is derived from photoionization models by \citet{2020MNRAS.492.5675C}.}
\label{tableA4}
\resizebox{\textwidth}{!}{%
\begin{tabular}{@{}lcccc@{}} 
\hline

Object        &  $\rm 12+\log(Ne^{+}/H^{+})_{\rm IR}$  &   $\rm 12+\log(Ne^{2+}/H^{+})_{\rm IR}$ &   $f$  &  $\rm 12+\log(Ne/H)_{\rm IR}$ \\
\hline                    		        			        	  
NGC\,3081     &  $8.05 \pm 0.07$     & 	$8.19 \pm 0.07$  &    1.13    & $8.48 \pm 0.07$	                  \\

NGC\,4388     & $8.40 \pm 0.06$   & $8.21 \pm 0.06$	  &    1.11    &	$8.64 \pm 0.06$        	  \\
NGC\,4507     & $7.99 \pm 0.09$  & $7.60 \pm 0.09$			       &    1.03    &	 $8.15\pm0.09$       	  \\
NGC\,5135     & $8.29 \pm 0.10$	& 	$7.68 \pm 0.10$     &    1.03    &	  $8.40\pm0.08$      	  \\
NGC\,5643     & $8.35 \pm 0.11$    & $8.20 \pm 0.12$       &    1.03    &	 $8.60\pm0.09$       	  \\
NGC\,5728     & $8.84 \pm 0.03$  & $8.88 \pm 0.01$   &    1.04    &	     $9.18\pm0.01$   	  \\
IC\,5063      & $7.84 \pm 0.03$    & $7.93 \pm 0.03$	       &    1.03    &	 $8.25\pm0.03$        	  \\
IC\,5135      & $8.24 \pm 0.03$     & $7.63 \pm 0.03$       &    1.04    &	      $8.35\pm0.03$    	  \\
MRK\,3        & $8.03 \pm 0.07$    & $8.09 \pm 0.07$       &    1.06    &	      $8.39\pm0.07$    	  \\
MRK\,273      & $8.19 \pm 0.04$  & 	$7.75 \pm 0.04$   &    1.04    & $8.35\pm0.04$	        	  \\
MRK\,348      & $8.38 \pm 0.03$   & $8.19 \pm 0.03$       &    1.04    &	         $8.62\pm0.03$	  \\
MRK\,573      & $8.35 \pm 0.09 $    & $8.29 \pm 0.09$       &    1.11    &	        $8.67\pm0.06$ 	  \\
NGC\,1068     & $8.03 \pm 0.06$	 & 	$8.14 \pm 0.06$		       &    1.32    &	         $8.51\pm 0.06$	  \\
NGC\,2992     &  $8.18\pm 0.03$ &  $7.92 \pm 0.03$   &    1.04    & $8.39\pm0.03$	   		  \\
NGC\,5506     &  $7.73\pm 0.05$   & $7.63\pm 0.05$	 &    1.06    &	  $8.01\pm0.05$  		  \\
NGC\,7674     & $8.33\pm0.03$	 & 	$8.41\pm0.02$  &    1.27    &	$8.78\pm0.02$ 	  \\
IZw\,92      & $7.85\pm0.12$	   & $7.36\pm0.12$			       &    1.04    &	   	$7.99\pm0.10$ 	  \\
NGC\,2110     &  $8.91\pm0.01$   & $8.49\pm0.01$			       &    1.04    &	  $9.07\pm0.01$  		  \\
NGC\,5929     & $8.58 \pm 0.03$			      & $8.14 \pm 0.03$				       &    1.06    &	   $8.74\pm0.03$ 		  \\
MRK\,463E     & $7.91 \pm 0.07$ & $8.16 \pm 0.07$			       &    1.10    &	   	$8.40\pm0.07$ 	  \\
MRK\,622      & $8.11 \pm 0.10$	  & 	$7.91 \pm 0.09$		       &    1.03    &  	$8.33\pm0.09$	  \\
NGC\,1386     & $8.38 \pm 0.02$		 & 	$8.37 \pm 0.02$		       &    1.06    &	   	$8.70\pm0.02$	  \\
NGC\,7582     & $8.48 \pm 0.06$    & $7.78 \pm 0.06$			       &    1.04    &	   	$8.58\pm0.06$	  \\
NGC\,1275     & $8.19\pm 0.02$   & 	$7.56\pm 0.02$   &    1.03    &	        	 $8.29\pm0.02$ \\
Circinus      & $8.54 \pm 0.05$   & $8.17 \pm 0.05$     &    1.13    &	   	$8.75\pm0.05$	  \\
Centaurus\,A  & $8.38 \pm 0.05$     & 	$7.87 \pm 0.05$       &    1.04    &	   	$8.52\pm0.05$	  \\
Cygnus\,A     & $8.62 \pm 0.02$     & 	$8.49 \pm 0.02$		       &    1.04    &	   $8.87\pm0.02$		  \\
MRK\,266SW    & $8.49 \pm 0.11$    & 	$7.86 \pm 0.11$		       &    1.04    &	$8.60 \pm  0.10$	  \\
MRK\,1066     & $7.62 \pm 0.01$     & 	$8.00 \pm 0.01$       &    1.03    &	   $8.16\pm0.01$		  \\
NGC\,1320     & ---	 & $8.22 \pm 0.07$			       &    1.09    &	 ---  		  \\
NGC\,1667     & $9.32 \pm 0.11$    & $8.86  \pm 0.11$ 	       &    1.05    &	  $9.47\pm0.11$ 		  \\
NGC\,3393     &   ---		  & $8.55 \pm 0.09$			       &    1.20    &	   	---	  \\
NGC\,5953     & $9.43 \pm 0.05$	    & $8.41 \pm 0.05$			       &    1.03    &	  $9.48\pm0.05$  		  \\
NGC\,7682     & 	$8.19\pm 0.04$     & $8.04 \pm 0.03$ 			       &    1.05    &	   	$8.44\pm0.03$ 	  \\
ESO428$-$G014 &  ---	   & 	$8.58 \pm 0.09$     &    1.07    & --- \\

\hline
\end{tabular}%
}
\end{table*}

\begin{table*}
\renewcommand\thetable{A5}
\renewcommand{\arraystretch}{1.6}
\setlength{\arrayrulewidth}{0.8pt}
\caption{Estimates of Ne ionic and total abundances based on the electron temperatures $t_3$ and $t_{\rm e}({\rm \ion{Ne}{iii}})$ for the Seyfert~2 sample.}
\label{tableA5}
\begin{tabular}{lccccc}
\hline
Object       &  ${\rm 12+\log(Ne^{2+}/H^{+})}_{t_{3}}$   &  $ 12+ {\rm \log(Ne^{2+}/H^{+})}_{t_{\rm e}({\rm \ion{Ne}{iii}})}$	   &	ICF(Ne$^{2+}$)  & ${\rm 12+\log(Ne/H)}_{t_{3}}$  &  $ 12+{\rm \log(Ne/H)}_{t_{\rm e}({\rm \ion{Ne}{iii}})}$   \\
\hline			 
NGC\,3081    &  	$ 7.57 \pm 0.14 $			      & 			$ 7.99 \pm 0.07 $				  &      1.93	      &	$ 7.85 \pm 0.14 $		     &  $ 8.28 \pm 0.07 $			       \\
NGC\,4388    &  $ 7.46 \pm 0.14 $			      & 	$ 7.81 \pm 0.08 $				  &	 2.81	      & $ 7.91 \pm 0.14 $ 		     &  $ 8.26 \pm 0.08 $				       \\
NGC\,4507    &  $ 7.10 \pm 0.14 $			      & 		$ 7.80 \pm 0.08 $				  &	 3.55	      & 	$ 7.65 \pm 0.14 $		     &  	$ 8.34 \pm 0.08 $				       \\
NGC\,5135    &  $ 7.18 \pm 0.14 $			      & 	$ 7.71 \pm 0.06 $		  &	 5.19	      & 	$ 7.90 \pm 0.14 $			     &  	$ 8.43 \pm 0.06 $					       \\
NGC\,5643    &  	$ 7.43 \pm 0.14 $				      & 	$ 8.02 \pm 0.06 $					  &	 2.48	      & $ 7.82 \pm 0.14 $		     &  	$8.41 \pm 0.06 $				       \\
NGC\,5728    &  	$ 7.13 \pm 0.14 $				      & 		$ 7.84 \pm 0.10 $					  &	 2.24	      & $ 7.48 \pm 0.14 $			     &  $8.19 \pm 0.10 $			       \\
IC\,5063     &  $ 7.34 \pm 0.13 $				      & 		$7.91 \pm 0.06 $					  &	 1.84	      & $ 7.61 \pm 0.13$		     &  	$8.18 \pm 0.06 $			       \\
IC\,5135     &  	$ 7.34 \pm 0.14 $			      & 	 $ 8.02 \pm 0.08 $				  &	 5.21	      &	$ 8.05 \pm 0.14 $		     &  	$ 8.74 \pm 0.08 $				       \\
MRK\,3       &  $ 7.63 \pm 0.14 $				      & 		 $ 8.07 \pm 0.07 $				  &	 1.97	      & 	$ 7.93 \pm 0.14 $			     &  $ 8.36 \pm 0.07 $			       \\
MRK\,273     &  	$ 7.42 \pm 0.13 $			      & 	 $7.75 \pm 0.08 $				  &	 3.90	      & 	$ 8.01 \pm 0.13 $			     &  	$ 8.34 \pm 0.08 $				       \\
MRK\,348     &  $ 7.61 \pm 0.14 $				      & 	$ 8.09 \pm 0.07 $			  &	 2.65	      & $ 8.04 \pm 0.14 $			     &  	$ 8.51 \pm 0.07 $      \\
MRK\,573     &  $ 7.73 \pm 0.13 $				      & 	 $ 8.09 \pm 0.08 $				  &	 2.35	      & $ 8.11 \pm 0.13 $ 		     &  $ 8.46 \pm 0.08 $       \\
NGC\,1068    &  $ 7.69 \pm 0.14 $				      & 	$ 8.06 \pm 0.08 $				  &	 2.34	      & $ 8.06 \pm 0.14 $	 		     &  $ 8.43 \pm 0.08 $  \\
NGC\,2992    &  $ 7.30 \pm 0.14 $				      & 		$7.87 \pm 0.06 $				  &	 2.94	      & 	$ 7.77 \pm 0.14 $		     &  $ 8.34 \pm 0.06 $				       \\
NGC\,5506    &  $ 7.35 \pm 0.14 $				      & 	 $7.74 \pm 0.07 $				  &	 2.38	      & 	$ 7.72 \pm 0.14 $			     &  $ 8.12 \pm 0.07 $       \\
NGC\,7674    &  $ 7.93 \pm 0.13 $			      & 	 $8.20 \pm 0.09 $				  &	 2.29	      & 	$ 8.29 \pm 0.13 $			     &  $ 8.56 \pm 0.09 $	       \\
IZw\,92      &  $ 7.25 \pm 0.14 $				      & 		 $7.84 \pm 0.06 $				  &	 4.30	      & 	$ 7.88 \pm 0.14 $			     & $ 8.48 \pm 0.06 $  \\
NGC\,2110    &  	$ 7.37 \pm 0.14 $				      & 	 $7.93 \pm 0.06 $				  &	 3.79	      & 	$ 7.96 \pm 0.14 $			     &  $ 8.51 \pm 0.06 $			       \\
NGC\,5929    &  $ 6.89 \pm 0.14 $				      & 	 $ 7.60 \pm 0.09 $				  &	 4.03	      & $ 7.49 \pm 0.14 $			     &  $ 8.20 \pm 0.09 $    \\
Mrk\,463E    &  $ 7.39 \pm 0.14 $				      & 		$ 7.73 \pm 0.08 $				  &	 1.70	      & 	$ 7.62 \pm 0.14 $	 		     & $ 7.96 \pm 0.08 $     \\
Mrk\,622     &  $ 8.07  \pm 0.12 $			      & 		$ 8.28  \pm 0.10 $				  &	 2.63	      & 	$ 8.49 \pm 0.12$		     &  $ 8.70 \pm 0.10 $		       \\
NGC\,1386    &  $ 7.45 \pm 0.14 $				      & 			$ 7.92 \pm 0.07 $				  &	 2.12	      &  $ 7.78 \pm 0.14 $		     &  $ 8.24 \pm 0.07 $     \\
NGC\,7582    &  $ 7.15 \pm 0.13 $			      & 			$ 7.48 \pm 0.08 $				  &	 6.24	      &  $ 7.94 \pm 0.13 $		     &  $ 8.27 \pm 0.08 $   \\
NGC\,1275    &  $ 7.48 \pm 0.14 $				      & 		$ 8.14 \pm 0.07 $			  &	 5.47	      &  $8.22  \pm 0.14 $		     &  	$ 8.88 \pm 0.07 $   \\
Circinus     &  $ 7.22 \pm 0.14 $				      & 		$ 7.70 \pm 0.07 $			  &	 3.80	      &  $ 7.80 \pm 0.14 $			     &  $ 8.28 \pm 0.07 $    \\
Centaurus\,A &  $ 7.33 \pm 0.14 $				      & 		$ 7.84 \pm 0.06 $				  &	 4.47	      &  $ 7.98 \pm 0.14 $		     &  	$ 8.49 \pm 0.06 $       \\
Cygnus\,A    &  $ 7.59 \pm 0.14 $				      & 		$ 8.00 \pm 0.07 $				  &	 2.43	      &  $ 7.97 \pm 0.14 $			     & $ 8.39 \pm 0.07 $    \\
Mrk\,266SW   & $ 7.48 \pm 0.12 $			      & 			$ 7.90 \pm 0.07 $				  &	 5.47	      &  $ 8.22 \pm 0.12 $		     & $ 8.63 \pm 0.07 $     \\
Mrk\,1066    &  $ 7.27 \pm 0.14 $				      & 			$ 7.73 \pm 0.07 $			  &	 1.52	      &  $ 7.46 \pm 0.14 $			     &  $ 7.91 \pm 0.07 $	       \\
NGC\,1320    &  $ 6.92 \pm 0.14 $				      & 			$ 7.61 \pm 0.08 $			  &	 ---	      & 	--- 		     &  	---				       \\
NGC\,1667    &  $ 7.35 \pm 0.14 $				      & 			$ 8.05 \pm 0.08 $				  &	 4.11	      &  $ 7.97 \pm 0.14 $			     &  $ 8.66 \pm 0.08 $   \\     
NGC\,3393    &  $ 7.73 \pm 0.13 $			      & 			$ 8.01 \pm 0.09 $				  &	 ---	      & 	--- 		     &  	 ---				       \\     
NGC\,5953    &  $	7.21 \pm 0.14 $			      & 			$ 7.77 \pm 0.06 $			  &	11.83	      &  $ 8.28 \pm 0.14 $			     &  $ 8.84 \pm 0.06 $			       \\     
NGC\,7682    &  $	7.66 \pm 0.14 $				      & 			$ 8.19 \pm 0.06 $			  &	 2.52	      &  $ 8.06 \pm 0.14 $			     &  $ 8.60 \pm 0.06 $	       \\     
ESO428$-$G014&  $	7.39 \pm 0.14 $					      & 			$ 7.89 \pm 0.07 $				  &	 ---	      & 	--- 		     &  	---				       \\     
\hline						 				
\end{tabular}
\end{table*}

\begin{table*}
\renewcommand\thetable{A6}
\renewcommand{\arraystretch}{1.6}
\setlength{\arrayrulewidth}{0.8pt}
\caption{Estimates for the Seyfert~2 sample of electron temperature $t_3$ (in units of $10^{4}$ K), ionic and total oxygen abundances, ionization correction factor (ICF) for the oxygen, and the logarithm of Ne/O assuming $t_3$ and $t_{3}(\ion{Ne}{iii})$ in the Ne derivations.}
\label{tableA6}
\resizebox{\textwidth}{!}{%
\begin{tabular}{@{}lccccccc@{}}
\hline
Object       & $t_{3}$& $\rm 12+\log(O^{+}/H^{+})$  &  $\rm 12+\log(O^{2+}/H^{+})$ & ICF(O) & $\rm 12+\log(O/H)$ & log(Ne/O)$_{t_{3}}$ & log(Ne/O)$_{t_{3}(\ion{Ne}{iii})}$  \\
\hline
NGC\,3081    & $1.48\pm 0.15$ & $8.06\pm 0.05$ 		    &	  $	8.14\pm 0.13$		   & $1.45\pm 0.07$   &	$8.57 \pm 0.08$ 		 &     $-0.71\pm 0.08$  	       &	  $-0.29 \pm 0.06$  		      \\			      
NGC\,4388    & $1.34 \pm 0.12$ & $8.17\pm 0.06$ 		    &	  $	8.17\pm 0.12$		   & $1.21\pm 0.03$   &	$8.56 \pm 0.08$		 &     $-0.65\pm 0.07$            &	  $-0.30\pm 0.05$  		      \\			      
NGC\,4507    & $2.18 \pm 0.29$ & 	$8.33\pm 0.11$ 	  &	  	$7.62\pm 0.13$		   & 1.00   &	$8.41 \pm 0.07$ 		 &     $-0.76\pm 0.19$            &	  $-0.07\pm 0.13$  		      \\			      
NGC\,5135    & $1.71\pm 0.20 $ & 	$8.06 \pm 0.06$ 		    &	  $	7.53\pm 0.13$ 		   & $1.22\pm 0.03$   &	$8.27 \pm 0.05$ 		 &     $-0.28\pm 0.15$            & $+0.25\pm 0.07$  		      \\			      
NGC\,5643    & $1.85\pm 0.23$  & $8.51  \pm 0.08$ 		    &	  	$7.98 \pm 0.13$ 		   & 1.00   &	$8.63 \pm 0.05$ 		 &     $-0.81 \pm 0.16$            &	  $-0.22\pm 0.09$  		      \\			      
NGC\,5728    & $2.29\pm 0.32$ & $8.46 \pm 0.11$ 		    &	$7.67 \pm 0.13$ 		   & $1.22\pm 0.03$   &	$8.63\pm 0.08$		 &     $-1.06\pm 0.19$            &	  $-0.34 \pm 0.14$ 		      \\			      
IC\,5063     & $1.80\pm 0.20$ & $8.48\pm 0.07$ 		    &	  $7.85\pm 0.13$ 	   & $ 1.15 \pm 0.02$  &	$8.64\pm 0.05$		 &     $-0.97\pm 0.15$           &	  $-0.40\pm 0.08$  		      \\			      
IC\,5135     & $2.16\pm 0.29$ & 	$8.50\pm 0.11$		    &	$7.50\pm0.13$ 	& $1.41\pm 0.06$   &	$8.70\pm 0.09$	 &     $-0.50\pm 0.21$ & $+ 0.19\pm 0.14$  		      \\			      
MRK\,3       & $1.50\pm 0.15$ & $8.31\pm 0.05$ 		    &	$8.11\pm 0.13$ 	   & $1.25\pm 0.04$   &	$8.62\pm 0.06$	 & $-0.59\pm 0.10$&	 $-0.16 \pm 0.05$  		      \\			      
MRK\,273     & $1.29\pm 0.11$ & $8.21\pm 0.06$ 	& $	7.89\pm 0.12$		   & $2.24 \pm 0.18$   &	$8.73\pm 0.08$		 &     $-0.37 \pm 0.08$            &	  $-0.05\pm 0.05$  		      \\			      
MRK\,348     & $1.58\pm 0.17$ & $8.35\pm 0.05$  		    &	$  	8.03 \pm 0.13$ 		   & $1.19 \pm 0.03$  &	$8.61\pm 0.06$ 		 &     $-0.49\pm 0.11$ &$-0.02\pm 0.06$      \\			      
MRK\,573     & $1.34\pm 0.12$ & $8.21\pm 0.06$  		    &	$8.22\pm 0.12$ 		   & $1.40 \pm 0.06$   & $	8.67\pm 0.09$ 		 &     $-0.41\pm 0.07$  & $-0.06\pm 0.05$  		      \\			      
NGC\,1068    & $1.38\pm 0.13$ & $7.78\pm 0.06$ & $8.21\pm 0.13$    & $1.28\pm 0.04$   &	$8.43\pm 0.08$  &     $-0.29\pm 0.06$&  $+0.08\pm 0.06$  		      \\			      
NGC\,2992    & $1.80\pm 0.22$ & $8.54\pm 0.07$  & $7.65\pm 0.13$  & $1.20\pm 0.03$    &	$8.68\pm 0.06$ 		 &     $-0.82\pm 0.18$ & $-0.25\pm 0.10$  		      \\			      
NGC\,5506    & $1.43\pm 0.14$ & $8.15\pm 0.05$  & $7.94\pm 0.13$    & $1.19\pm 0.03$    &	$8.44\pm 0.07$ 		 &     $-0.64\pm 0.09$            &	  $-0.25\pm 0.05$ 		      \\			      
NGC\,7674    & $1.18\pm 0.09$ & $8.01\pm 0.07$ &	 $8.37\pm 0.12$ 		   & $1.24\pm 0.04$    &$	8.63\pm 0.10$ 		 &   $-0.24\pm 0.06$           &	  $+0.03\pm 0.05$		      \\			      
IZw\,92      & $1.86\pm 0.23$  & $8.14\pm 0.08$  & $7.82	\pm 0.13$ 	   & $1.19\pm 0.03$    &	$8.40\pm 0.05$ 		 &     $-0.44\pm 0.15$            &	    $+0.16\pm 0.07$  		      \\			      
NGC\,2110    & $1.77\pm 0.21$  & $8.59\pm 0.07$   & $7.72\pm 0.13$  & 1.00   &	$8.65\pm 0.06$  &     $-0.69\pm 0.17$            &	 $-0.13\pm 0.09$ 		      \\			      
NGC\,5929    & $2.25\pm 0.31$  & $	8.62\pm 0.11$ &	 $7.08\pm 0.13$ 		   & $1.08\pm 0.01$    &	$8.67\pm 0.10$ 		 &     $-1.14\pm 0.22$            &	  $-0.43\pm 0.16$		      \\			      
Mrk\,463E    & $1.33\pm 0.12$  & $8.04\pm 0.06$ & $8.02\pm 0.12$  & $1.13\pm 0.02$    &	$8.39\pm 0.08$ 	&     $-0.71 \pm 0.07$            &	 $-0.37\pm 0.05$		      \\			      
Mrk\,622     & $1.04\pm 0.07$  & $8.95\pm 0.08$ & $8.20\pm 0.11$  & $1.46\pm 0.07$  &$9.18\pm 0.09$  &     $-0.53\pm 0.07$           &$-0.32\pm 0.06$ 		      \\			      
NGC\,1386    & $1.57\pm 0.17$  & $8.17\pm 0.05$ & $8.00\pm 0.13$  & 1.00   &	$8.40\pm 0.06$  &     $-0.62\pm 0.10$            &	$-0.16\pm 0.05$		      \\			      
NGC\,7582    & $1.31\pm 0.12$  & $7.83\pm 0.06$ & $7.50\pm 0.12$  & $1.11 \pm 0.02$   & $8.04\pm 0.07$ &     $-0.05\pm 0.08$ & $+0.28\pm 0.06$    \\	

NGC\,1275    & $2.05\pm 0.27$  & 	$8.55\pm 0.10$  & $7.81\pm 0.13$  & $1.27\pm 0.04$   & $8.74\pm 0.07$ &     $-0.41\pm 0.20$   &	$+0.25\pm 0.12$ \\			      
Circinus     & $1.60\pm 0.17$  & $7.88\pm 0.05$ & $7.93\pm 0.13$  & $1.65\pm 0.09$   &	$8.43\pm 0.07$  &     $-0.41\pm 0.09$ & $+0.07\pm 0.05$   \\	

Centaurus\,A & $1.66\pm 0.19$  & $	8.49\pm 0.06$  		    &	$  	7.69\pm  0.13$ 	   & $1.11\pm 0.02$  &	$8.61\pm 0.05$ 		 &     $-0.58\pm 0.15$            &	$-0.07\pm 0.08$   \\			      
Cygnus\,A    & $1.47 \pm 0.15$ & $8.43 \pm 0.05$  & $8.11\pm  0.13$ & $1.45\pm 0.06$    & $8.77\pm 0.06$  &     $-0.64\pm 0.10$ &	$-0.22\pm 0.06$     \\			      
Mrk\,266SW   & $1.46\pm 0.13$  & $	8.48\pm 0.05$  		    & $7.69\pm  0.13$ 		   & $1.23\pm 0.03$    & $8.64\pm 0.05$  &     $-0.32\pm 0.11$            &	 $+0.08\pm 0.07$ \\			      
Mrk\,1066    & $1.55\pm 0.17$  & 	$8.26\pm 0.05$  		    &	 $ 	7.57\pm  0.13$		   & $1.22\pm 0.03$    &	$8.43\pm 0.05$ 		 &     $-0.89\pm 0.13$            &$-0.43\pm 0.07$    \\			      
NGC\,1320    & $2.17 \pm 0.29$ & $	7.66\pm 0.11$  		    &	 $ 	7.65\pm  0.13$		   & $1.41 \pm 0.06$    &$	8.12\pm 0.05$ 		 &       ---           &	   ---  		      \\			      
NGC\,1667    & $2.19\pm 0.30$  & $8.98\pm 0.11$    &	$ 	7.72\pm  0.13$ 	 & 1.00   &	$9.01\pm 0.09$ 		 &     $-1.04\pm 0.22$& $-0.34\pm 0.15$    \\				   
NGC\,3393    & $1.19\pm 0.09$  & $8.06\pm 0.07$  & $	8.29\pm 0.12$  & 1.00   &	$8.49\pm 0.10$ 		 &       ---           &	   ---  		      \\			      
NGC\,5953    & $1.78\pm 0.21$  & $8.16 \pm 0.07$ 	&	 $7.50\pm  0.13$		   & 1.00   &	$8.25\pm 0.05$ 		 &      $+0.03\pm 0.16$            &	$+0.59 \pm 0.08$  		      \\			      
NGC\,7682    & $1.72\pm 0.20$  & $8.69 \pm 0.06$ 		    &	  $8.45\pm  0.13$		   & 1.00   &	$8.90\pm 0.05$ 		 &     $-0.84\pm 0.12$            &	  $-0.30\pm 0.06$  		      \\			      
ESO428$-$G014& $1.64\pm 0.18$  & $8.20\pm 0.06$  & $8.04\pm  0.13$ 		   & $1.16\pm 0.02$    &	$8.50\pm 0.06$ 		 &       ---           &	   ---  		      \\			      
\hline
\end{tabular}%
}
\begin{minipage}{1.0\linewidth}
{~~~~Note: ICF(O) is assumed to be equal to 1.00 where the \ion{He}{i} $\lambda 5846$ {\AA} emission line was not presented in the original work.}
\end{minipage}
\end{table*}


\bsp	
\label{lastpage}
\end{document}